\documentclass[arguments]{aastex631}
\shorttitle{AASTeX v6.31 RS Oph with NICER}
\graphicspath{{./}{figures/}}
\usepackage{xcolor}
\begin{document}

\title{The RS Oph outburst of 2021 monitored in X-rays with NICER}

\correspondingauthor{Marina Orio}
\email{orio@astro.wisc.edu}
\author[0000-0003-1563-9803]{Marina Orio}
\affiliation{Department of Astronomy, University of Wisconsin 
475 N. Charter Str., Madison, WI, USA}
\affiliation{INAF-Padova, vicolo Osservatorio 5,
35122 Padova, Italy.}
\author{Keith Gendreau}
\affil{Center for Exploration and Space Studies (CRESST), NASA/GSFC, Greenbelt, MD 20771, USA}
\affil{NASA Goddard Space Flight Center, Greenbelt, MD 20771, USA}
\author{Morgan Giese}
\affiliation{Department of Astronomy, University of Wisconsin
475 N. Charter Str., Madison, WI, USA}
\author[0000-0002-2647-4373]{Gerardo Juan M. Luna}
\affil{CONICET-Universidad de Buenos Aires, Instituto de Astronom\'ia y F\'isica del Espacio (IAFE), Av. Inte. G\"uiraldes 2620, C1428ZAA, Buenos Aires, Argentina}
\affiliation{Universidad de Buenos Aires, Facultad de Ciencias Exactas y Naturales, Buenos Aires, Argentina.}
\affiliation{Universidad Nacional de Hurlingham, Av. Gdor. Vergara 2222, Villa Tesei, Buenos Aires, Argentina}
\author{Jozef Magdolen}
\affil{ Faculty of Materials Science and Technology in Trnava, Slovak University of Technology in Bratislava, Bottova 25, 917 24 Trnava, Slovakia}
\author[0000-0001-7681-5845]{Tod E. Strohmayer}
\affil{Astrophysics Science Division and Joint Space-Science Institute, NASA Goddard Space Flight Center, Greenbelt, MD 20771, USA}
\author{Andy E. Zhang}
\affiliation{Department of Astronomy, University of Wisconsin
475 N. Charter Str., Madison, WI, USA}
\author[0000-0002-3422-0074]{Diego Altamirano}
\affiliation{School of Physics and Astronomy, University of Southampton, Southampton, Hampshire SO17 1BJ, UK}
\author{Andrej Dobrotka}
\affiliation{Advanced Technologies Research Institute, Faculty of Materials Science and Technology in Trnava, Slovak University of Technology in Bratislava, Bottova 25, 917 24 Trnava, Slovakia}
\author[0000-0003-1244-3100]{Teruaki Enoto}
\affiliation{RIKEN Cluster for Pioneering Research, 2-1 Hirosawa, Wako, Saitama 351-0198, Japan}
\author[0000-0001-7828-7708]{Elizabeth C.~Ferrara}
\affil{NASA Goddard Space Flight Center, Greenbelt, MD 20771, USA}
\affil{Department of Astronomy, University of Maryland, College Park, MD, 20742, USA}
\author{Richard Ignace}
\affiliation{Physics \& Astronomy, East Tennessee State University, Johnson City, TN 37615, USA}
\author{Sebastian Heinz}
\affiliation{Department of Astronomy, University of Wisconsin
475 N. Charter Str., Madison, WI, USA}
\author{Craig Markwardt}
\affil{X-ray Astrophysics Laboratory, NASA Goddard Space Flight Center, Greenbelt, MD 20771, USA}
\author[0000-0003-3298-7455]{Joy S.\ Nichols}
\affiliation{Harvard \& Smithsonian Center for Astrophysics, 60 Garden
  Street, Cambridge, MA 02138, USA }
\author{Michael L. Parker}
\affil{Institute of Astronomy, Madingley Road, Cambridge, CB3 0HA, UK)}
\author{Dheeraj R. Pasham}
\affiliation{Kavli Institute for Astrophysics and Space Research, Massachusetts Institute of Technology}
\author{Songpeng Pei}
\affil{Liupanshui Normal University, No.19,Yucai Lane, Minghu Rd,Zhongshan District, Liupanshui, Guizhou Province, People's Republic of China}
\author{Pragati Pradhan, Ron Remillard}
\affil{MIT Kavli Institute for Astrophysics and Space Research, Cambridge, MA 02139, USA}
\author{James F. Steiner}
\affiliation{Center for Astrophysics | Harvard \& Smithsonian, 60 Garden Street, Cambridge, MA 02138, USA}
\author{Francesco Tombesi}
\affil{Department of Astronomy, University of Maryland, College Park, MD, 20742, USA}
\affil{NASA Goddard Space Flight Center, Greenbelt, MD 20771, USA}
\affil{Department of Physics, Tor Vergata University of Rome, Via della Ricerca Scientifica 1, 00133 Rome, Italy}
%
\begin{abstract}
The 2021 outburst of the symbiotic recurrent nova RS Oph was monitored
 with the Neutron Star Interior Composition Explorer Mission ({\sl NICER})
 in the 0.2-12 keV range from day one after the optical maximum,
 until day 88, producing an unprecedented,
detailed view of the outburst development.
 The X-ray flux preceding the supersoft 
 X-ray phase peaked almost 5 days after optical maximum
 and originated only in shocked ejecta for 21 to 25 days. 
 The emission was thermal; in the first 5 days 
 only a non-collisional-ionization equilibrium 
 model fits the spectrum, and a transition to
 equilibrium occurred  between days 6 and 12.
 The ratio of peak  X-rays flux measured  in the {\sl  NICER} range 
 to that measured with {\sl Fermi} in the 60 MeV-500 GeV range
 was about 0.1, and the ratio to the peak flux measured with H.E.S.S.
 in the 250 GeV-2.5 TeV range was about 100.  
 The central supersoft X-ray source (SSS), namely the 
 shell hydrogen burning white dwarf (WD), became visible
 in the fourth week, initially with short flares. 
 A huge increase in flux occurred on day 41,
 but  the SSS flux remained variable. 
 A quasi-periodic oscillation  every $\simeq$35 s 
 was always observed during the SSS 
 phase, with variations in amplitude and a period drift
 that appeared to decrease in the end.  
 The SSS has characteristics of a WD of mass $>$1 M$_\odot$. 
 Thermonuclear burning switched off shortly after day 75, earlier
 than in 2006 outburst. We discuss implications for the
 nova physics. 
\end{abstract}


\section{Introduction \label{sec:intro}}
RS Oph is arguably the best known recurrent symbiotic nova. Classical
 and recurrent novae are binary systems hosting a WD, and their outbursts
 are attributed to a thermonuclear runaway (TNR) on the surface of the
 WD, that is accreting material from its binary companion. The model
 predict that TNR is 
 usually followed by a radiation driven wind, that is mainly
 responsible for depleting the
 accreted envelope \citep{Starrfield2012,
 Wolf2013}. The designation {\it recurrent} implies that the
 outburst {has been observed repeatedly over intervals shorter than 100
years, 
 although all novae ({\it classical} included) are thought to be
 recurrent, on longer, secular time scales that can greatly vary, depending
 on the mass accretion rate and the WD mass. The more massive the WD
 is, the smaller its radius, so the accumulated material is more
 degenerate and is ignited with lower accreted mass \citep{Yaron2005,
 Starrfield2012, Wolf2013}. Thus, the frequently
 erupting recurrent novae host rather
 massive WDs. RS Oph is also a {\it symbiotic} system, that is a system
 with a giant companion, specifically a M0-2 III mass donor
\citep{Dobrzycka1996, Anupama1999} in a
binary with a 453.6 day orbital period \citep{Brandi2009}.
\citet{Brandi2009, Miko2017} have studied optical
 and UV spectra of RS Oph in outburst and at quiescence, presenting
 compelling evidence that the WD is  very massive, in the 1.2-1.4 M$_\odot$
 range, and it is made of carbon and oxygen (CO WD). The effective temperature
 estimated in the supersoft X-ray phase by \citet{Nelson2008}
 following the previous eruption in 2006
 was about 800,000 K, which is indicative of a mass
 of at least 1.2 M$_\odot$.
 This implies that it must have grown in mass
 and not have ejected all the accreted material, since the largest
 mass of newly formed  CO WDs is below 1.2 M$_\odot$ even
 for very low metallicity M$_\odot$ \citep{Meng2008}. 
 This has spurred much interest in the possibility that RS Oph
 is a type Ia supernova progenitor. 
 
   RS Oph was observed in outburst in 1898, 1933, 1958, 1967, 1985, and 2006. 
  At least two outbursts may have been missed in  1907 and 1945 when
  RS Oph was aligned with the Sun \citep{Schaefer2004}.
\citet{Schaefer2004} reported a dimming magnitude in plates of the year
1907, just after the end of the seasonal observing
gap, and attributed it to a post-outburst dip observed
in other events.

  Novae are luminous at all wavelengths from
 gamma rays to radio, and X-rays have proven to be a very important 
 window to understanding their physics since the '80ies \citep[after
 the initial discovery by][]{Ogelman1984}.
Early in the nova outburst, the X-rays are attributed to 
powerful shocks in the outflow. The X-ray grating
spectra have been successfully modeled as thermal plasma in collisional 
ionization equilibrium \citep[see the discussions by][]{Orio2012, Peretz2016,
 Drake2016, Orio2020,
Chomiuk2021}. In most novae, the X-ray luminosity in the 0.2-10 keV range peaks
 at 10$^{34}$ erg s$^{-1}$, but in symbiotic novae, it is even a factor of 
100 larger, a fact that has been attributed to the 
 collision of the nova ejecta with the circumstellar red giant wind.
 The shocks are so powerful that they often cause secondary gamma ray emission
\citep{Franck2018}, either by a leptonic mechanism (inverse Compton
effect) or by a hadronic mechanism  (caused by the acceleration of protons). 
 In the 2006 outburst, the initial X-ray luminosity of RS 
Oph was close to 10$^{36}$ erg s$^{-1}$.

Later in the outburst, novae become even much more X-ray luminous, emitting X-rays
in the supersoft range  below 0.8 keV
 (see review by Orio 2012). In fact, 
 after the thermonuclear flash, the WD atmosphere contracts and returns
 almost to pre-outburst radius. The peak wavelength of the emission
 moves from the optical range to the UV and extreme UV and finally,
 to the soft X-rays within weeks, in a phase of constant
 bolometric luminosity, still powered by shell burning. The 
 central source appears as a supersoft X-ray source, with peak temperature
 up to a million K, for a time lasting from days to
 a few years \citep{Orio2012}.
 The first outburst that could be observed in X-rays was the one 
 of 2006, monitored with {\sl Swift} and RXTE 
\citep{Bode2006, Hachisu2007, Osborne2011,Sokoloski2006},
 and also observed with high spectral resolution with the gratings of
 {\sl Chandra} and {\sl XMM-Newton} \citep{Ness2007,
 Nelson2008, Drake2009, Ness2009}. 

 The distance estimates for RS Oph are converging around a value
 of 2.4-2.6 kpc: from  the expansion velocity and resolved radio imaging of
 2006 \citet{Rupen2008} derived 2.45$\pm$0.37 pc. The  GAIA DR3 distance
 is 2.44$^{+0.08}_{-0.16}$ kpc (geometric) and 2.44$^{+0.21}_{-0.22}$
 kpc (photogeometric) \citep[see][]{Bailer-Jones2021} and assuming that the giant fills its Roche Lobe, using the
orbital parameters \citet{Brandi2009}, the resulting distance is
3.1+/-0.5 kpc \citep{Barry2008}. The uncertainty on the GAIA distance 
 may be affected by larger statistical uncertainty than estimated, because of 
the surrounding nebula  and the wobble of the long binary period,  
 but with the well determined orbital parameters by \citet{Brandi2009}
 it seems that historical estimates around 1.6 kpc, based on the 
intervening neutral hydrogen column density to the source
\citep{Bode1987, Hjellming1986}, are now obsolete.
 Even with the lower distance estimate, 
 RS Oph would still be the most intrinsically X-ray luminous nova
 so far observed.

 RS Oph appeared in outburst again on  August 9 2021 
at 09.542 UT
 (as announced in \url{http://ooruri.kusastro.kyoto-u.ac.jp/mailarchive/vsnet-alert/26131}) and \url{http://www.cbat.eps.harvard.edu/iau/cbet/005000/CBET005013.txt} at visual magnitude 4.8. 
 Immediately afterwards, the nova was also detected
in hard X-rays with MAXI \citep{Shidatsu2021}, INTEGRAL \citep{Ferrigno2021}
 and {\sl Swift}-BAT \citep[see][]{Page2022}, and at gamma ray energy with 
 the Fermi-LAT \citep{Cheung2022}, H.E.S.S. \citep{HESS2022}, and
 MAGIC \citep{Acciari2022}.   Also in gamma rays, like in the X-ray range,
 RS Oph is the most luminous nova so far observed.
The highest energy range was that of H.E.S.S. and MAGIC,
 from 10 GeV to tens of TeV.
 The flux peaked only in the lower range of tens of GeV these instruments
 and the spectrum was fitted with a power law index $>$3,
 significantly higher than the 1.9 power law index
 in the spectrum measured with the Fermi-LAT in the 100 MeV-13 GeV range
\citep[]{Cheung2022}. 

 The 2021 early optical spectrum was described by \citet{Munari2021}
 as of ``He/N'' type, with strong Balmer, He I and N II lines. The
 full width at half maximum of the emission lines was 2900 km s$^{-1}$ and blue 
 shifted P-Cyg components of the lines appeared and disappeared within a 
few days \citep{Miko2021, Munari2021b}.
 Acceleration to up to $\simeq$4700 km s$^{-1}$ was observed after
 the first two days, and P-Cyg profiles appeared also in lines of Fe II, O I, 
 and Mg II \citep{Miko2021, Pandey2022}.
 A narrow emission component disappeared within the first few days,
 while
 a narrow absorption component persisted for longer \citep{Luna2021, Shore2021},
 and altogether
 the velocity of the lines indicated deceleration \citep{Munari2021b}
 a few days after the initial acceleration.
 Intrinsic linear optical polarization was observed $\sim$1.9 days after outburst
 \citep{Nikolov2021} while satellite components appeared in the optical spectra after two weeks
 in H $\alpha$ and H $\beta$, suggesting a bipolar outflow as observed
 in the radio in 2006 \citep{Rupen2008}. The bipolar
 outflow was confirmed at radio wavelength
  by \citet{Munari2022}, who found the leading lobes to be expanding
at the very high velocity of 7550 km s$^{-1}$.
High ionization lines
 appeared around day 18 of the outburst \citep{Shore2021}. A summary
 and a visual illustration of  the optical spectral changes in the  
 first 3 weeks after maximum can be found in \citet{Munari2021c}.

 The AAVSO
 optical light curve of RS Oph in different bands, from B to I, in 2021
appeared
 extremely similar to the AAVSO 2006 light curve, with no 
 significant differences \citep[see also][Fig. 2]{Page2022}. 
Here, we will assume the same time for the optical maximum as 
 in \citet{Page2022}, namely JD 2459435.042 (2021 August 9.542), 
 although the visual AAVSO optical lightcurve 
 shows a plateau that lasted for almost all the following day.
 The maximum magnitude was V=4.8, the time for a decay by 2 magnitudes t$_2$
 was 7 days and the time for a decay by 3 magnitudes t$_3$ was 14 days. 
 All the subsequent evolution was smooth, and in the last optical observations
 on November 14 2021 the nova was at V$\simeq$11.2, like in 2006 at the
 same post-outburst epoch. \citet{Page2022} already showed, however,
 that there are substantial differences in the X-ray light curves in
 the 0.3-10 keV range.

 In this article, we describe the evolution of the outburst in
 the 0.2-12 keV X-ray band of the Neutron Star Interior Composition Explorer camera
 ({\sl NICER}) from the second post-outburst day until it was too close
 to the Sun in November of 2021. The nova was then re-observed again once in
 2021 February, when it was returning to quiescence.  
 Section 2 describes the data and
 Section 3 the general development of the light curve and spectrum
 we observed,
 including a comparison with the lightcurve measured in 2006 with the
 {\sl Swift} X-ray telescope (XRT). In Section 4 we analyse in detail the 
 X-ray emission in the first month, when the X-ray flux was mainly due to 
 shocks in the nova outflowing material.
 In Section 5 we examine a phase of transition, in which the WD was emerging as a luminous supersoft X-ray source, but also the shocked material was
 emitting at softer and softer energy, causing superposition of
 the two X-ray spectra.  Section 6 describes the period of maximum
 X-ray light. Section 7 examines the aperiodic variability of
 the supersoft X-ray source (SSS) and Section 8 is dedicated to the analysis of
 an intriguing quasi-periodic modulation
 with a semi-period of $\approx$35 s. In Section 9 we examine the final decay. Section 10 is dedicated to a discussion of some interesting
 and unusual aspects of our results, and Section 11 to conclusions. 
\section{The NICER monitoring}
We started monitoring RS Oph with
 {\sl NICER} 1.27 days after the optical maximum. 
The {\sl NICER} camera is an external attached payload on the International
 Space Station (ISS).  Although its main task is
to perform a fundamental investigation of the extreme
 physics of neutron stars,  measuring their X-ray pulse
profiles in order to better constrain the neutron star equation
of state, {\sl NICER} is useful for a variety of astrophysical
 targets. The excellent response and calibration in the
 supersoft range is ideal to study the SSS.
 {\sl NICER} provides also unprecedented timing-spectroscopy
capability, with high throughput and low background \citep{Prigozhin2016}.
The instrument is the X-ray Timing Instrument (XTI), 
 designed
to detect the soft X-ray (0.2 - 12 keV) band emission from compact sources
 with  both high resolution timing and spectral information. 
It is a highly modular
collection of X-ray concentrator (XRC) optics, each with an associated
 detector, The XTI
collects cosmic X-rays using grazing-incidence, 
gold-coated aluminum foil optics, 
equipped with 56 pairs of XRC optic modules and a silicon-drift detector for high
timing observations (time-tagging resolution $\leq$ 300 nanoseconds). 

 The data extraction and analysis was performed with the HEASOFT version
 6.29c and its NICERDAS package, with current calibration files.
Frequent interruptions of the {\sl NICER} exposures are due to the obstruction of the Earth or by elements of the International Space Station,
and the maximum uninterrupted 
exposure capacity for {\sl NICER} is limited to $\simeq$1000 s.
 Moreover, during the exposures, space
 weather conditions can also impact the feasibility of the data when there are 
flares in the background, mostly near the South Atlantic Anomaly. 
We excluded high background periods using the  nicer$_{-}$bkg$_{-}$estimator tool,
 which excludes periods of inclement ``space weather''.
 space weather periods, we also used 
In the first analysis in 2021 we also used for some  good 
time intervals  (GTIis) 
 the alternative nibackgen3C50 tool, which offers a different method to
exclude high background. Finally, before submitting this paper we
 repeated the extraction of the first month of observations
 and of selected later dates, using the SCORPEON method to estimate
 the background, accounting for the position in the Galaxy and other factors 
 All tools are described in
\url{https://heasarc.gsfc.nasa.gov/docs/software/lheasoft/help/nicer.html}.

\section{The NICER lightcurve}
Fig. 1 shows the light curve
 in different energy ranges, with a logarithmic y-axis. After 2021 October 15 the light curve could only be extracted
 above 0.3 keV, because of soft flux contamination when
 it was already close to the Sun.
 We summarize significant phases and important post-outburst
 dates in Table 1, for a general outlook at the evolution.
 A section is dedicated to each stage or important phenomenon in
 the observed X-ray evolution.

The  {\sl NICER} count rate and flux in this initial phase during
 the month of 2021 August peaked close to day 5, 
as observed also with {\sl Swift} almost in the same
 energy range \citep{Page2022}. 
 In the gamma ray range, the flux measured with Fermi peaked
 instead after the second day (60 MeV - 500 GeV, with peak flux
 at energy of a few GeV), but the flux measured
 with H.E.S.S. \citep[energy range 250 GeV - 2.5
 TeV, with peak flux at a few hundred  GeV, see][]{HESS2022}
  peaked quite close to the peak in the {\sl NICER} range.

 After the first month, especially since day 37,
 the vast majority of the X-ray flux was measured below 1 keV.
  Fig. 2  shows the comparison
 between the light curve observed with {\sl NICER} and 
 the light curve observed with the {\sl Swift} X-ray telescope (XRT)
 in the previous outburst in 2006.
 The {\sl NICER} background has not been subtracted in
 this lightcurve, but  the signal was extremely high compared
 to it. The energy range in this
 figure is 0.3-10 keV, the same as the XRT,
 although {\sl NICER} is calibrated from 0.2 keV (there was
 little signal in the 10-12 keV range). 
 Unlike the optical light curve, the X-rays reveal substantial differences
 from the 2006 outburst, especially in the luminous SSS phase.
  Fig. 3  shows the comparison of the
 {\sl NICER} 1-10 keV range and 3-10 keV range lightcurve.
 Both lightcurves peaked after 5 days,
 but the count rate above 3 keV decreased  very rapidly 
 and was almost null after day 30, while residual flux
 in the 1-3 keV range was still measured at the end of the monitoring.
\begin{figure}
\begin{center}
\includegraphics[width=140mm]{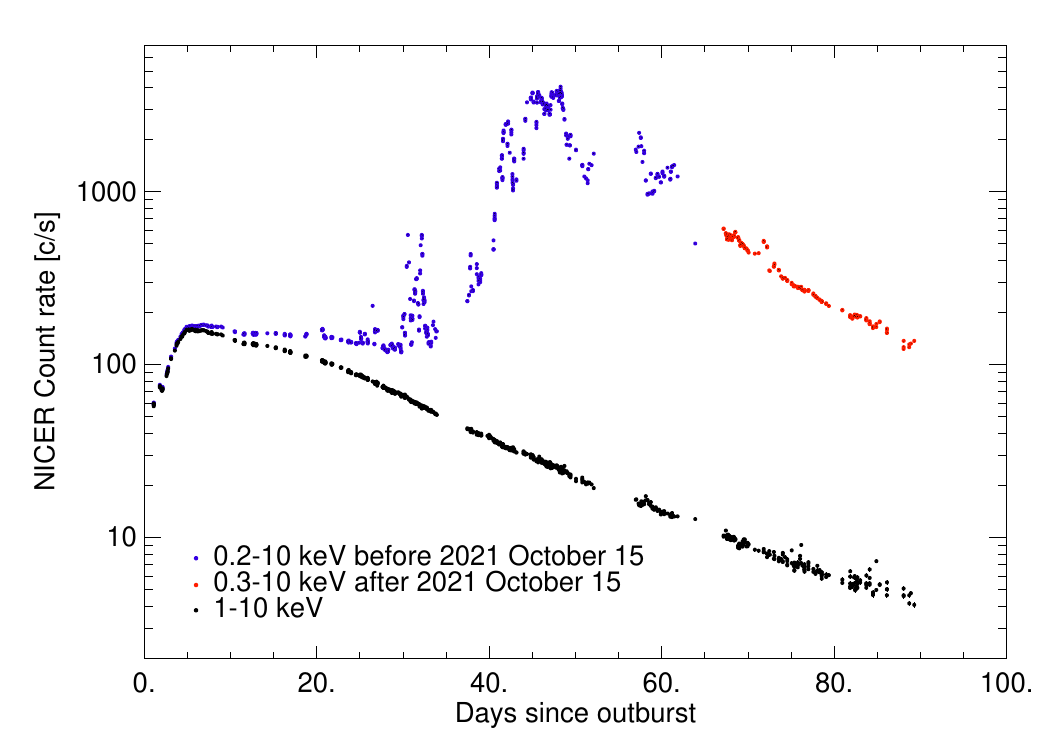}
\end{center}
\caption{
The NICER light curve since optical maximum in the 0.2-10 keV range (black symbols, and red symbols for light curve in the 0.3-10 keV range 
 from 2021 October 15 to avoid contamination from the Sun) and in the 1-10 keV range (blue
 symbols).
}
\label{nicer_comp_lc}
\end{figure}
\begin{figure}
\begin{center}
\includegraphics[width=140mm]{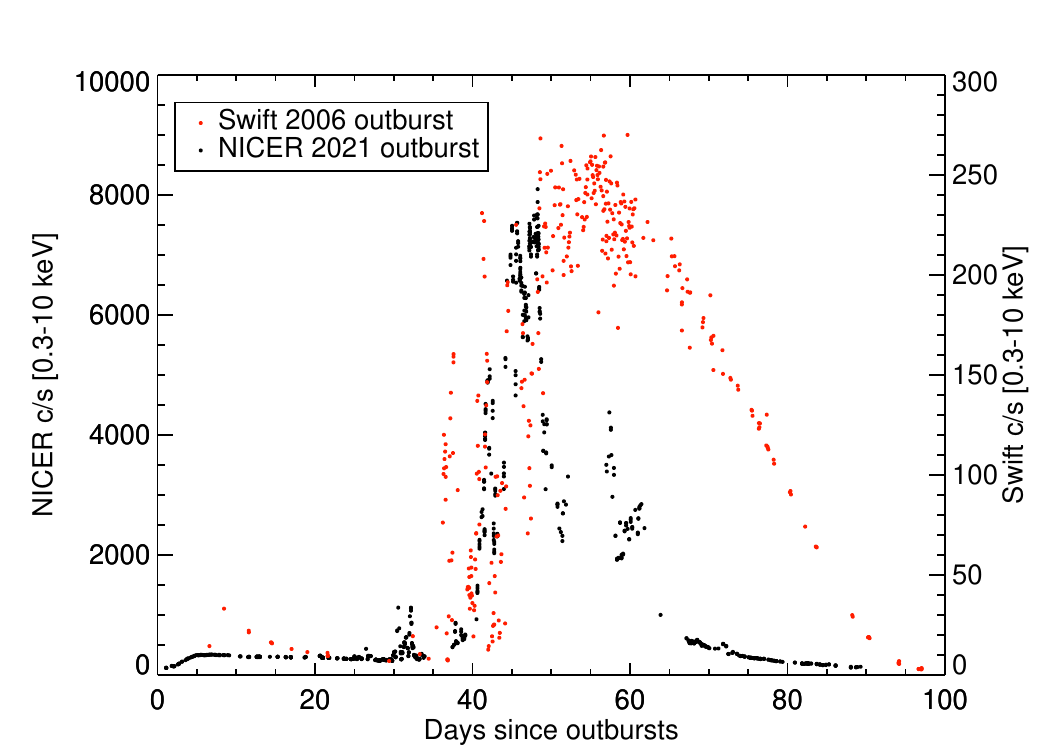}
\caption{NICER 0.3-10 keV light curve during the 2021 outburst (black dots) compared with the {\it Swift} 0.3-10 keV light curve from the previous outburst in 2006 (red dots). The {\it Swift} curve was obtained from the online tool at \url{https://www.swift.ac.uk/user\_objects/} \citep{Evans2007}. We used PIMMS for the count rate comparison.}
\label{swift_nicer_lc}
\end{center}
\end{figure}

During the first 21 days after the outburst, the emission originated in
 shocked plasma that we attribute to the ejecta
 and/or ejecta colliding with the red giant wind. 
 Flares of supersoft flux occurred
 since  day 21 post-maximum, and initially they were only short-lasting.
 The nova became a very luminous, albeit variable, supersoft
 X-ray source after day 37, with a sharp rise on day 40. 
 The flux of the SSS was overwhelming compared
 to that of the shocked material,  but there was always
 significant emission also from shocks
 (the flux up to three orders of magnitude smaller than that
 of the SSS, but definitely non-negligible).
 Although there was a similar beginning of the SSS flux in 2006, 
 Fig. \ref{swift_nicer_lc} shows that in
 in 2006 there was an  extremely luminous flare already on day 35.
\begin{figure}
\begin{center}
\includegraphics[width=140mm]{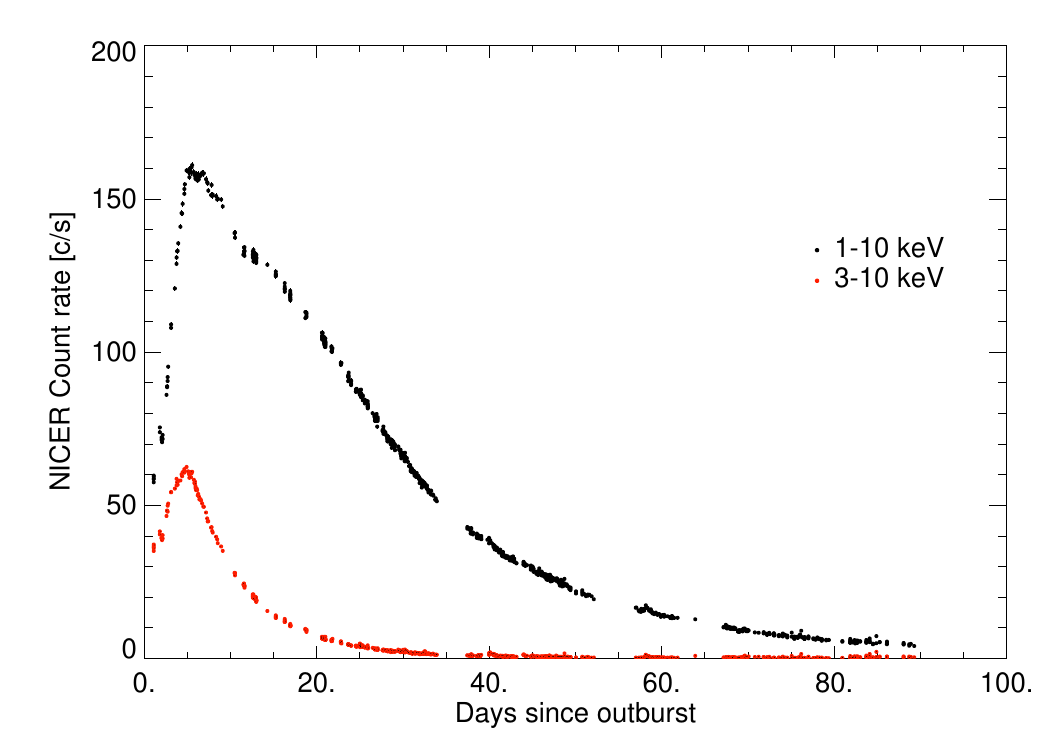}
\end{center}
\caption{
 The 1-10 keV NICER light curve is plotted in black and the 3-10 keV light curve in red.} 
\end{figure}
%
\begin{deluxetable*}{lrl}
\tablecaption{Stages of evolution of the outburst as seen in X-rays}
\tablewidth{700pt}
\tabletypesize{\scriptsize}
\tablehead{
\colhead{ {\bf X-ray phase} } & \colhead{ {\bf When}} & \colhead{ {\bf Characteristics}} 
}
\startdata
\\
  NEI shock phase & 5 $\leq$ t $\leq$ 11 days & Thermal plasma NOT yet in equilibrium  \\
 ``Only shocks'' phase &  day 1 to 21 & Thermal plasma emission, L$_{\rm X,max}=6.8
\times 10^{36}$ erg cm$^{-2}$ s$^{-1}$ \\ 
 Hot thermal plasma    & t $\leq$ 6 days & Initially dominant hot component,
 kT$\leq$ 40 keV  \\
 ``Multi-phase'' thermal plasma & t $\leq$ 6 days  & A second, ``cooler'' component emerges,
  with kT$\leq$ 1 keV  \\ 
 First periodic modulation  &  day 13    & Clear modulation with 66.7 s period 
 lasting  $\geq$ half hour\\
Emergence of very luminous soft flux & day 21-25 & Additional very ``soft'' component emerges \\
 & & Soft flux possibly only in unresolved emission lines \\
 Soft flares  & day 21 to 37 & Soft flux increases with sparse, irregular ``bursts'' \\
 & & Soft flux either in emission lines or continuum \\
 Short period modulation & day 26 to 65 & $\simeq$ 35 s QPO with decreasing ``drift'' \\
 Luminous blackbody/atmospheric emission & day 37 to 65 & SSS emission of central source (continuum)  \\
  Irregular variability & days 40 to 60-65 &  Changing ionization in absorbing ejecta? \\
 Cooling of central source & days 60-90 & Decline to turn-off \\
 Turn-off, end for  ``sun constraint'' & days 70-90 & Shocked  plasma still measurable \\ 
\enddata
\end{deluxetable*}
%
\section{Spectral lines and spectral fits: the initial shocks}
 Fig. 3 illustrates the cooling process of the shocked plasma by 
 comparing the light curves in the 1-10 keV and in the 3-10 keV range.
 The average spectra of each day during the first week are shown in
Fig. 4, while Fig. 5 shows a snapshot of the average spectrum every 4 days for 
 the following 24 days. 
 The panel on the top right in Fig. 4, shows how 
 at the softer energy the flux increased in the first week
 as new emission lines emerged: 
 Mg XI on the second day, Ne X only the fourth.
 The plot on the bottom left, illustrates the flux increase 
 between 2.7 keV and 4.7 keV until the sixth day, particularly on
 the third, followed by cooling only on the 7th day. Cooling is even
 more evident in the bottom right
 panel, zooming on the iron feature of the Fe XXV He-like
 triplet around 6.7 keV. There
 was a a standstill between the third and fifth day,
 followed by flux decrease from day 6. 
 The flux in the unresolved Fe XXV helium-like triplet was
 always much larger than that in the H-like Fe XXVI feature at 6.97 keV. 
 Fig. 5 shows that  cooling continued  
 during the whole second week. A complex of lines in the softest
 range emerged, including unresolved 
 ``soft'' emission lines, which
 we attribute mostly to N VI and N VII ({\sl NICER}'s spectral
resolution in the softest range is not sufficient to resolve  
 lines). 

Spectral fitting was done by us with the XSPEC tool \citep{Arnaud1996}, 
initially with version 12.10 and in the conclusive
 phase checking the results with version 12.13, after the data
 were binned with the GRPPHA tool \citep[see][]{Dorman2001}.
While the general evolution of the X-ray spectrum has also been monitored
 with the {\sl Swift} XRT and the general trends illustrated
 in our Fig. 6 were found also by \citet{Page2022}, with {\sl NICER} we obtained a much more detailed picture than with {\sl Swift},
 measuring several emission lines.
 The {\sl NICER} monitoring was also denser than the {\sl Swift} one, 
with several $\approx$1000 s long GTIs almost every day.
Like in the analysis of the {\sl Swift} XRT spectra \citep{Page2022}, and in that of the
high resolution X-ray spectra of the third week \citep{Orio2022a}, 
we rule out an additional power law  model component 
that could have been produced by non-thermal emission.
 The unusual strength of the He-like 
 lines in a gas that seems so hot that it should be almost
 completely ionized, prompted us to explore 
 the possibility of departure from collisional ionization equilibrium,
 with the VPSHOCK model in XSPEC of 
 parallel shock plasma at constant temperature 
 \citep[initially studied for supernova remnants, see][]{Borkowski2001}. 
We also explored the addition of a partially covering absorber,
 that we added to the models
 to obtain a better fit until day 11. This means
 that we assumed that one
 fraction of the emitting surface had additional column density,
 as may be expected from intrinsic
 absorption near the source, due to a non-spherically symmetric outflow.

Fig. 6 illustrates the evolution
 of the absorbing column density N(H) and of the maximum plasma temperature
 obtained with our spectral fits using
 XSPEC. The apparent discontinuity in the absolute
 flux after day 5 is due to switching to the XSPEC
 BVAPEC model of plasma in collisional ionization
 equilibrium (CIE) \citep{Brickhouse2005}. 
 The change in the spectrum from NEI to CIE was not abrupt; on the same day
 around the transition from one model to another, two different
 models fit almost equally well and this is the case until day 12. 
 However, around day 5.5 the NEI models
 are no longer necessary, because an equally good result is obtained
 with two BVAPEC component and a partially covering
 absorber extended to a fraction that decreased from 60\% to about 40\%.
 Starting on day 8, a second soft component appears necessary in both NEI 
 and CIE fits.    In the  NEI fits, the less hot component turns out to be
 about 0.3 keV, compared to a value of 0.9-1.0 keV resulting in
 the CIE case. The hotter component turns out, in fact,
  to be about 20\% hotter in the NEI case, compensating for this difference.
 The ``global'' column density and that of the partially covering absorber
 are both higher in the NEI model than in
 the CIE one, resulting in a larger absolute flux,
 up to of few times 10$^{-8}$ erg cm$^{-2}$ s$^{-1}$
 and increasing until day 11.  

 In Table 2 and Fig. 7 we show the parameters of different models
 fits to the data
 for a selected GTI of day 7, when also the NEI fit is possible with only
 a ``hot'' component. Both fits have the same statistical probability,
 with a reduced $\chi^2\simeq$1.2. The NEI fit, with higher temperature, better predicts
 the strength of some lines, but we begin to see discrepancy
 at the softest energy. This discrepancy increased in the fit  to the spectra
 of the following $\simeq$24 hours, when the second, soft
 component is needed in both the NEI and CIE fits.
 
 Between day 5.5 and day 11 we obtain fits with 
 similar values of $\chi^2$/(degrees of freedom) - hereafter reduced  $\chi^2$ -
 between 1.0 and 1.4, with both the NEI and CIE model. 
 Guided by the principle of an Occam's razor, in Fig. 6 we present the main
 parameters of the 
 BVAPEC fits from day 5.5, because  they indicate a decreasing absolute flux
 with  the temperature, in contrast with
 the result of the VPSHOCK model, which is 
 less intuitive and implies that the absolute luminosity
 was even above Eddington level during the shock phase.
 The BVAPEC fits also have fewer free parameters.
 Moreover, a puzzling issue of the NEI fits is the low value of
 the electron density and an increasing discrepancy 
 between the value obtained form the ionization time scale
 and the lower limit obtained from the electron density.
All this said, we cannot rule out that the transition
 to equilibrium occurred only around days 11-12.

 It is also interesting that 
 the fraction of the partially covering absorber decreases, and
 on day 12 it is no longer necessary to obtain a good fit.
 On day 18,  {\sl Chandra High Energy Transmission Gratings} spectra, in which
 detailed line ratios were obtained and He-like triplets were
 resolved,  was fitted well with an equilibrium model with two components \citep{Orio2022a}.
 The same model used for the {\sl HETG} fits,  with almost the same parameters,
  the {\sl NICER} spectrum of the GTI closer in time to the {\sl Chandra}
 exposure. 
  
  While the total flux in the 0.2-12 keV band increased until the end
 of the 5th day, the maximum
 plasma temperature peaked already by the third day. 
 However,  due to the absorbing column density (complete and partial),
 the unabsorbed total flux  also seems to have  increased for the first 4
 days (or for even longer assuming a NEI model).  
 We saw in Fig. 4 that  the flux in the prominent Fe XXV
  decreased in intensity only after the fifth day
 (transition from cyan to pink curve in the figure),
 consistently with the maximum temperature returned by the fits.
 In the second and third week,
 the plasma was constantly cooling, with a  rapid
 ``softening'' of the spectrum (Fig. 5). 
\begin{figure}
\begin{center}
\includegraphics[width=87mm]{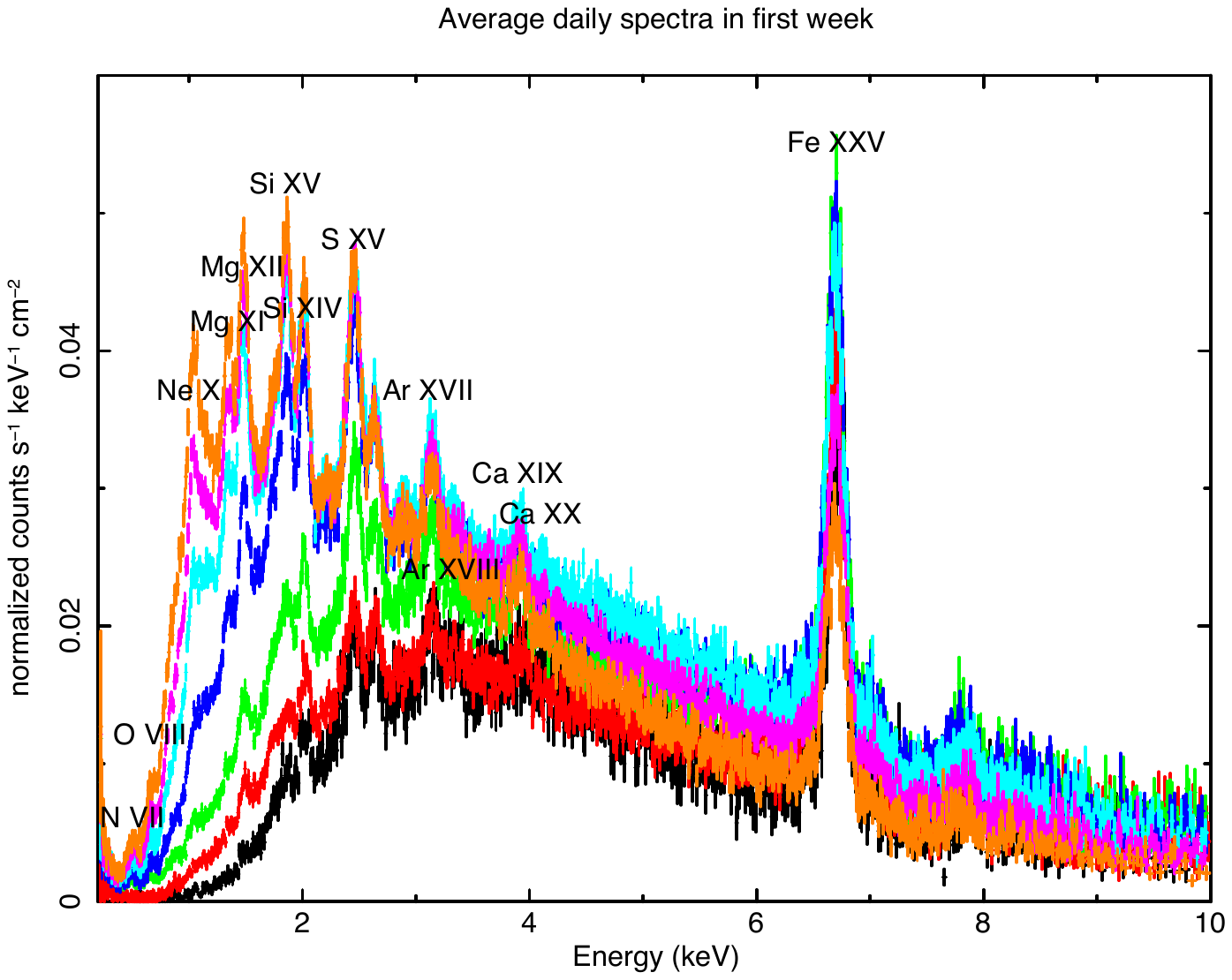}
\includegraphics[width=87mm]{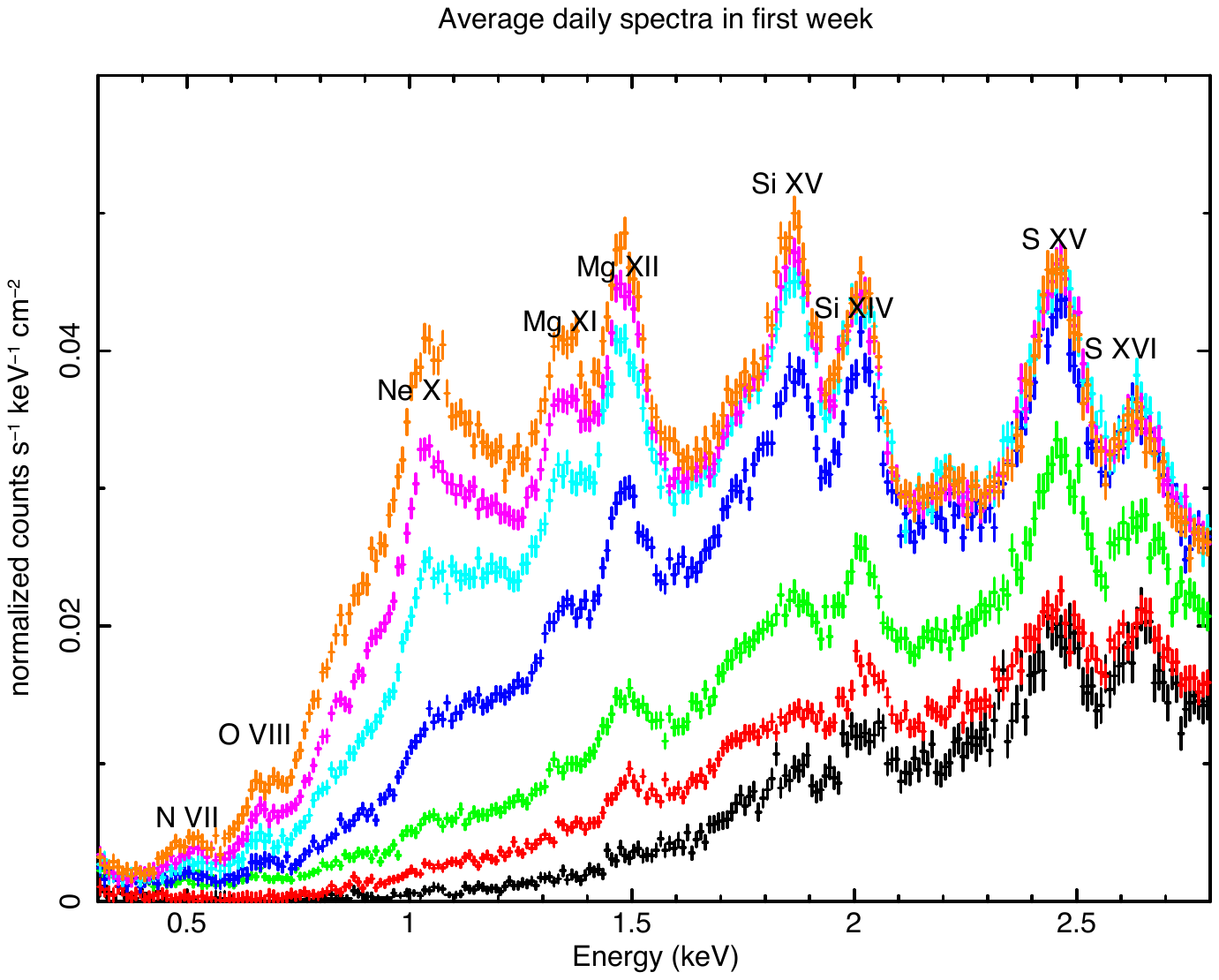}
\includegraphics[width=87mm]{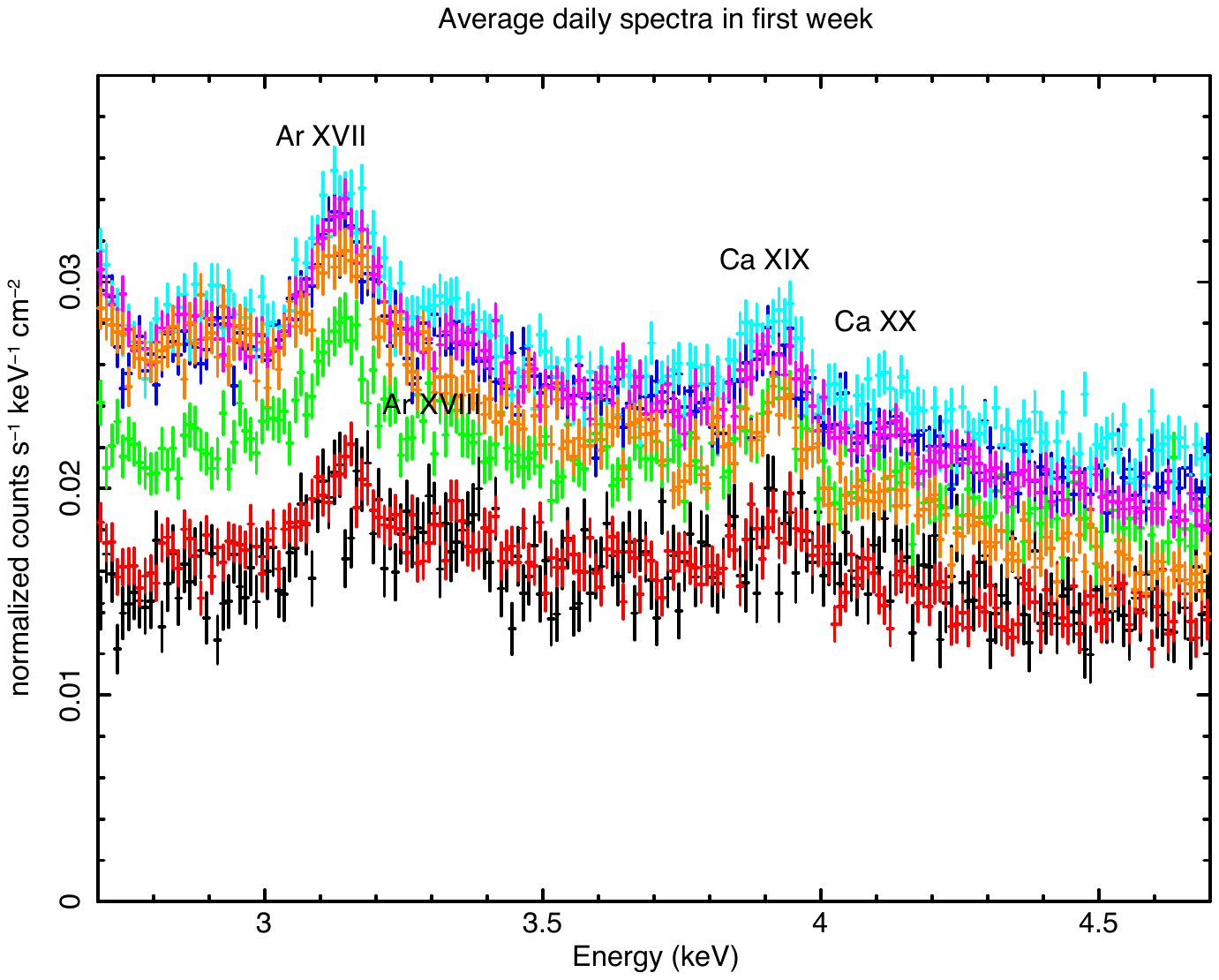}
\includegraphics[width=87mm]{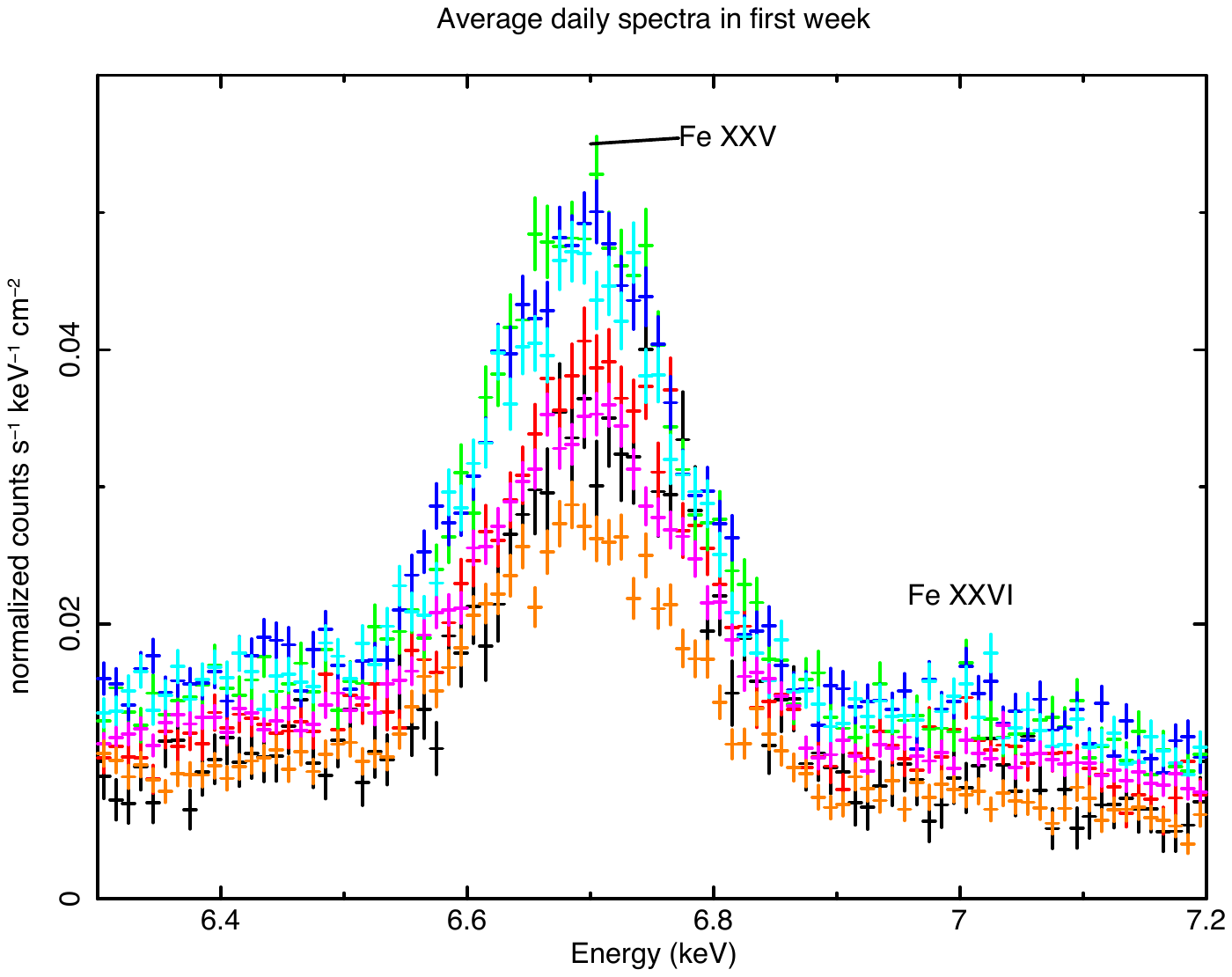}
\end{center}
\caption{The average count-rate spectra observed with NICER 
 from  day 1 to 7.5. The dates are 2021 August 10 (black,
 observation 4202300101),
August 11 (red, obs. 4202300102), August 12 (green, obs. 4202300103),
 August 13 (dark blue,  obs. 4202300104),
August 14 (cyan,  obs. 4202300105), August 15 (pink,
 obs. 4202300106) and August 16 (orange,  obs. 4202300107). 
 The prominent emission lines are marked.}
\label{W1}
\end{figure} 
\begin{figure}
\begin{center}
\includegraphics[width=120mm]{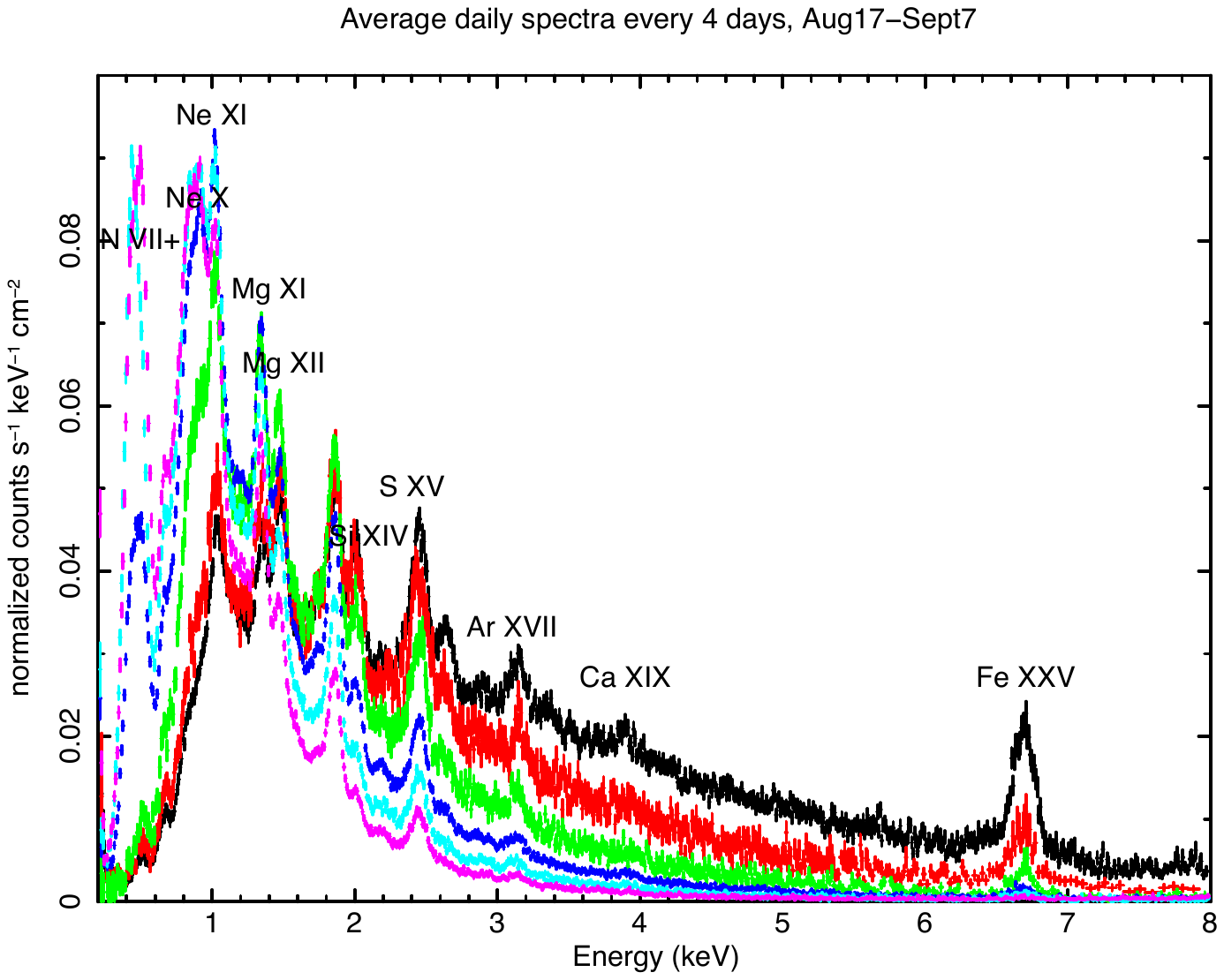}
\end{center}
\caption{The average daily spectra observed with NICER in the
 period between days 8 and 29, plotted every 4 days.  
 The colors sequence in order of date is black (2021
 August 17), red (August 21),
 blue (August 25), green (August 29), cyan (September 3), pink (September 7).}
\label{W2}
\end{figure}
  In Fig. 7 we show the ratio of the emissivities of \ion{Si}{13}/\ion{Si}{14} and \ion{S}{15}/\ion{S}{16} as a function of temperature derived from the ATOMDB \footnote{\url{http://www.atomdb.org/index.php}} database in collisional
 ionization equilibrium (CIE) and
 Non Equilibrium Ionization (NEI) regimes.
 In the case of non-equilibrium plasma (NEI), we assumed
  $\tau=10^{11}$ s $\times$
cm$^{-3}$  (a value closed to those obtained
 in the  fits, see discussion below).
These ratios are more sensitive to temperature in the CIE  
 regime than in the NEI regime. The measurement show that the values 
 returned by fitting Gaussian profiles and subtracting an ``ad hoc'' continuum
 for different maximum temperatures obtained with the fits described
 below, seem to be always above the NEI values for Si and for temperatures
 higher than about 5 keV also for S. 

 While the 
 departures from collisional ionization equilibrium explain 
 the unusual line ratios,
 more than one plasma component at different temperature
 can also interplay to explain this phenomenon.
 From the middle of the sixth day, the temperature is in a range
 such that the spectrum can be fitted in both ways.
 From day 12, we ruled out that the NEI model is adequate, and we conclude that
 most likely, equilibrium had been reached.

The abundances of elements from carbon to iron were
 allowed to vary in these fits,
 but to avoid too many free parameters, we assumed
 that they abundances were the same in the two plasma components.
 Although in both VPSHOCK and BVAPEC,
 we allowed the abundances to deviate from solar, the results are within
 rather large,
 (20-30\%) statistical errors. Only nitrogen results constantly enhanced,
 while iron turns out to be depleted \citep[see][]{Orio2022a}.  
The abundances of elements of atomic number between carbon and calcium,
 except nitrogen and iron, in most fits are somewhat enhanced above solar,
 depending on each element.

There is a caveat in the NEI fits. The 
upper limit to the ionization time scale, a parameter of the fit, 
 is consistently only a few times 10$^{11}$ s cm$^{-3}$. This
 implies
 unusually low electron density, varying from a few 10$^5$ to a few
 10$^6$ cm$^{-3}$
 for the medium in which the hotter component is produced. 
 Table 2 reports also an lower limit, close to 10$^7$ cm$^{-3}$, that
 we would obtain  instead from the emission measure assuming
 that the medium has filled a spherical volume expanding
 at $\simeq$7550 km s$^{-1}$ \citep[from the radio][]{Munari2022} and
 that flux has origin in the whole volume.
 This is consistent with the electron density inferred
 from the material emitting the flux at
 radio wavelengths \citep{Munari2022}.  This limit is
 obtained with the highest velocity measured during the outburst, but  
 with the lower velocity of $\simeq$2800 km s$^{-1}$ 
 inferred from the optical H$\alpha$ line at the beginning
 of the outburst \citep{Fajrin2021}, the resulting lower limit on the electron
 density would be even higher by a factor of 20, so clearly  
 there is a tension between the two values derived from
 the emission measure and from the ionization time scale.

 On day 18 the X-ray flux of RS Oph
 seemed to originate in very dense clumps of matter, with much
 higher electron density than the average \citet{Orio2022a}, like in 
 V959 Mon, \citep{Peretz2016} and
 V3890 Sgr, \citep{Orio2020}. The line ratios diagnostics
 in the X-ray high resolution spectrum  RS Oph on day 18
 reveal that the emission originated in clumps of
 material with electron density 
 possibly as high 10$^{12}$ cm$^{-3}$ \citep{Orio2022a}.
 Typical values of the electron density in the nova ejecta in the first
 days  have 
 been estimated to be even close to  10$^{11}$ cm$^{-3}$  both in
 X-rays \citep[e.g. the recent nova V1674 Her,][]{Sokolovsky2023} and in 
 optical spectra \citep[e.g.][who found
 a value of 7.4 $\times 10^{-11}$ cm$^{-3}$ for V1500 Cyg]{Neff1978}.
 Other published values from optical lines in novae span orders of magnitudes, but
 were mostly obtained at later post-outburst epochs, in the nebular
 phase.  The  electron density distribution was estimated  between 
10$^5$ cm$^{-3}$ and  $\times 10^7$ cm$^{-3}$ for the
 symbiotic nova V 407 Cyg \citep{Shore2012a}; it was
 still between 10$^6$ cm$^{-3}$
 and 4 $\times 10^7$ cm$^{-3}$ on day 186 of the outburst 
   of V5668 Sgr \citep{Shore2018}, and it was 
 a few times $10^7$ cm$^{-3}$ about a month after the outburst for 
 the recurrent nova T Pyx \citep{Shore2012b}. For the latter, 
 we also know it was decreasing on time as t$^{-3}$, so it must have been significantly higher in the early days. 
 Thus the NEI fits parameters, 
 taken at face value, imply that the initial shocks producing
 X-ray flux occur not only in a much lower density medium than the thermal
 plasma we measured at later dates,  but also quite lower than estimated
  in most other novae. 

 Another intriguing issue is  the very high absolute luminosity:
  in the BVAPEC model it is large, reaching a few times
 10$^{37}$ erg s$^{-1}$, but it is predicted to be
 even  super-Eddington in the NEI model. 
 When we are able to fit the spectra with equilibrium
 models and without adding a partially covering absorber, 
 namely around day 12, the resulting unabsorbed flux is much 
 smaller, but it still exceeds 10$^{36}$ erg s$^{-1}$
 for at least another week. The same order of magnitude of absolute flux
 is extrapolated from the fits'
 parameters  of \citet[][see Table 1 on line.]{Page2022}   

 We note that from day 12 our model is
 essentially the same as the one in \citet{Page2022}, with corresponding
 temperatures of the two plasma components, but we 
 obtain the fit
 with only 5.95 $\times$ 10$^{21}$ cm$^{-2}$ already on day 12, less than half
 the value in \citet{Page2022} for the same day. 
 Because {\sl NICER} is calibrated from 0.2 keV, the data presented here
 are more sensitive to the absorption and we suggest that the lower 
 in column density is real, although the decrease may have been less 
 sharp and sudden than what is apparent in the plot.
 We note that
 the  {\sl BVAPEC} model used by us differs from the  {\sl APEC} model used
 by \citet{Page2022} for {\sl Swift}, because it includes abundances, 
  line width and velocity as parameters.
 Although the fits were not very sensitive to the blue shift velocity, 
 a small broadening velocity of $\simeq$500-800 km s$^{-1}$, consistent with the
 precise measurements with the X-ray gratings  \citep[see][]{Orio2022a},
 generally improved the fits. However, the variable
 abundances are probably the main cause of the difference between our fits
 and the {\sl Swift} ones of \citet{Page2022}.
\begin{figure}
\begin{center}
\includegraphics[width=87mm]{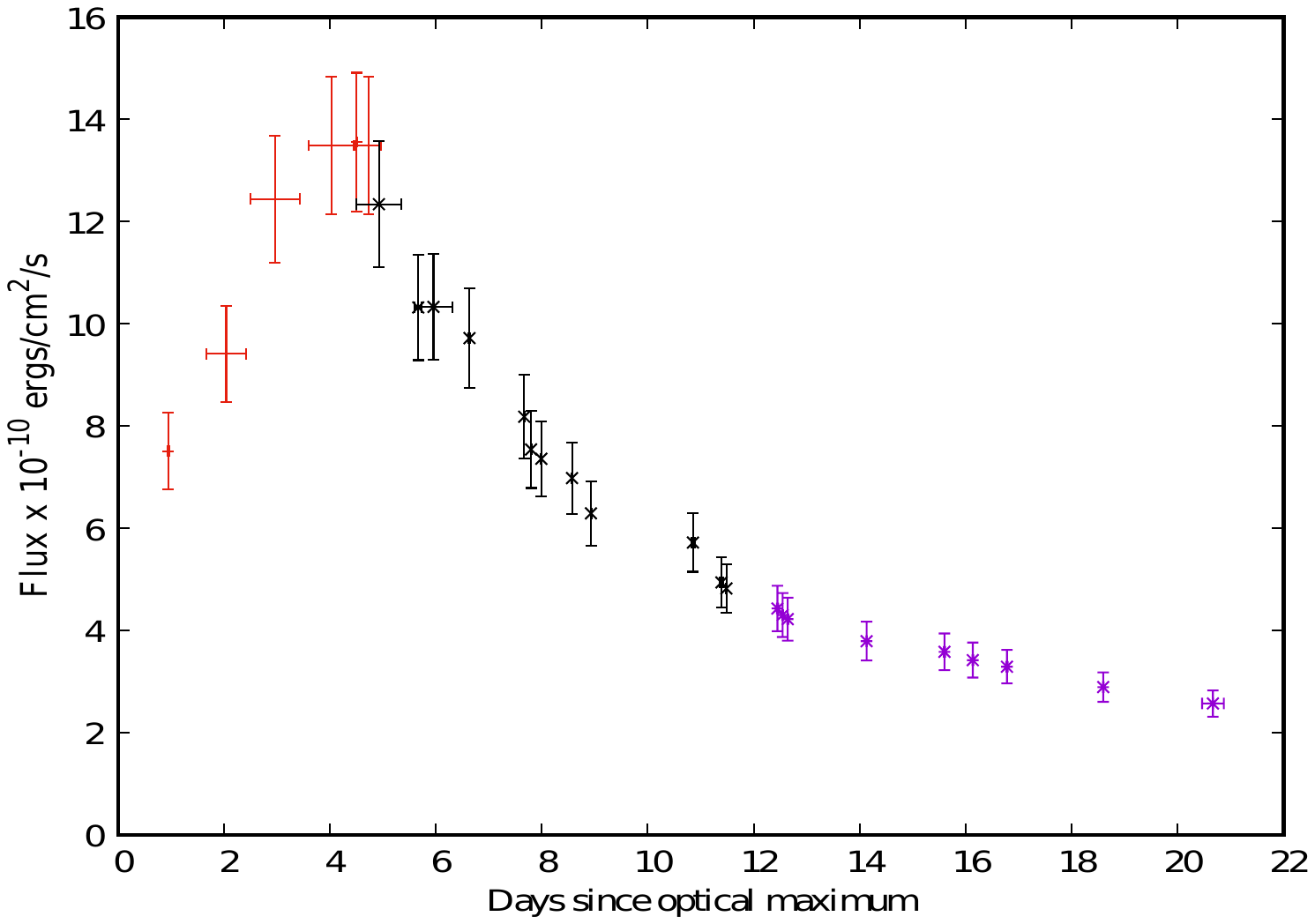}
\includegraphics[width=87mm]{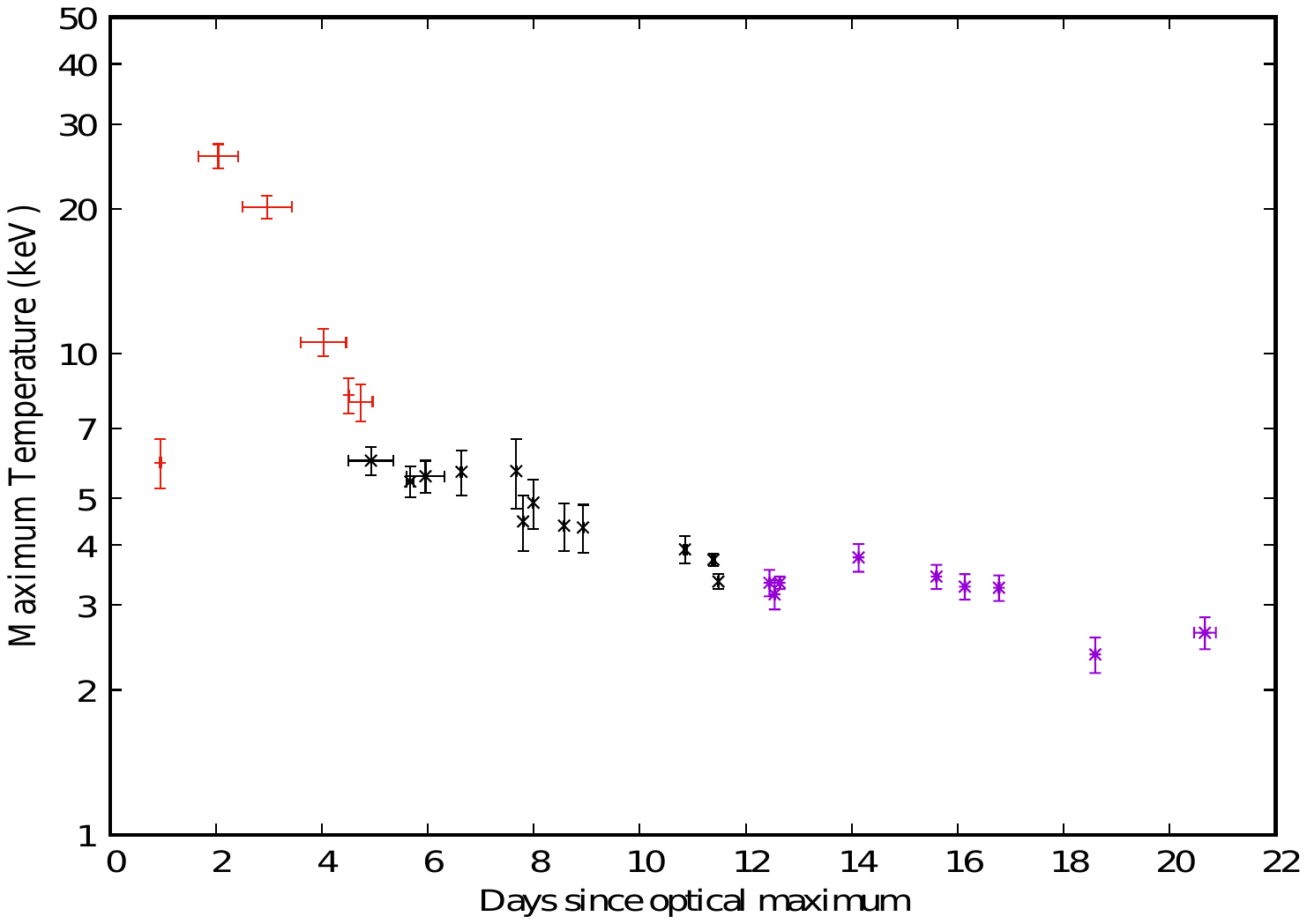}
\includegraphics[width=87mm]{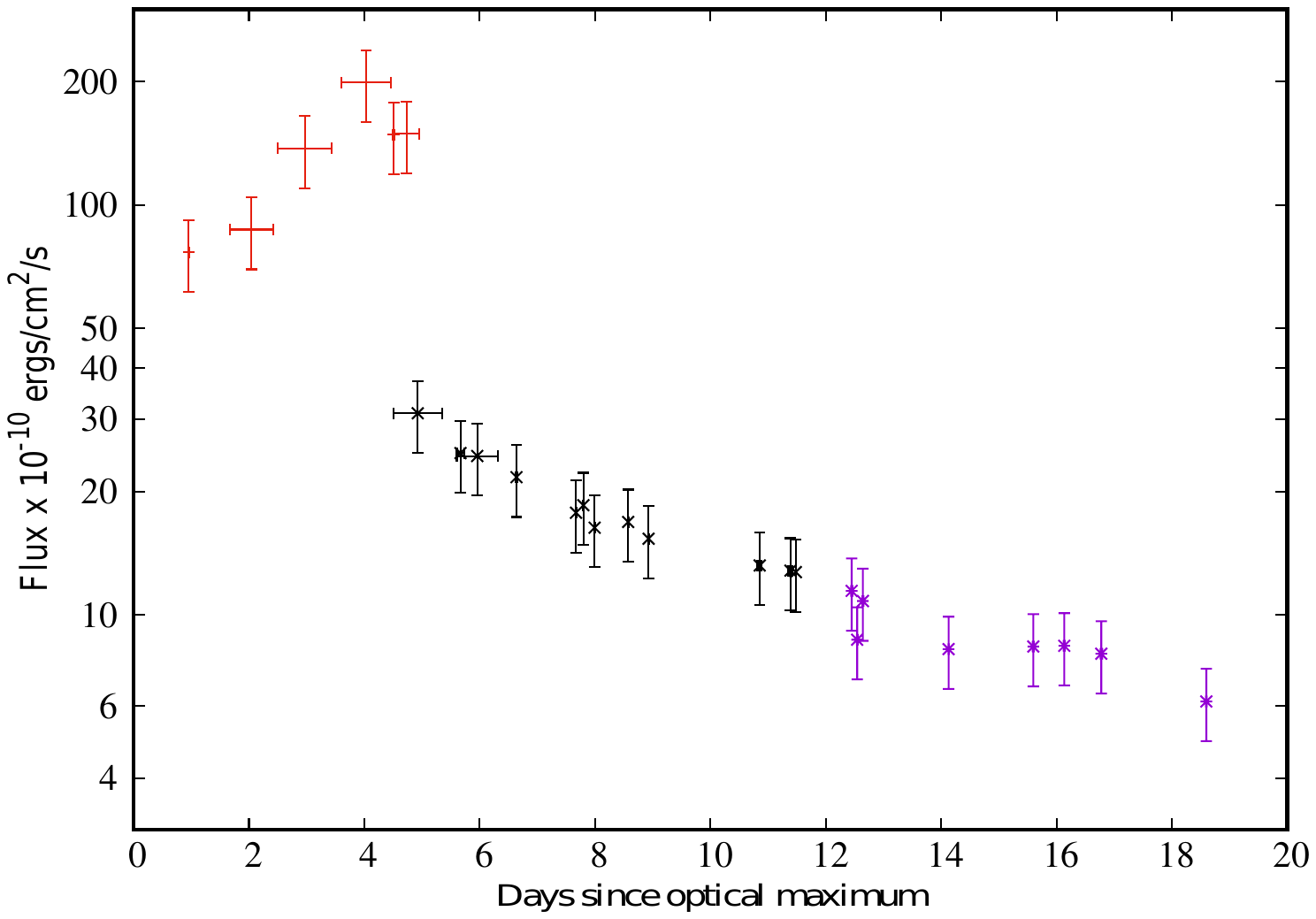}
\includegraphics[width=87mm]{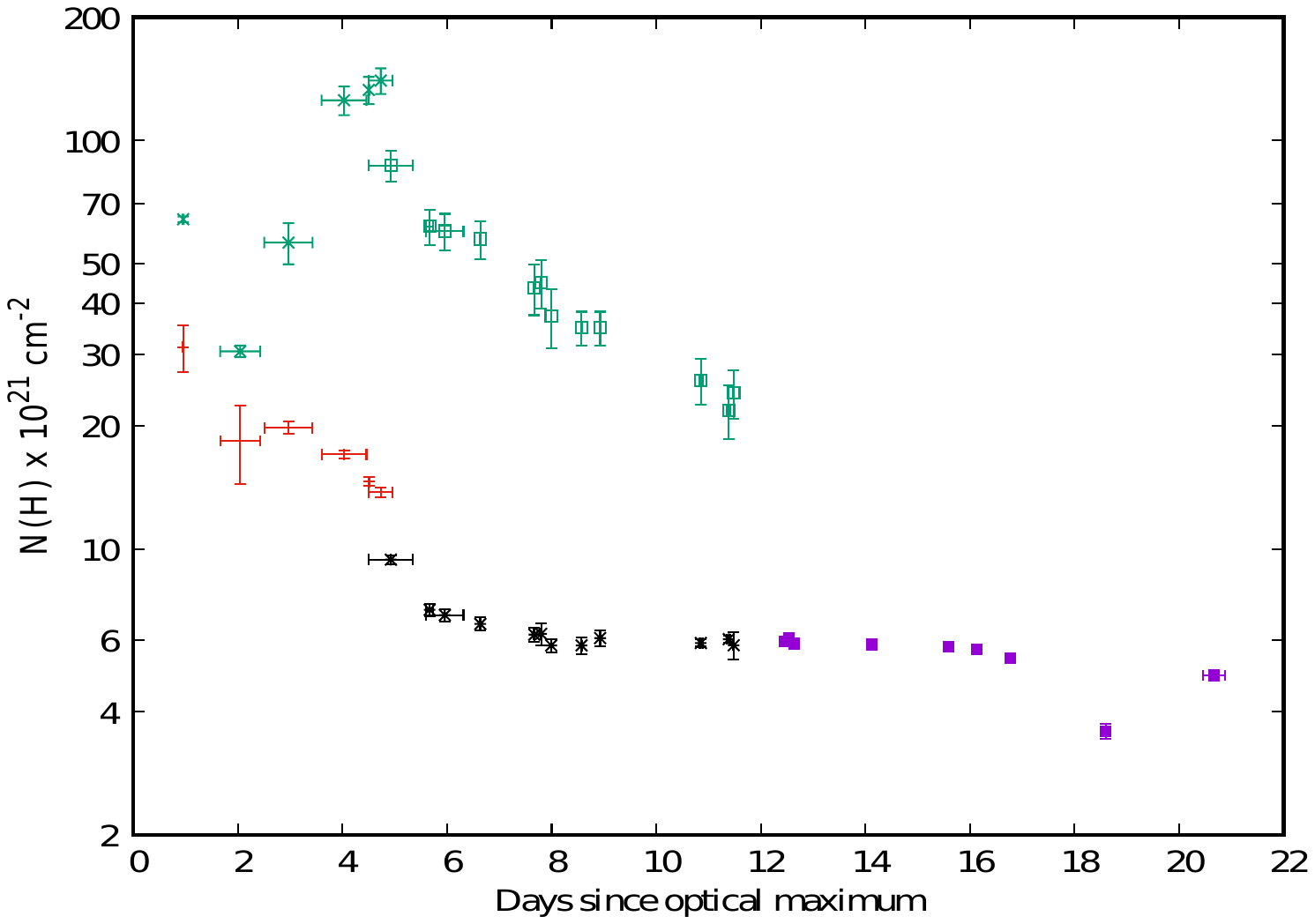}
\end{center}
\caption{For the first 20 days: in the left upper panel, absorbed flux 
derived by model fitting, in the left lower
 panel, absolute flux, in the upper right, 
 plasma temperature of the hotter component (or only component for
 the red points), in the lower right panel, column density
 N(H).  NEI model fits were used in the
 first 5.5 days of observations and
 their parameters are plotted in red. 
  A second component from day 5.5 has a temperature 
  around 0.9 keV in all exposures.
 The N(H) values in green are the ones of the partially covering
 absorber (see text for  covering fraction) and the 2-component
 BVAPEC parameters have been plotted in black if there was the 
 partially covering absorber, in purple if this addition
 was not necessary.
 The y-axis error bars show the largest of the 2$\sigma$ error in the positive
 and negative direction, calculated assuming that the other parameters do not 
 vary. The x-axis error bars show the time interval over which 
 the spectrum was integrated. Individual short GTIs were chosen for
 spectra extraction and fitting when there was significant variability during
 the day.}
\end{figure}
\begin{figure}
\begin{center}
\includegraphics[width=87mm]{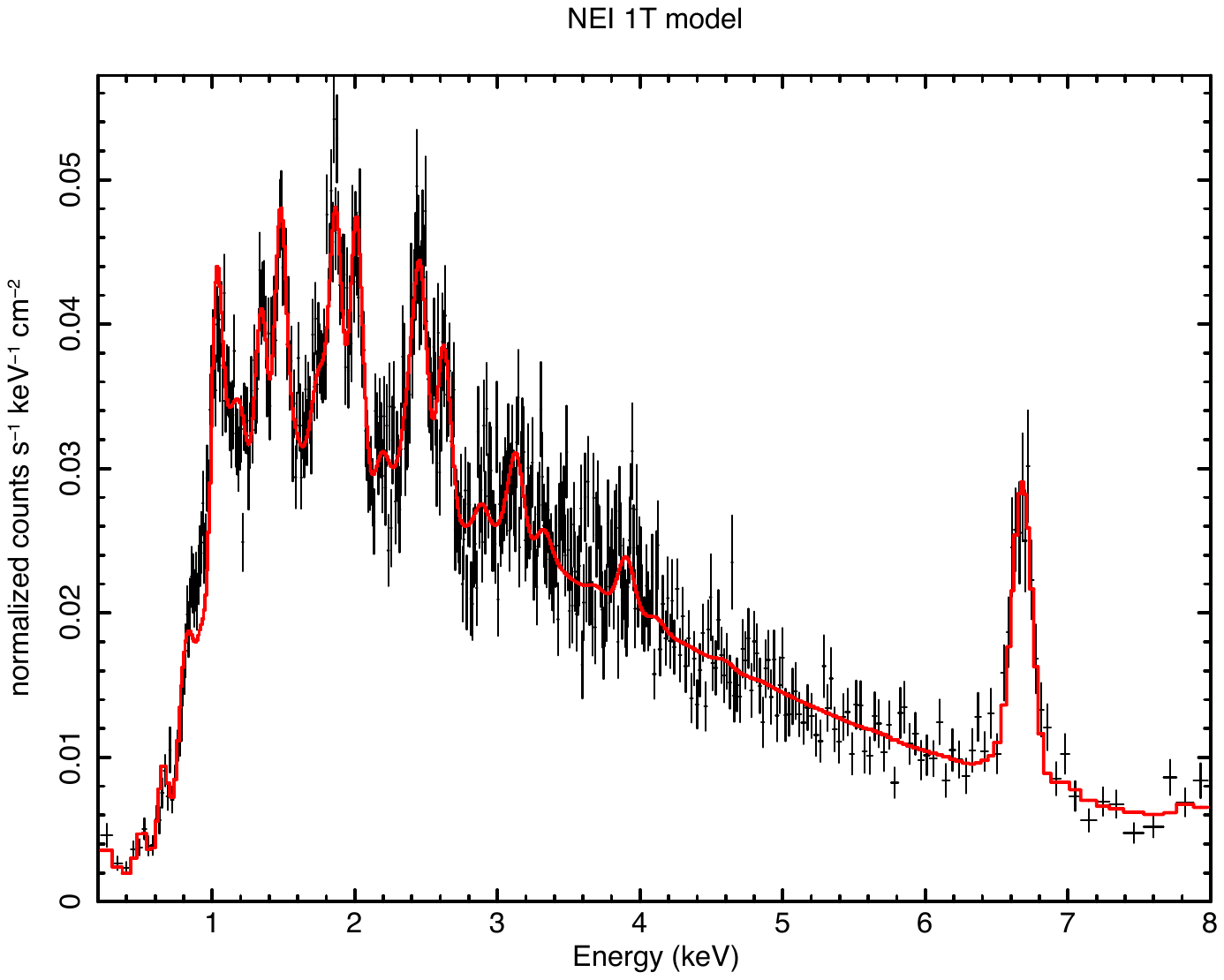}
\includegraphics[width=87mm]{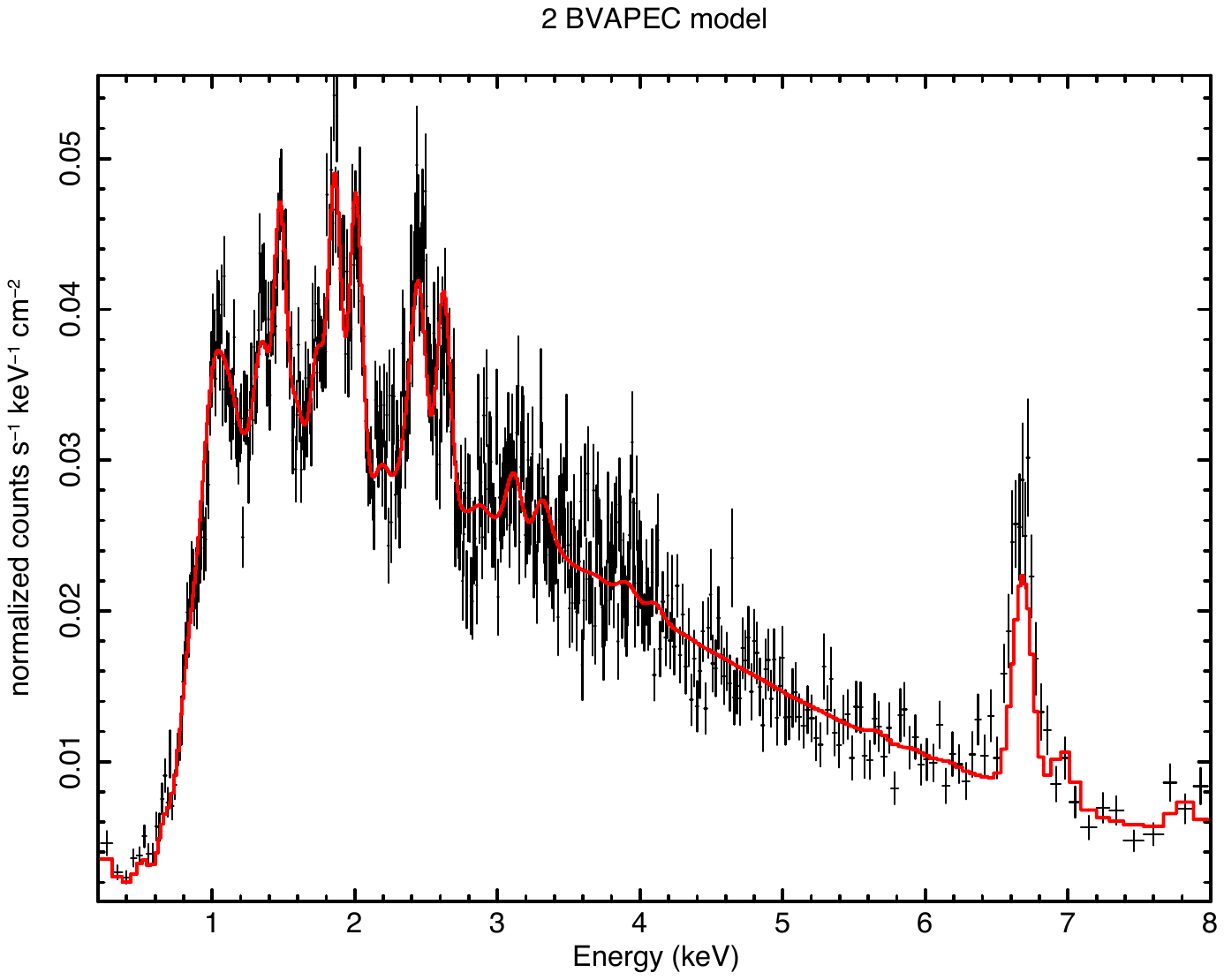}
\includegraphics[width=87mm]{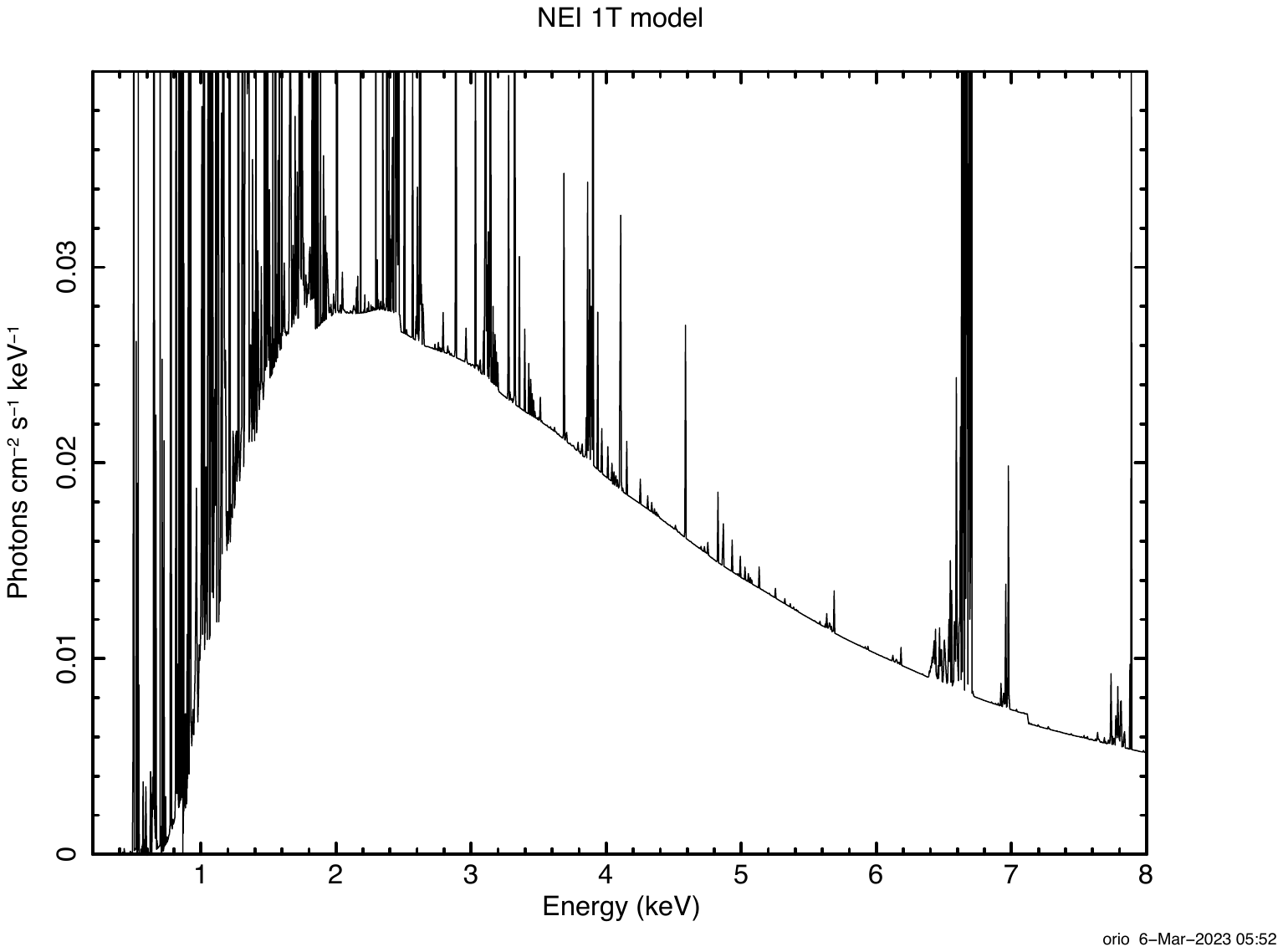}
\includegraphics[width=87mm]{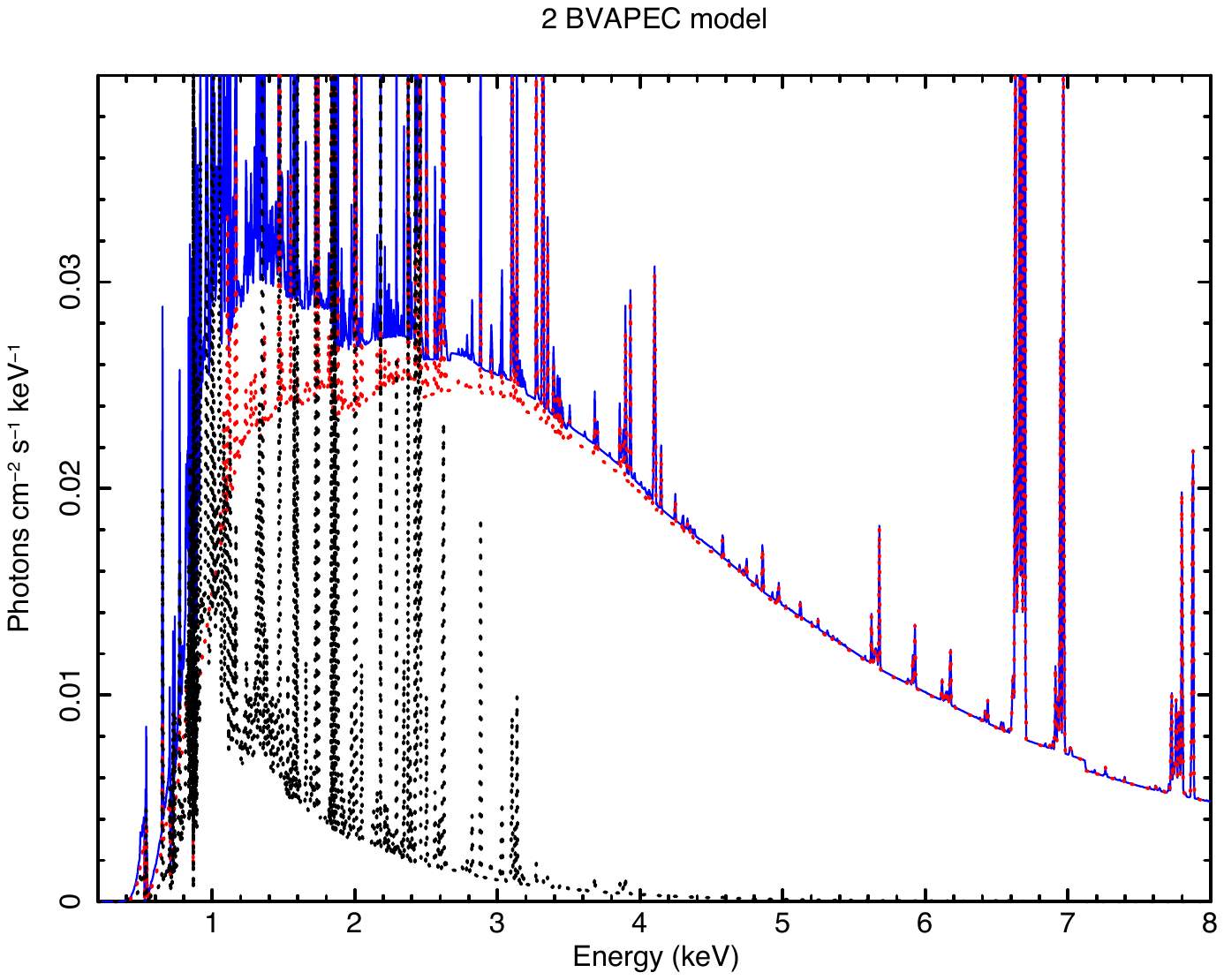}
\end{center}
\caption{For the 4th GTI of day 7, August 16, the top panels show the fits
 with a plasma that is  not in equilibrium, and with two plasma components
 at different temperature and 
 in collisional ionization equilibrium, respectively on the left and the right.
 The bottom panels show the corresponding models before convolution
 with the instrument response. The fit parameters are in Table 2.
 The fit on the right shows the ``hot'' component in red, the ``cool''
 one in black, and the result is shown in blue.}
\end{figure}
\section{The soft flares}
 Like in 2006, the initial rise of the SSS was marked by short flares of 
 extremely soft flux.
 On August 28 (day 18), a new phenomenon started.  During a very brief GTI
 that lasted for only 45 s and was interrupted for technical reasons, 
 there was a large increase in flux below 0.6 keV. 
The quality of the data was checked with more than one method,
 and the large brightening in this short GTI appears to be due to the
source, not to space weather or background. 
 On August 30 (day 21), as Fig. 9 shows, at
 the beginning of the exposure, the count rate in the 0.2-0.6 keV range
 was much higher than 
 the average count rate measured two days earlier, but 
 decreased again on that day. An 
 {\sl XMM-Newton} exposure was started $\simeq$ 3 hours after
 the end of the {\sl NICER} observations: the   {\sl XMM-Newton} EPIC pn light curve,
 shown by \citet{Orio2022a}, continued to show a  
 steady decrease.  The lightcurve and the spectra at
 maximum and minimum on day 21 are plotted 
 in the left panels of Fig. 9. The soft energy excess can
 be fitted either like in \citet{Page2022}, by adding a third blackbody
 component, or instead with an additional BVAPEC component at
 lower temperature.  At energy $>$0.5 keV the spectrum
 was unchanged and only the softest portion appeared to 
 flare. The grating
 spectra analyzed by \citet{Orio2021a} indicate
 that this initial soft excess was more ``structured''
 and complex than a blackbody, and was most likely  due to new
 emission lines appearing in the soft range.

Another, large soft flare was observed the beginning of
 an exposure on September 5 (day 27), as shown
 in the plots on the right in Fig. 9. The spectrum
 observed after the flare on the following day is shown in Fig. 10
 and the possible models that fit it are in Table 3. In the following days, 
 there were several more soft flares lasting for
 up to a few hours.  Both during and after the flares, the
 spectra can be modeled with an equally good fit either by adding to the two
 CIE plasma components (BVAPEC model) a blackbody affected by the same
 column density as the thermal plasma and a temperature of 80-90 eV,
 like in \citet{Page2022}, or a  third low-temperature BVAPEC thermal component like in  \citet{Orio2022a},
 initially with temperature around 200 eV and cooling
 to $\simeq$90 eV in the following 10 days.
 Most fits required an additional absorbing column density for
this soft component.  Figs. 11 shows  that the soft excess
 on day 37
 peaked around 0.5 keV, which corresponds to the N VII H-like line,
 however  adding a third plasma component at low temperature with
 elevated nitrogen does not fit the spectrum. In fact,
the apparent line is too broad to be a single
 emission  line.

 The high resolution spectra
obtained with {\sl XMM-Newton} on day 21 (2021 August 30) were taken during the 
 decline, but they do indicate that, at least in this early phase,
the larger soft flux was in several emission lines that later
 seemed to fade or disappear (see Fig. 7 of that article).
{\sl NICER} cannot resolve well emission lines at 
 energy $\leq$0.7 keV keV. 

 We fitted also the spectra of this phase with enhanced abundances
with respect to solar, mostly 2-7 times for all elements
 except for nitrogen (overabundant by up
 to a factor of 70) and depleted iron, but
 the abundances values have large uncertainty. The iron abundance
 turns out to be consistently depleted by a factor of a few. 
 The nitrogen overabundance indicates mixing with ashes
 of the CNO burning,
 but given errors up to 50\% in the determination of the abundances
 of this element,  we cannot draw a firm conclusions.
\citet{Orio2022a}, using high resolution
 X-ray grating spectra, derived enhanced abundance of nitrogen
 on day 30 and depleted iron on days 18 and 21.
  We note
 that enhanced abundances of the other elements
 are somewhat unexpected, because the ejecta are being
 diluted by the interaction with the red giant wind.

  We give examples of fits' parameters in Tables  3 and 4;
 some of the model fits
 are also plotted in in Fig. 9,  10 and 11.  Table 3 presents as many
 as three different fits for
 an out-of-flare GTI, one with a blackbody,
 the second and the third with different N(H) for one of
 the  components. Table 4 show
 the parameters that fit the last flare before a more
 permanent, steep rise on day 37.
  If we assume additional absorption for the coolest
 component, its temperature is higher, and 
  the resulting absolute flux is  
 as large as 6 $\times 10^{-9}$ erg s$^{-1}$ cm$^{-2}$. The absorbed flux
 in the same exposure was a little over 1.7 $\times 10^{-10}$ 
 erg s$^{-1}$ cm$^{-2}$.
 The parameters 
 can be compared  with \citet{Page2022}. The {\sl NICER} spectra, with
 a lower energy range (calibrated and reliable from 0.2 keV instead of
 0.3 keV as the {\sl Swift} XRT) constrain N(H) to be significantly 
lower than the value obtained by \citet{Page2022} in the same period.   
\begin{figure}[ht!]
\begin{center}
\includegraphics[width=130mm]{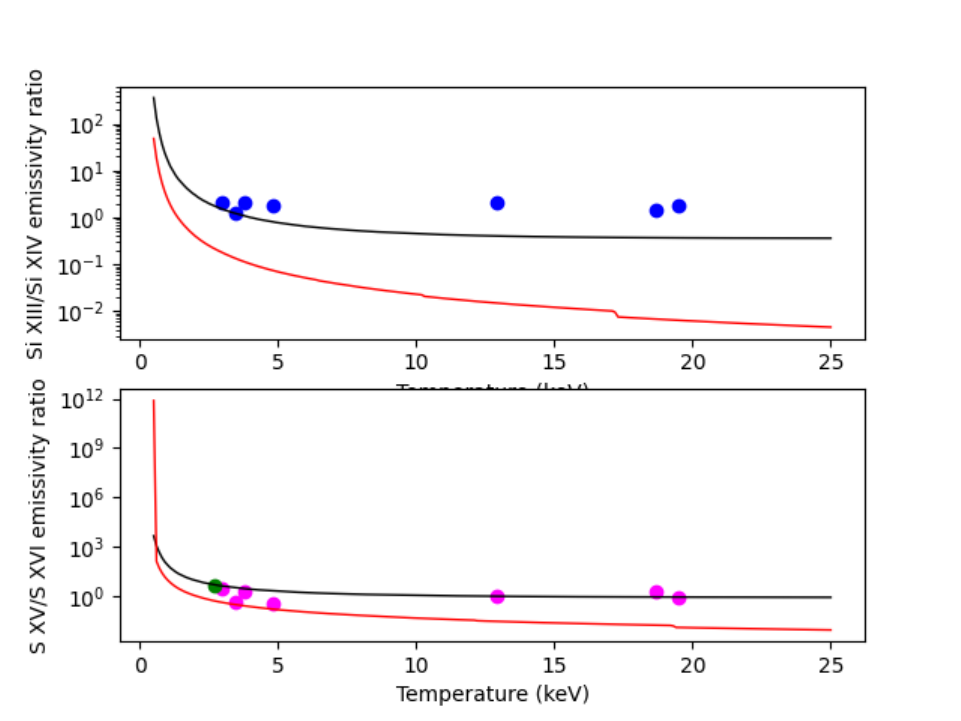}
\end{center}
\caption{Predicted flux line ratios of Si XIII/Si XIV and S XV/S XVI 
 (including all the merged lines for the He-like triplet)
for a single temperature plasma region, under the assumption
 of collisional ionization equilibrium (CIE, in red) 
and assuming non-equilibrium 
 (NEI, in black, like in the ``VPSHOCK'' XSPEC model, with ionization
 time scale parameter 10$^{11}$ cm $\times$ s). The dots 
 indicate actual measurements in correlation with the maximum temperature
 of an acceptable fit to {\sl NICER} spectra; the green dot in
 the lower panel shows a measurement with a resolved triplet,
 with the {\sl Chandra} HETG grating from \cite{Orio2022a}.
 In that paper a very good fit to the overall
spectrum was obtained assuming two regions at different temperature and in CIE, instead of a single region in NEI.}
\end{figure}

 Until the large increase on day 37 (September 16) all fits result in  
 $\chi^2$/d.o.f. in the 1.1-1.5 range, either
 using  either a ``central source'' (blackbody or atmosphere)
 and two thermal components, 
 (like in  \citet{Page2022}) or using three BVAPEC  components. 
For most GTIs, however, the spectra are fitted by 
 assuming at least two different values of the 
absorbing column densities N(H),
 and as shown in Table 3, different combinations of
 temperature and (NH) can give an equally good fit.
  Table 4 compares fits for 
 the spectrum of a  flare of 2021 September 15, day 36.  We notice
 that, around day 33 (September 11),
 substituting the blackbody with an atmospheric model 
 from \citet{Rauch2010} gives a much better fit, but only
 if we add also another very soft thermal plasma component 
 with a temperature of 80-90 eV. Thus it is likely that the soft flux
 at this stage  actual due to both emission in a thermal
 plasma and to the appearance of the SSS continuum. 
 The fits with the atmospheric models indicate that
 the WD would initially have an effective temperature of 450,000 K, with
 a bolometric luminosity around 7 $\times 10^{37}$ erg s$^{-1}$
 (see Table 4).  When the source is not flaring, the
 very soft blackbody or atmospheric
 component is still necessary in the fit, but and it contributes  to the observed
 flux by only a few \%. During the flares,  already  
 after day 30 this component contributes by more than 60\% to the unabsorbed flux.
\begin{table}
\caption{Parameters of the fit of the fourth GTI of day 7 (2021 August 16)
with two different models: a VPSHOCK model in XSPEC of non-equilibrium
 thermal plasma (Model 1), and two  two BVAPEC plasma components 
 (Model 2). This is the last day in which a fit with one temperature
 can be obtained; a second component then becomes necessary also in the
 NEI assumption. NEI models no longer fit from day 12.
  If there was not a large
 difference in the 2$\sigma$ statistical error in the positive and negative direction,
 we only report the largest one;  the errors are calculated assuming
 that the other parameters are fixed. n*$_e$ is the value
 of the electron density resulting from the ionization time scale,
 while below it we report a lower limit derived from
 the emission measure.}
\begin{center}
\begin{tabular}{ccc}
\hline
     & Model 1 & Model 2 \\ 
\hline
$\chi^2$/d.o.f. & 1.2   & 1.2 \\
N(H) $\times 10^{21}$ cm$^{-2}$              & 10.7$\pm$0.9   & 6.8$\pm$0.3\\
N(H)$_{\rm part.} \times 10^{21}$  cm$^{-2}$ & 47.92$\pm$18.9 & 58.3$\pm$5.9 \\
Cov. Fract. & 0.49$\pm$0.08  & 0.71$\pm$0.03 \\
T$_1$ (keV) & 7.24$\pm$0.54  & 5.52$\pm$0.51 \\
T$_2$ (keV) & ----           &  1.04$\pm$0.03 \\
 N/N$_\odot$ & 45$^{+61}_{-15}$ & 36$^{+30}_{-19}$ \\
 Fe/Fe$_\odot$ & 0.6$\pm$0.2 & 0.4$\pm$0.1\\
 $\tau_{\rm max}$ (s $\times$ cm$^{-3}$) & 3.42$\pm0.85 \times 10^{11}$ & -- \\
 EM$_1$ cm$^{-3}$ &  5.71 $\times 10^{58}$  & 3.19  $\times 10^{58}$\\
 EM$_2$ cm$^{-3}$ & -- & 5.96 $\times 10^{57}$    \\
n*$_{\rm e}$ cm$^{-3}$  & 6.1  $\times 10^5$ & -- \\
n$_{\rm e}$ cm$^{-3}$  & $\geq 1.2 \times 10^6$ & $\geq 9.7 \times 10^6$ \\
 F$_{\rm x,abs}   \times 10^{-10}$ erg cm$^{-2}$ s$^{-1}$ & 10.12  & 9.73 \\
 F$_{\rm x,unabs} \times 10^{-10}$ erg cm$^{-2}$ s$^{-1}$ & 76.12  & 22.22 \\
\hline
\end{tabular}
\end{center}
\end{table}
\begin{table}
\caption{Parameters of the fit of the first GTI of day 27 (September 6), 
 with three different composite models: a blackbody and two BVAPEC
 plasma components (Model 1),
 or three BVAPEC thermal plasma components (Model 2 and 3). In Model 3 
 all components are affected by the same column density, while
 in Model 2 the coolest and hottest component are absorbed by
 N(H)$_{1,2,3}$+N(H)$_{1,3}$. The flux of the single components is
 numbered in order of rising temperature. If there was not a large
 difference in the 2$\sigma$ statistical error in the positive and negative direction, 
 we only report the largest one;  the errors are calculated assuming
 that the other parameters are fixed. The electron density with a star
 is obtained from the maximum ionization time scale, while the lower
 limit on the electron density is obtained from the emission measure
 assuming an expansion velocity of 7550 km s$^{-1}$..} 
\begin{center}
\begin{tabular}{cccc}
\hline
     & Model 1 & Model 2 & Model 3 \\
\hline
 $\chi^2$/d.o.f. & 1.5   & 1.3 & 1.3  \\
 N(H)$_{1,2,3}$ $\times 10^{21}$ cm$^{-2}$ & 3.4$\pm$0.2 & 2.6$\pm$0.1 & 3.1$\pm$0.3 \\
 N(H)$_{1,3}$ $\times 10^{21}$ cm$^{-2}$ & 0  & 7.9$\pm$0.7 & 5.7$\pm$0.2 \\
 T$_{\rm bb}$ (eV)                   & 45.2$\pm$0.6  & --  & -- \\
 T$_1$ (eV)                         &    --          & 290$\pm$80 & 90$\pm$18 \\
 T$_2$ (keV)                         & 0.66$\pm$0.02 & 0.73$\pm$0.08 & 0.80$\pm$0.03 \\
 T$_3$ (keV)                         & 2.27$\pm$0.30 & 1.85$\pm$0.30 & 4.89$\pm$2.30 \\ 
 F(tot) $\times 10^{-10}$  erg cm$^{-2}$ s$^{-1}$ & 1.74$\pm$0.91 & 1.73$\pm$0.32 & 1.77$\pm$0.14 \\
 F(tot,unabs) $\times 10^{-9}$ erg cm$^{-2}$ s$^{-1}$ & 1.41$\pm$0.59 & 6.25$\pm$0.77 & 3.32$\pm$1.17 \\ 
 F$_{\rm bb}$   $\times 10^{-11}$ erg cm$^{-2}$ s$^{-1}$ & 0.30$\pm$0.02  &  -- & -- \\
 F(bb,unabs) $\times 10^{-10}$ erg cm$^{-2}$ s$^{-1}$   & 9.17$\pm$0.73  & -- & -- \\
 F(1) $\times 10^{-11}$ erg cm$^{-2}$ s$^{-1}$ & -- & 2.95$\pm$1.48 & 0.87$\pm$0.80 \\
 F(1,un) $\times 10^{-9}$ erg cm$^{-2}$ s$^{-1}$ & --  & 5.87$\pm$2.90 & 2.84$\pm$2.70 \\
 F(2) $\times 10^{-11}$ erg cm$^{-2}$ s$^{-1}$ & 5.53$\pm$0.45 & 9.53$\pm$3.83 & 13.85$\pm$10.00 \\
 F(2,un)   $\times 10^{-10}$ erg cm$^{-2}$ s$^{-1}$ & 1.00$\pm$0.08 & 2.59$\pm$1.04 & 4.35$\pm$0.35 \\
 F(3)   $\times 10^{-10}$ erg cm$^{-2}$ s$^{-1}$ & 1.16$\pm$0.12 & 0.48$\pm$0.24 & 0.32$\pm$0.15 \\
 F(3, un)  $\times 10^{-10}$ erg cm$^{-2}$ s$^{-1}$ & 3.89$\pm$0.39 & 1.22$\pm$0.28 & 0.43$\pm$0.20 \\ 
 \hline
\end{tabular}
\end{center}
\end{table}
\begin{table}
\caption{}{Parameters of the fits to the last, and highest GTIs of 
 day 36 (2021 September 15), 
 with two different composite models: a WD atmosphere and two BVAPEC
 plasma components (Model 1),
 or three BVAPEC thermal plasma components (Model 2). The flux of the single components is
 numbered in order of rising temperature. The second component, of intermediate
 temperature, has an added column density N(H)$_{2}$. The errors
 are calculated and presented as in Table 2.}
\begin{center}
\begin{tabular}{ccc}
\hline
 & Model 1 & Model 2 \\
\hline
 $\chi^2$/d.o.f. & 1.1 & 1.1  \\
 N(H)$_{1,2,3}$ $\times 10^{21}$ cm$^{-2}$ & 4.92$\pm$0.04 & 1.96$\pm$0.10 \\
  N(H)$_{2}$ $\times 10^{21}$ cm$^{-2}$  & -- & 2.26$^{+0.70}_{2.20}$ \\ 
 T$_{\rm atm}$ (K)    &  450,000$\pm$5,000 & -- \\
T$_1$ (eV)            &   90.5$\pm$89.0    & 93.1$\pm$5.9 \\
T$_2$ (keV)                         & -- & 0.57$\pm$0.04 \\
T$_3$ (keV)                         & 0.62$\pm$0.02 & 1.38$^{+0.68}_{-0.21}$ \\
 F(tot) $\times 10^{-10}$  erg cm$^{-2}$ s$^{-1}$ & 2.15 & 2.17 \\
 F(tot,unabs) $\times 10^{-8}$ erg cm$^{-2}$ s$^{-1}$ & 14.7$^{14.0}_{-4.0}$ & 1.48$^{+1.45}_{-0.52}$ \\
 F$_{\rm atm}$   $\times 10^{-11}$ erg cm$^{-2}$ s$^{-1}$ & 1.10$\pm$0.60 & -- \\
 F(atm,unabs) $\times 10^{-7}$ erg cm$^{-2}$ s$^{-1}$ & 1.08$\pm$0.60 & -- \\ 
  F(1) $\times 10^{-10}$ erg cm$^{-2}$ s$^{-1}$ & 1.27$^{+0.42}_{-1.20}$  & 0.77$\pm$0.09 \\ 
  F(1,un) $\times 10^{-9}$ erg cm$^{-2}$ s$^{-1}$ & 38.3$^{+35.0}_{-10.7}$ & 0.48$\pm$0.06 \\
 F(2) $\times 10^{-11}$ erg cm$^{-2}$ s$^{-1}$ &  - & 8.85$_{-3.74}^{+23.8}$ \\
 F(2, un) $\times 10^{-10}$ erg cm$^{-2}$ s$^{-1}$ &  - & 3.93$_{-1.66}^{+3.7}$  \\
 F(3)   $\times 10^{-10}$ erg cm$^{-2}$ s$^{-1}$ & 0.77$\pm$0.17 & 0.53$\pm$0.11\\
 F(3, un)  $\times 10^{-10}$ erg cm$^{-2}$ s$^{-1}$ & 379$\pm$65 & 1.23$\pm$0.02 \\
 \hline
\end{tabular}
\end{center}
\end{table}

In the 2006 outburst, flaring was observed on day 27 post-maximum
 during a long exposure with {\sl XMM-Newton} in the
 supersoft region, and the high resolution
 spectrum showed that it was due to the appearance of
 prominent emission lines \citep{Nelson2008}.
 Also in 2021, it is likely that even for
 several days, the first manifestation of
 the SSS presence
 is in an emission line spectrum, due to material
 that is either photoionized or shock-ionized quite close the WD. However,
 due to the degeneracy in the combination of column density and temperature
 in the fit,
 we cannot exactly determine when the WD becomes visible.
 It is important to notice that the flares do not appear to be due
 to increasing blackbody/atmospheric temperature, neither to decreasing
 column density; actually the fits improve with even
 increased column density in the flares' spectra, as if
 new material has been emitted in parallel to the emergence
 of the SSS, and is contributing to some more intrinsic absorption.
\section{The ``stormy'' luminous supersoft X-ray phase}
The supersoft X-ray phase was quite different from the 2006 outburst.
 It was less luminous, shorter lived and with more episodes
 of irregular, large variability, that occurred even
 within hours, as shown in Fig. 12.  Fig. 13  shows the 
 spectrum at peak. The less soft portion of the spectrum still
 shows the presence of the shocked thermal plasma, that by this time
 had cooled to temperatures below 0.5 keV. 

 One challenge posed by the luminous supersoft X-ray spectra of novae
 in general, and specifically
 of RS Oph in 2006 and 2021, is that most high resolution X-ray
 spectra obtained  
 with the {\sl Chandra} and {\sl XMM-Newton} gratings show
 blue-shifted absorption lines \citep[e.g.]{Rauch2010, Ness2011,
 Orio2021, Ness2022a},  indicating that we are not
 simply observing an atmosphere at rest. The blue-shift can
 be large, corresponding to even 2000 km s$^{-1}$ for
 many features of some novae; this is in
 contrast with a  key assumption
 in the models, that by the time the central WD becomes 
 observable, mass loss has ceased \citep[e.g.]{Kato2020}.
 Reasonable fits to the SSS spectra have been obtained in the following 
 semi-empirical ways:

$\bullet$  We can 
 assume that we observe a small shell of material near the WD
 atmosphere that has recently been detached and is expanding, thus 
 modeling the spectrum with a photoionization code for X-ray astronomy,
 assuming that the photoionizing source is a blackbody, whose
 temperature is a fit parameter. This has been done with varying results
 \citep{Ness2011, Orio2021, Ness2022b, Milla2023}. Sometimes, 
 many shells have to be included to fit the spectra, introducing
 many variable parameters \citep[see]{Ness2011}, so 
 the parameterization becomes more uncertain. 
 This approach relies entirely on modeling the absorption features and
 edges, but the features are not resolved in this range
 with a broad band spectrum like {\sl NICER}'s.

 $\bullet$ \citet{vanRossum2012} computed a model inspired by the
 physics of the winds of massive stars for nova V4743 Sgr and made
 a grid of models publicly available. In this model the blue-shift
 of the lines is parameterized by mass outflow rate and effective
 effective temperature. The model predicts how the  
 the line profiles change with blue-shift (wind velocity), but it does 
 not give satisfactory results for other novae \citep{Orio2018}. We cannot 
 not measure absorption line profiles with {\sl NICER}, so we
 did not try this model. 

 $\bullet$
 A third, common approach is an atmospheric model 
 \citep{Rauch2010}, assuming in first approximation 
  that all the absorption features keep the basic profile of the  
 atmosphere ``at rest'', even if they are generated in
 a wind. The features
 are then assumed to be blue-shifted with the velocity as a free parameter.
 In other words, one assumes that the photoionized source is the WD
 and that the structure of the absorption spectrum does not change
 in material that is being detached, but is very still close to
 the WD. This approach  often yields a good
 fit, including  the departure from a blackbody continuum that is always
 observed. It has been used in the literature to fit
 the  broad-band X--ray spectra of novae, using the shape of the  continuum 
 and sometimes
 adding absorption edges. \citet{Osborne2011, Page2022} for instance 
 experimented with it fitting RS Oph XRT's spectra of 2006 and 2021. 
 \citet{Nelson2008} fitted the high resolution 2006 spectra of RS Oph
 with a peak temperature of about 800,000 K; however, not all spectral
 features were explained, also  
because there is a limited available grid of abundances, 
that may not be suitable for RS Oph or other novae. 

 $\bullet$ For broad band spectra, another possibility is to
 assume a blackbody and include absorption edges that change or ``cut''
 the shape of the continuum. This has been done by \citet{Osborne2011, Page2022}
 for RS Oph. \citet{Page2022} found that the {\sl Swift} XRT spectra of 2006
 have much deeper absorption edges than the 2021 ones. 
 They fitted the SSS phase adding two components of thermal plasma to a blackbody, modified by ``ad hoc'' absorption edges,
  similar to those of a WD atmosphere. 
 However, also in the luminous SSS phase the {\sl NICER} spectra 
 give a more complex picture than the {\sl Swift} XRT ones.
 One reason is that, as an imaging telescope, 
 {\sl Swift} XRT yields data that are strongly affected
 by pile-up for such a luminous and soft source. The spectra
 examined in \citep{Page2022} are obtained by cutting off the central
 region of the point spread function,
 to avoid the effects of pile-up. We did not find spectra of
 exactly overlapping GTIs for {\sl Swift} XRT and {\sl NICER}, but
 if we consider a spectrum taken on 2021 October 5 approximately 3
 hours after and before exposures by {\sl NICER}, we observe
 in Fig. 14 how the spectrum extracted from the ``non-piled up annulus''
 is much ``flatter'' than  the the {\sl NICER} ones, so if one
 simply assumes that the flux has been reduced by a constant factor,
 the translation makes the wings of the spectral
 distribution excessively  broad.
 Thus, a straightforward spectral fit tends to converge
 towards a  higher value of the
 column density N(H) and a higher blackbody (or 
 atmospheric) temperature than the value we obtained
 fitting  the {\sl NICER} spectrum. The inclusion of absorption edges
 corrects this effect, but it is empirical 
 and may not necessarily have physical meaning. 
 Correcting the pile-up corrected {\sl Swift} spectrum  assuming
 that the flux is reduced by an energy-dependent
 factor may be a complex task.  

 Given all the above considerations,
 we limited our attempts to fit the {\sl NICER} spectra to 
 a blackbody and to the atmospheric model \citep{Rauch2010},
 with the addition of 
 a CIE (BVAPEC) model thermal plasma.  This thermal plasma has flux
 at least 3 orders of magnitude smaller than the central
 source, but it modifies the SSS continuum.
 In fact, in the portion of the spectrum 
 in which the continuum is low, emission lines are still prominent,
 and resolved above 1 keV (see panel
 on the right in Fig. 14). We  experimented with  
 several models in the public grid of the T\"ubingen 
 non-local-thermodynamic-equilibrium atmosphere TMAP(see
 \url{http://astro.uni-tuebingen.de/#rauch/TMAP/TMAP.html} and 
 \citet{Rauch2003, Rauch2010}). T\"ubingen grid 003 model was
 found to be a reasonable fit to the 2006 high resolution spectra
 \citep{Nelson2008}.
 Similarly, \citep{Osborne2011} found that this particular model reproduced
 the continuum for several {\sl Swift} XRT broad band spectra of 2006.
 However, we did not obtain a rigorous fit for the {\sl NICER}
 spectra of 2021 with this model, neither with the others
 in the grid.  One reason is
 the extreme difficulty in making the  fit converge in XSPEC
 with two components at such different flux magnitude. 
 Like \citet{Page2022} we found that including the thermal plasma
 (with much lower flux than the ``central source'') is essential
 also where it  overlaps with the strong SSS  continuum, because it 
modifies it.

 At peak luminosity the best of the fits with this model to
 several spectra around the peak (second half of September 2022)
 returns an effective temperature of about 750,000 K,
 much higher than obtained for day 37 in Table 4,  but 
  slightly lower than obtained by fitting the high
 resolution spectra in 2006 \citep{Nelson2008}.
 We obtained
  column density N(H)=4.7 $\times$ 10$^{21}$ cm$^{-2}$ and bolometric 
 luminosity 2 $\times 10^{38}$ erg s$^{-1}$. 
 At least one plasma component at 217 eV is necessary to 
 improve the fit;  a second thermal component  around 80 eV
 improves the fit even more.

 \citet{Ness2022a} find that an absorbing medium with varying ionization states
 can explain the SSS irregular variability.  For an {\sl XSPEC} global fit
 we did not have a model with varying ionization states, but
 the TBVARABS model allows to test 
 instead varying the abundances of the absorbing medium. 
 If there is circumstellar absorption of the red giant wind and wind
 mixed with ejecta, it is reasonable to expect non-solar abundances.
 The fit with TBVARABS,  a blackbody and additional thermal plasma
 is closer to the observed spectrum and is shown in Fig. 14.
 In TBVARABS we let the abundances of 
 carbon and oxygen vary and  found that we much improve
 the fit if they are enhanced (the first
 4.9 times solar, the second 2.5 times solar, and slightly depleted
 oxygen (0.82 times solar). The resulting column density
 N(H) is still higher than for the same day
 in \citet{Page2022}'s fits,  4.3 $\times 10^{21}$ cm$^{-2}$,
 and the temperatures of the two thermal components are 
 81 eV and 462 eV, respectively. The blackbody has a 
 temperature of 36.6 eV (only $\simeq$425,000 K), but its bolometric 
 luminosity would have to be much over Eddington level, namely
 about 2 $\times 10^{40}$ erg s$^{-1}$ \citep[a blackbody
 is known to overestimate the luminosity, e.g.][]{Heise1994}.
 We do not consider this
 a realistic fit, but it is an experiment to guide  a future analysis
  in which 
 the {\sl XMM-Newton} RGS grating spectra \citep[see][]{Ness2022a}
 will be the starting point.
 A rigorous model for the SSS spectrum
 should first rely on the high resolution spectra,
  and we leave it for future work.
 We focus in the following sections on
 other results we obtained with {\sl NICER}
 alone for the SSS phase, its duration and
 variability. 

\section{The aperiodic large-amplitude variability} 
An interesting characteristic of the SSS phase is the large irregular variability
 (even by a factor of 100\%, see Fig. 12) over time scales of hours and days.
 Even if our
 fits are only qualitative, we notice that we cannot fit the spectra 
taken in exposures with high count rate by simply increasing temperature or emission
 measure, or decreasing column density N(H), used for the fit to low count rate spectra.
 The {\sl NICER} data are not compatible with changes in the emitting region
size or in the temperature; our qualitative fits with atmospheric models
 to the spectra of days 40-67 always converge towards T$_{\rm eff}\simeq$750,000 K,
 no matter what the count rate is.
However, given that our model fits for this phase are still qualitative and
 not rigorous, we refer here to the discussion
of the high resolution X-ray spectra of RS Oph by \citet{Ness2022a}, who
 compared a high resolution {\sl XMM-Newton} RGS spectrum of RS Oph obtained
 in 2021, with  those of 2006 at different times after the outburst.
 They  discussed three different possibile scenarios:

 1. The spectrum during the SSS phase has variable flux, either because 
 of an ``eclipsing'', opaque intervening body, or because
 the WD photospheric  radius changes at constant temperature,

2. The  central source has varying temperature.

3. The absorption varies on short time scales along the line of sight.
 Given that varying the column density N(H)
does not explain the variability, \citet{Ness2022a} explored the effect of 
 changing ionization stages in the ejecta and in the pre-existing
 wind.

 The last scenario, according to \citet{Ness2022a}, is the correct one.
 In fact, these authors' physical model 
 of multi-ionization photoelectric absorption does explains 
 the observed variability, while they can fit the high resolution spectra with a 
central source of constant temperature and flux. 
  This  is
consistent either with a turbulent, clumpy outflow in which different clumps 
may have different ionization stages,
or perhaps it may be explained by varying conditions in the red giant wind.

\begin{figure}
\begin{center}
\includegraphics[width=87mm]{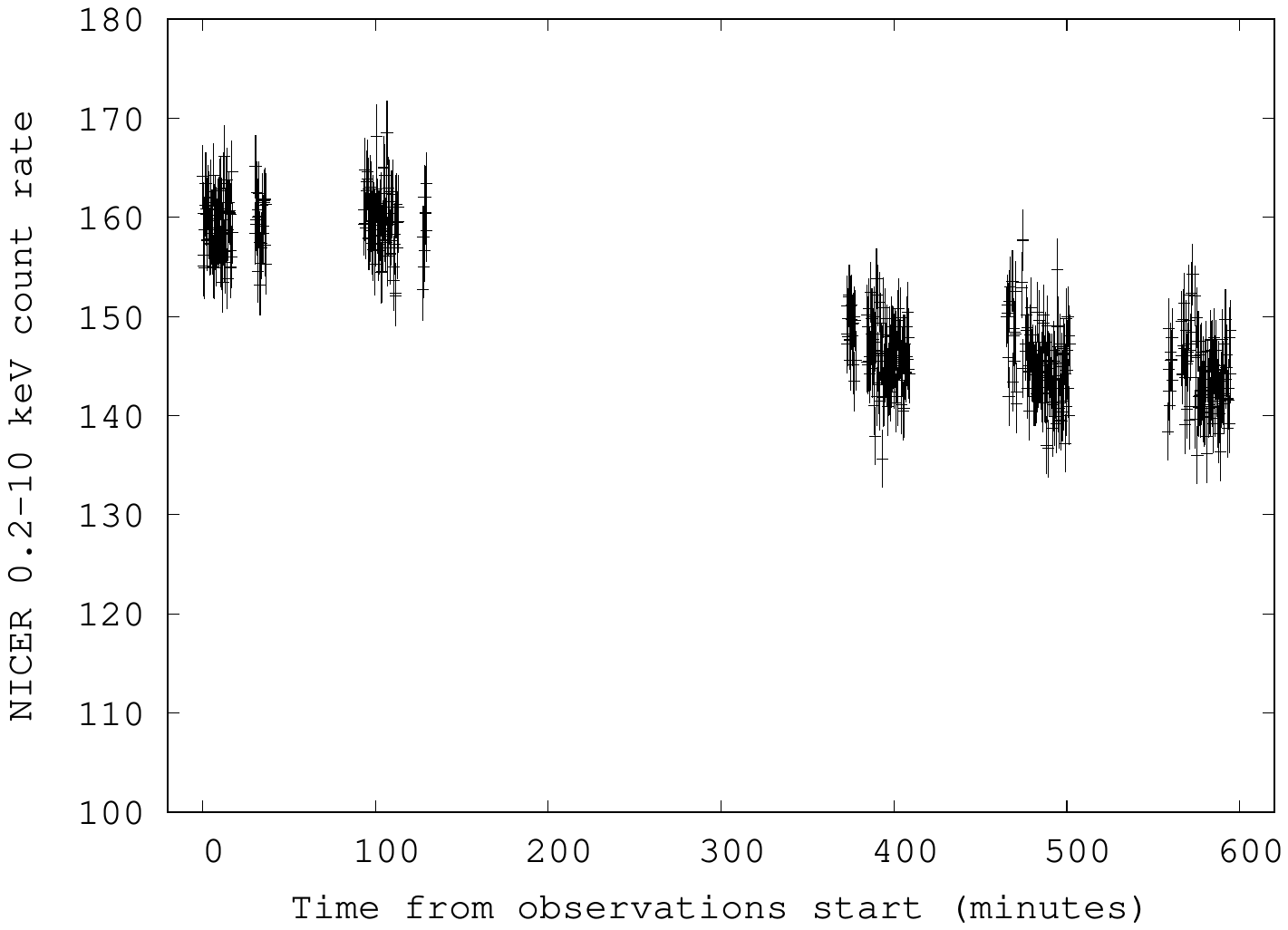}
\includegraphics[width=87mm]{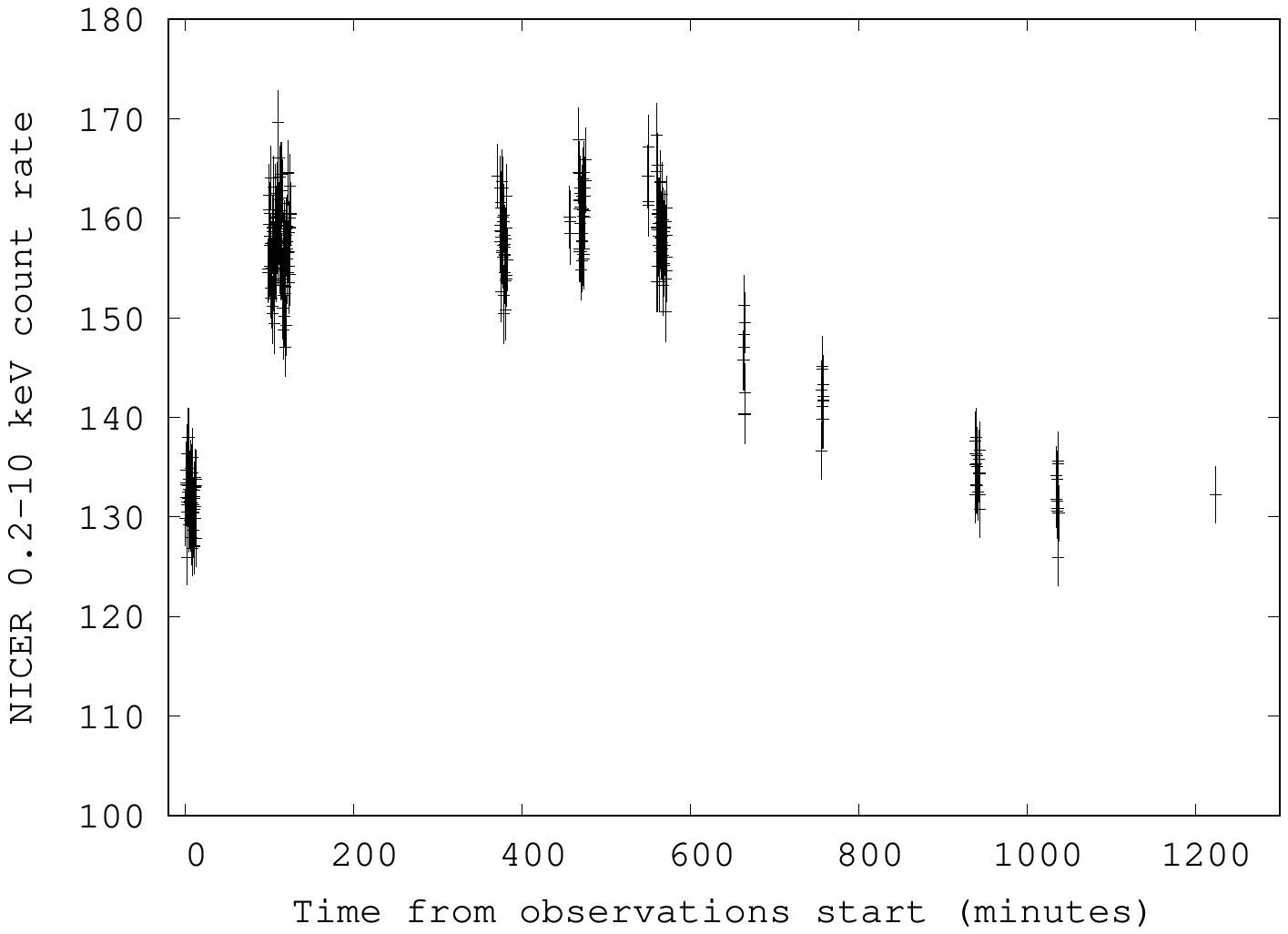}
\includegraphics[width=87mm]{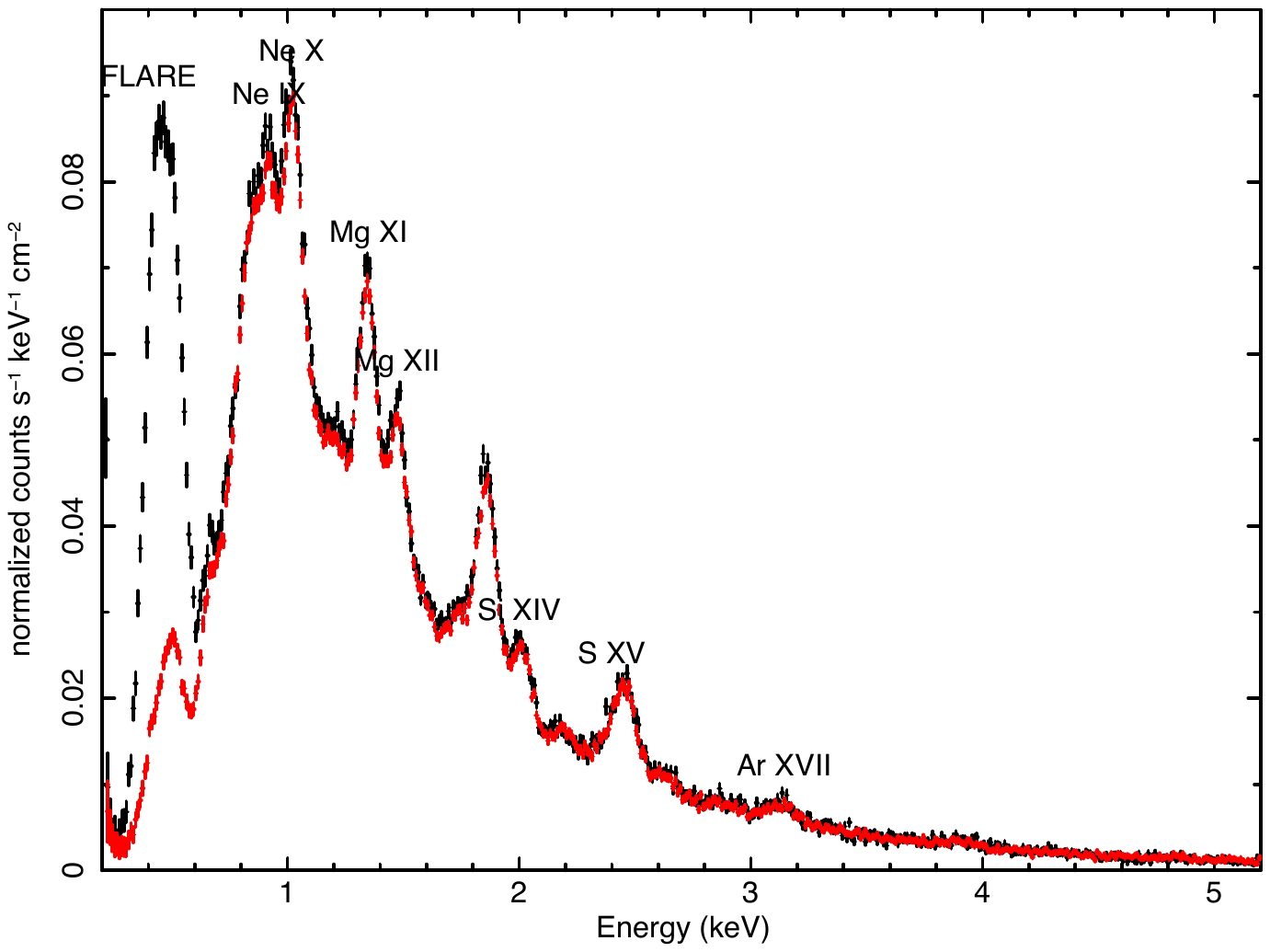}
\includegraphics[width=87mm]{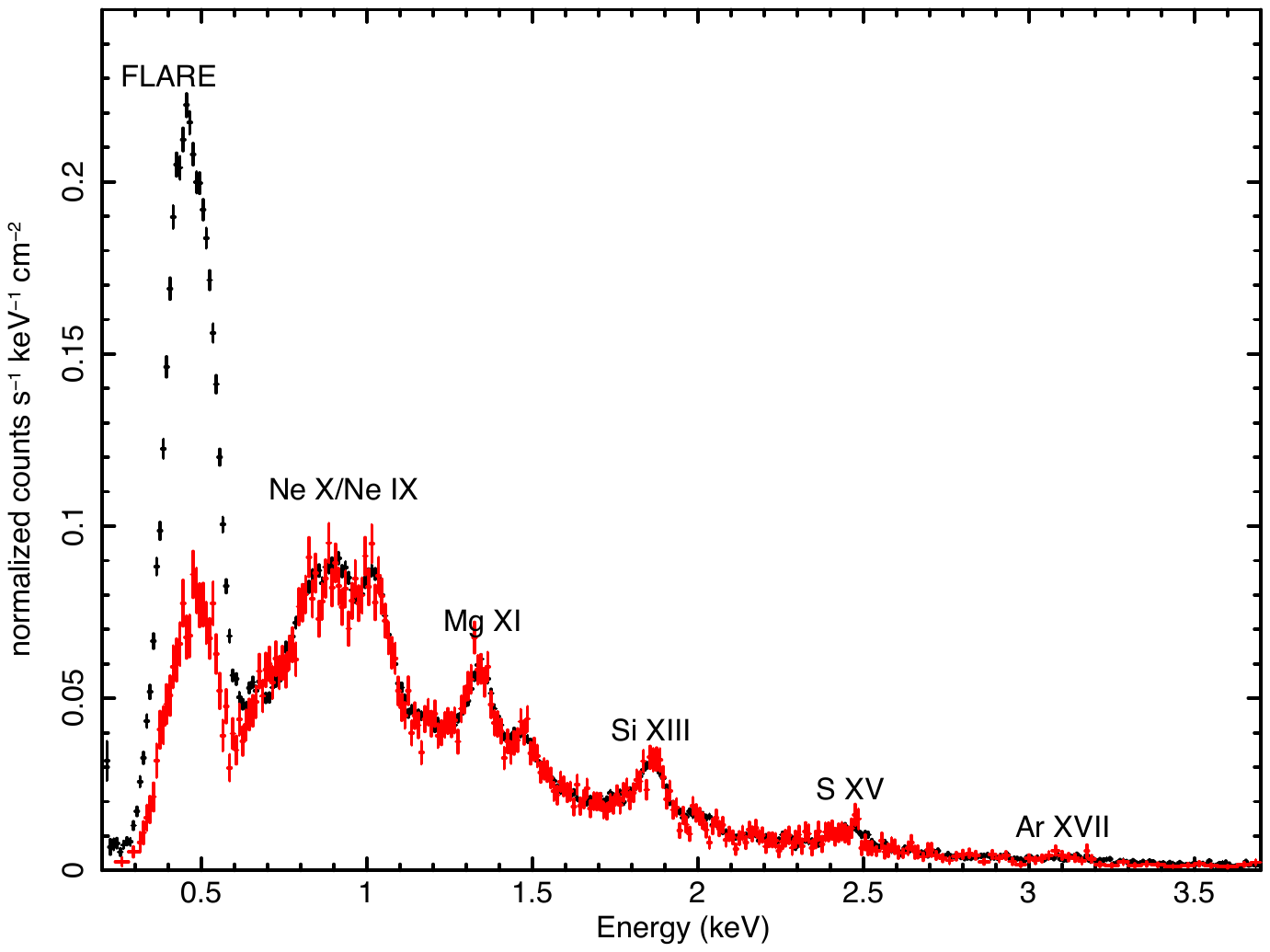}
\end{center}
\caption{Top: the light curves
 observed over few hours on day 21 (2021 August 30; left) and day
 27  (2021 September 5; right). On day 21, new observations were resumed with XMM-Newton
 after about 3 hours. The light curve during that exposure  
shows a continuing decline in flux for the whole day \citep{Orio2022a}.
The $\simeq$ 35 s QPO on day 27 was very pronounced 
 during this day. The bottom panels
 compare spectra during high and low count rate
 intervals.  On the left the spectrum
 measured on August 30 during the initial GTIs with higher
 count rate (black, first 130 minutes) compared with that of
 the last GITs with lower count rate on day 21 
(red, last $\simeq$360 minutes)  and on the right, the spectrum
 of the third GTI of day 27 (in black, it lasted for about 400 s) 
 and that of the GTI before the last (in red).} 
\end{figure}
\begin{figure}
\begin{center}
\includegraphics[width=140mm]{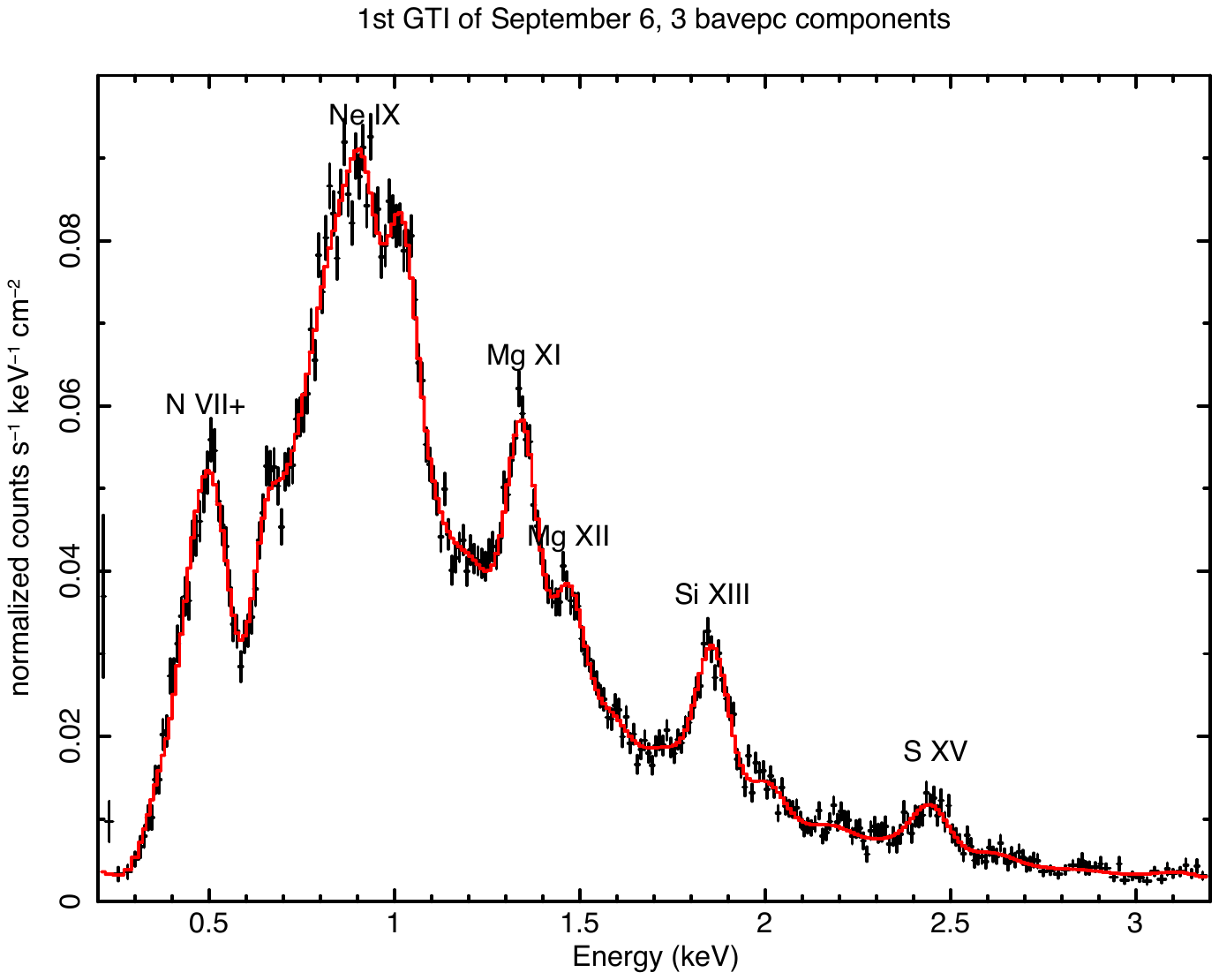}
\end{center}
\caption{The spectrum of the first GTI of day 28 (2021 September 6), during a 
``quiet'' day after the flare described above. The red line traces
 the fit with Model 3 of Table 3 (in red). 
 }
\end{figure} 
\begin{figure}
\begin{center}
\includegraphics[width=87mm]{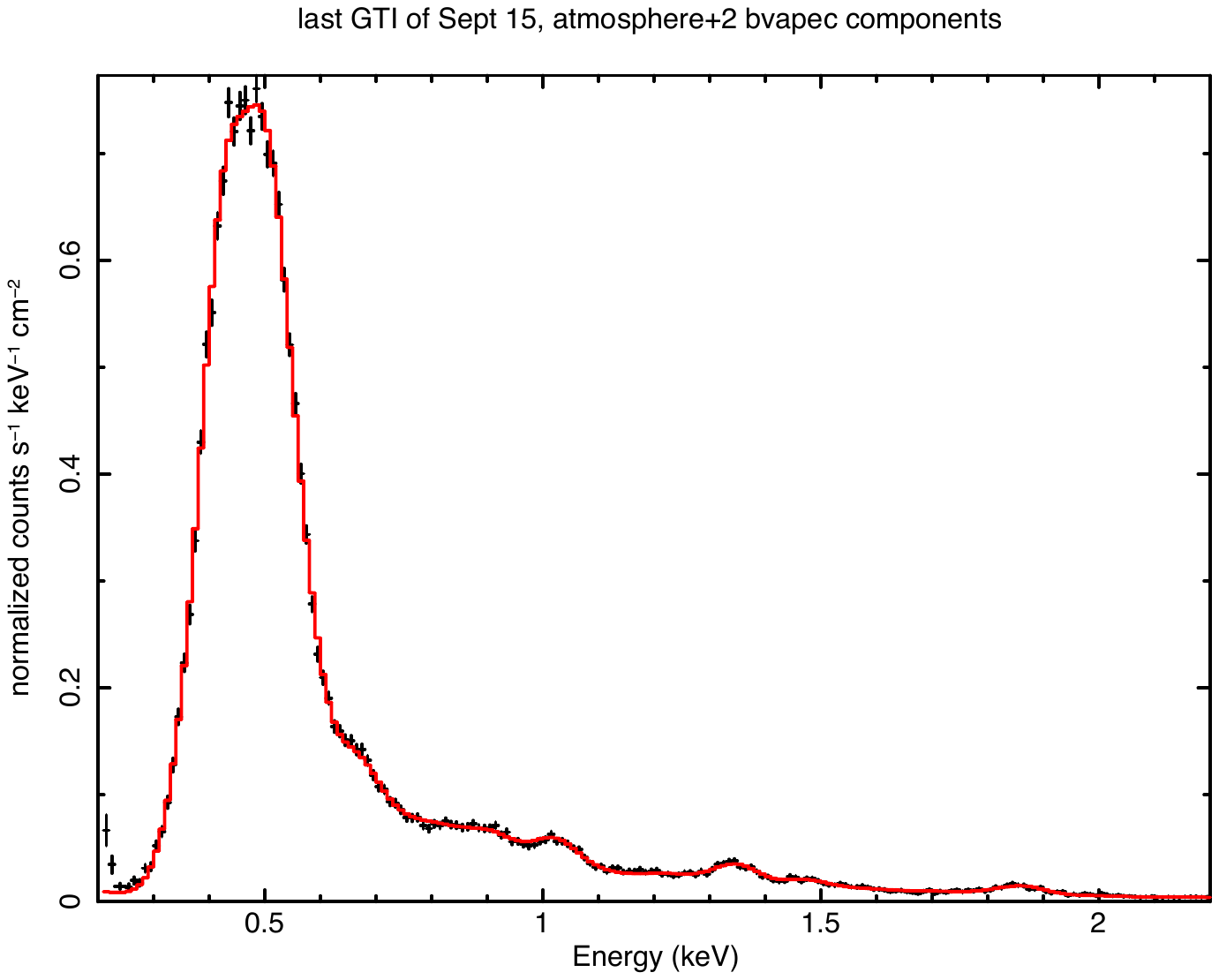}
\includegraphics[width=87mm]{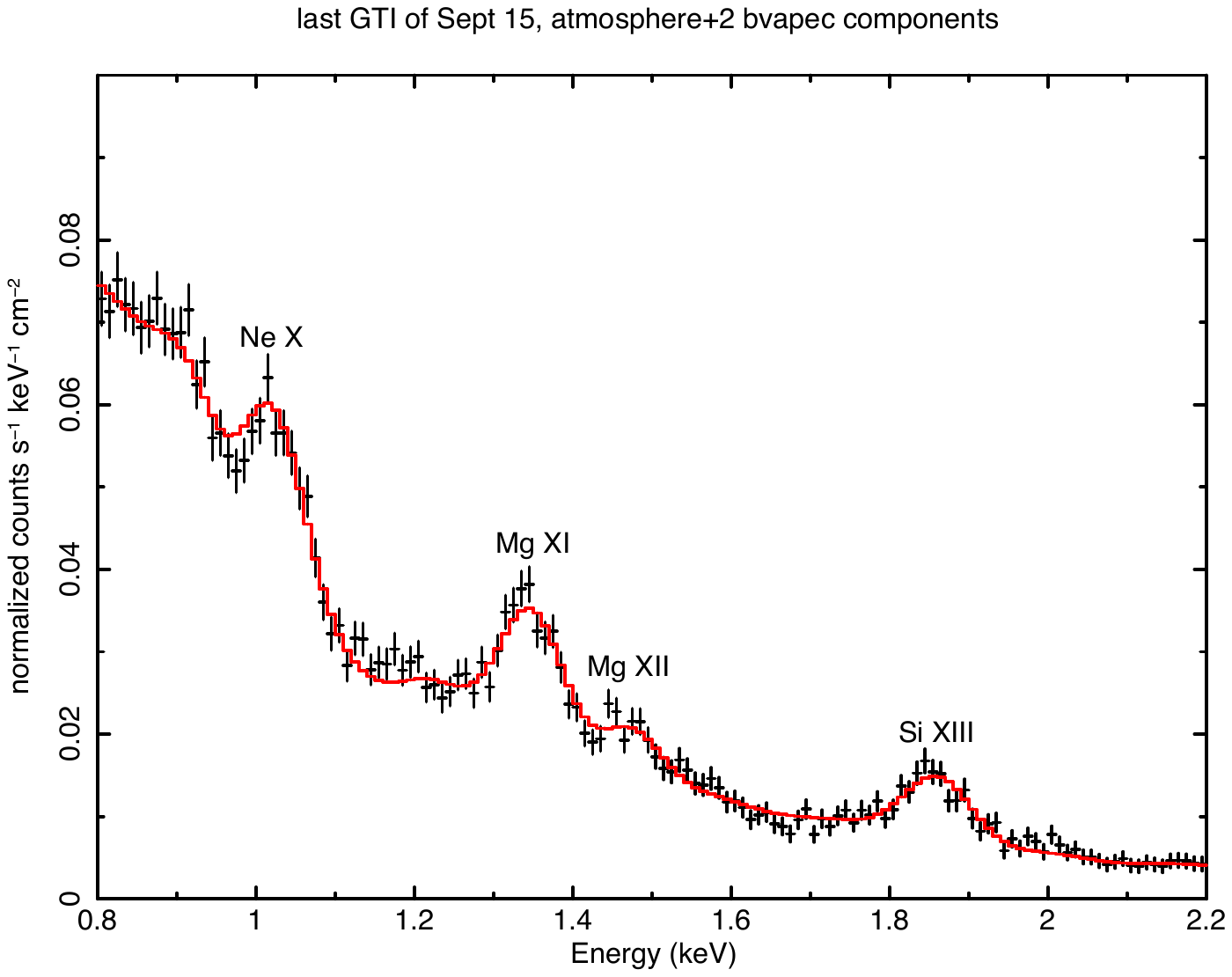}
\end{center}
\caption{Two portions of the spectrum of a flare on day 37 (2021 September 15). 
The red line traces the fit with Model 1 in Table 4.}
\end{figure}
\begin{figure}
\begin{center}
\includegraphics[width=87mm]{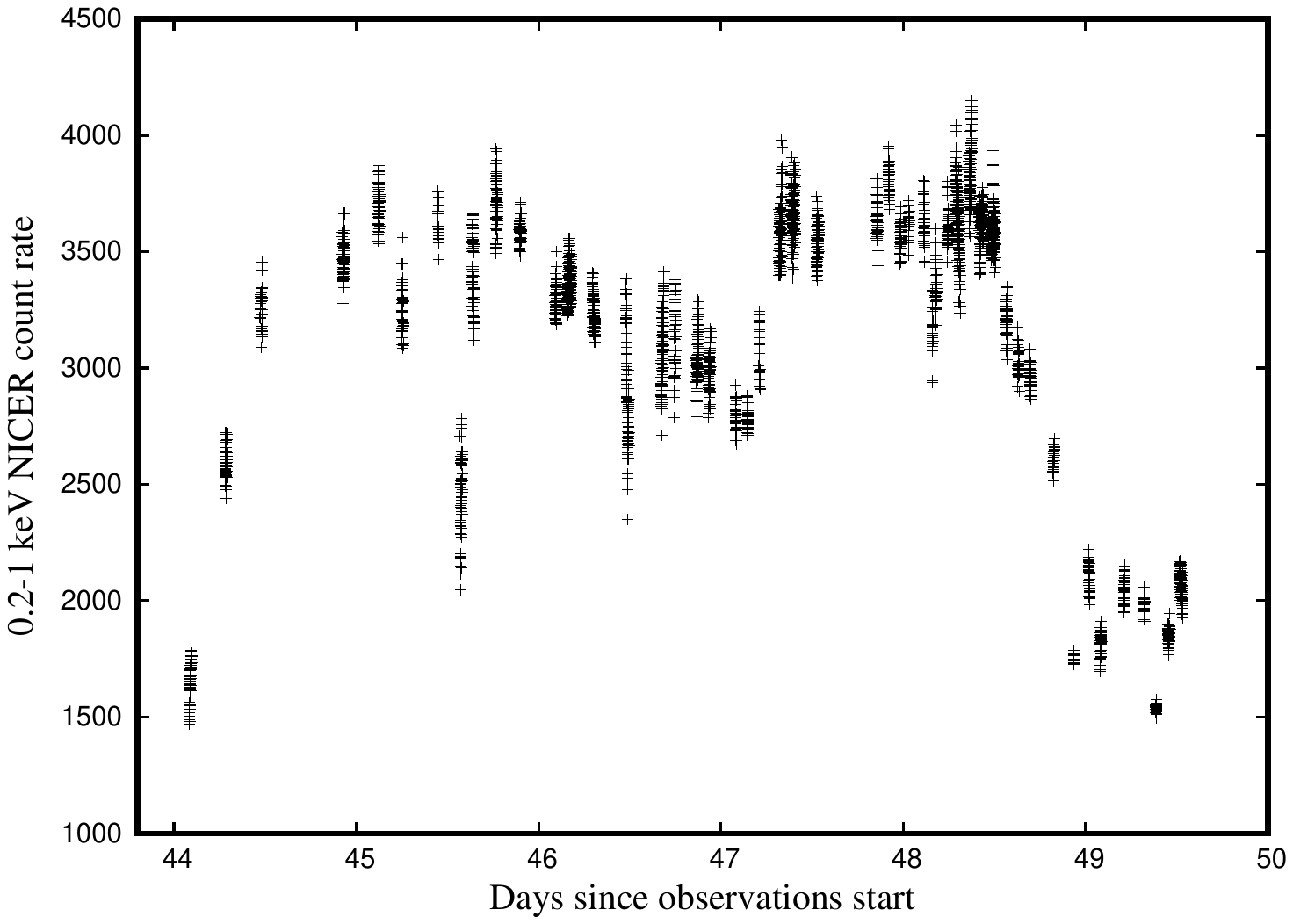}
\includegraphics[width=87mm]{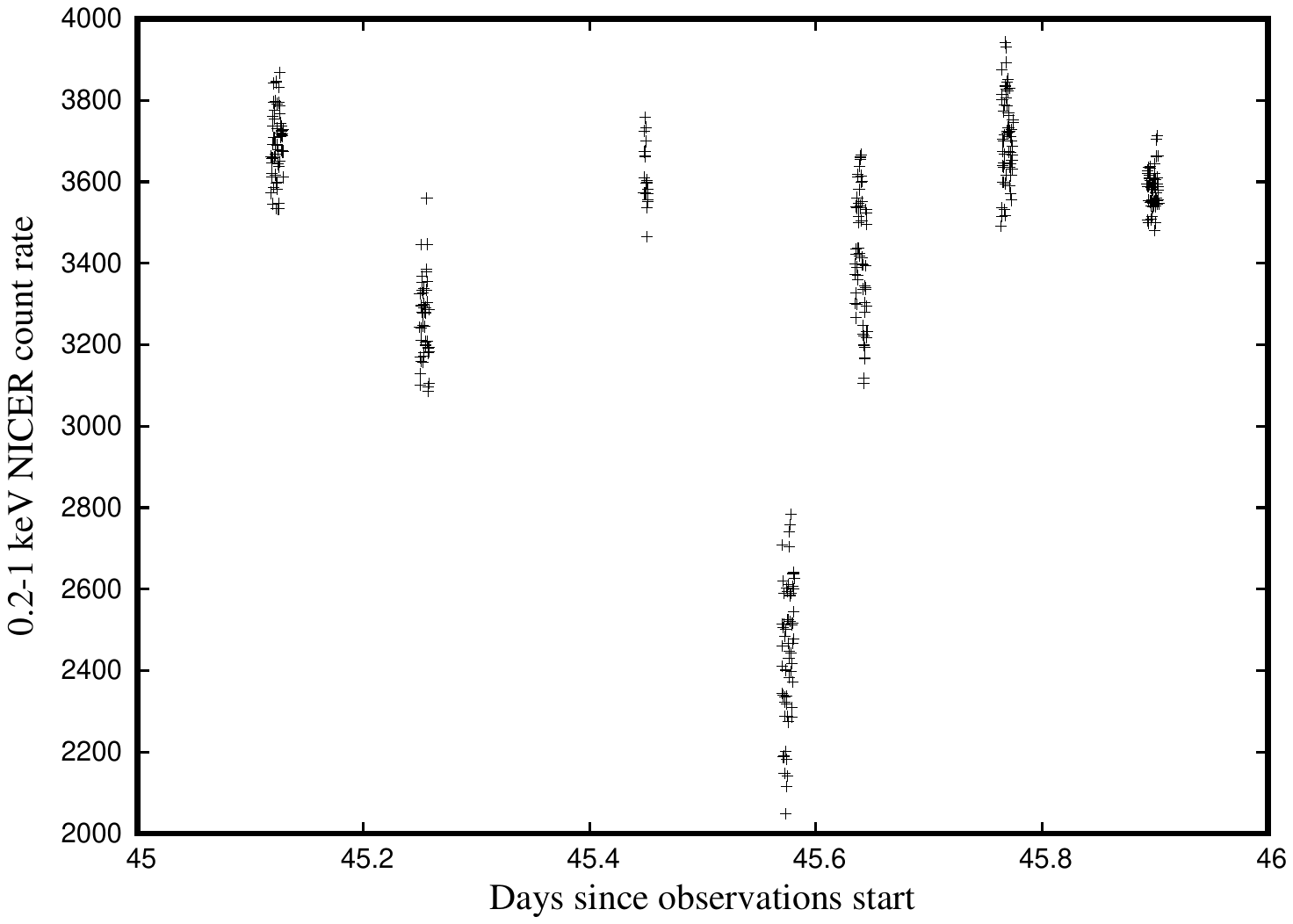}
\end{center}
\caption{The light curve (0.2-1 keV range, non-background subtracted)
 for the days of largest supersoft luminosity, between days
 44 and 50, and  in more detail
 for 2021 September 18-19 (day 46 of
 the outburst.}
\end{figure}
\begin{figure}
\begin{center}
\includegraphics[width=83mm]{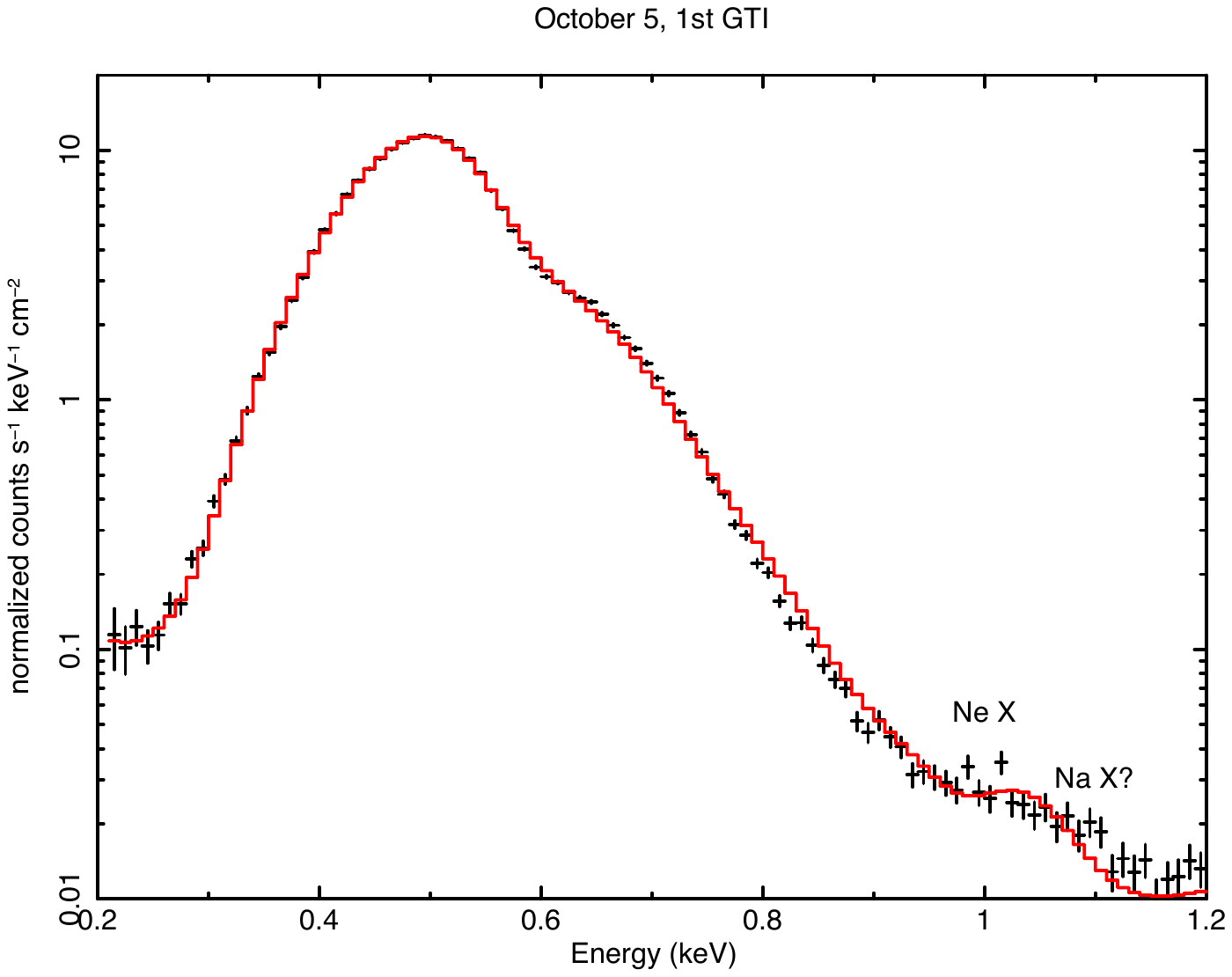}
\includegraphics[width=83mm]{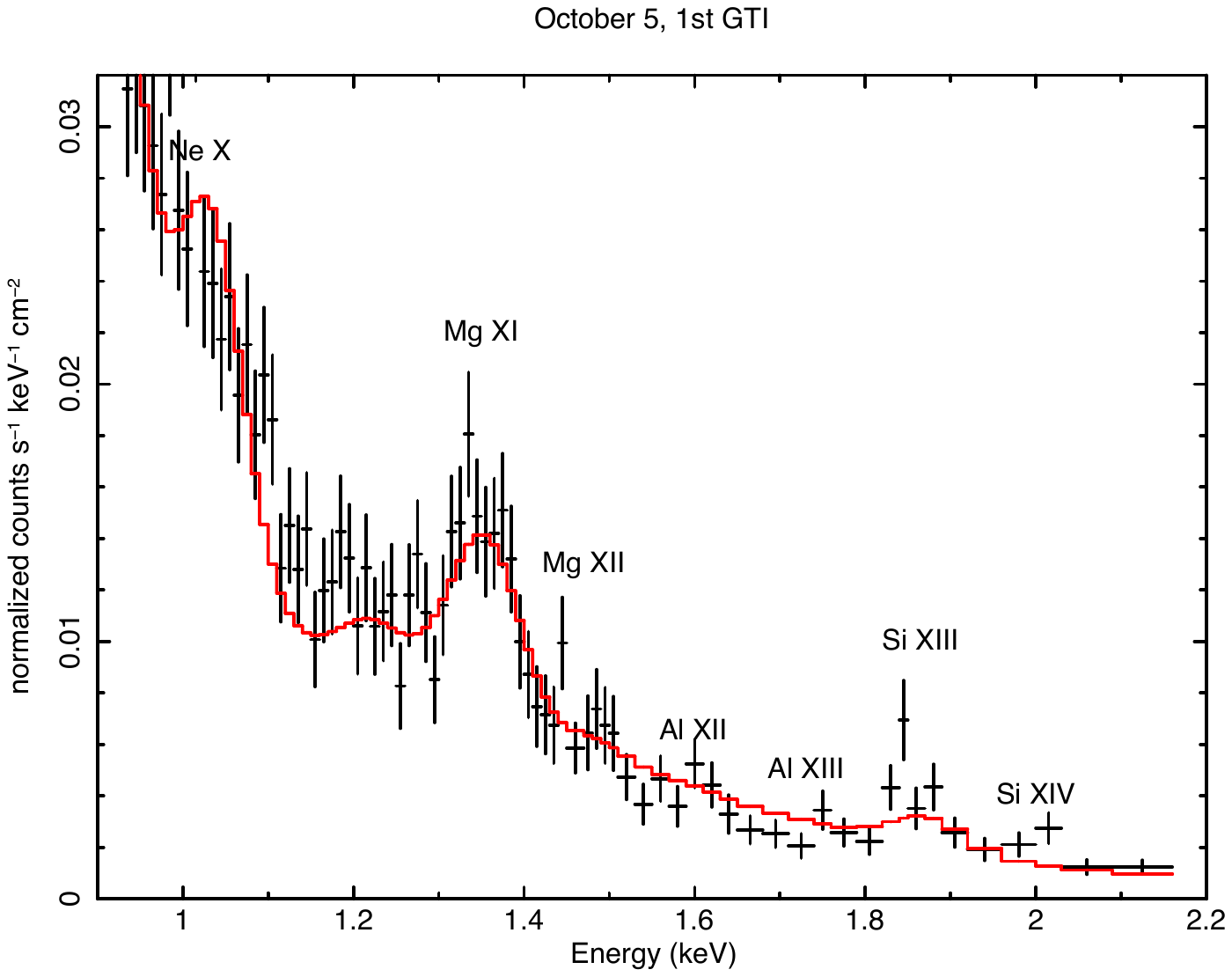}
\end{center}
\caption{The spectrum of RS Oph observed
 with NICER in the first $\simeq$1000 s long GTI of day 57 (2021 October 5),
 fitted with a blackbody at a temperature of 37 eV, 
 absorbing column density N(H)=4.3 10$^{21}$ cm$^{-2}$
with non-solar abundances of C and O
 (C/C$_\odot$=4.9, N/N$_\odot$=2.5), two components of BVAPEC thermal plasma
 at  81 eV and 462 eV. The flux is 1.74  $\times 10^{-9}$ erg s$^{-1}$,
of which the BVAPEC thermal plasma contributes only less than 2\%. 
 The total unabsorbed luminosity would be about 2 $\times 10^{40}$ erg s$^{-1}$ (a blackbody overestimates the luminosity, see text for the discussion).}
\end{figure}
\begin{figure}
\begin{center}
\includegraphics[width=140mm]{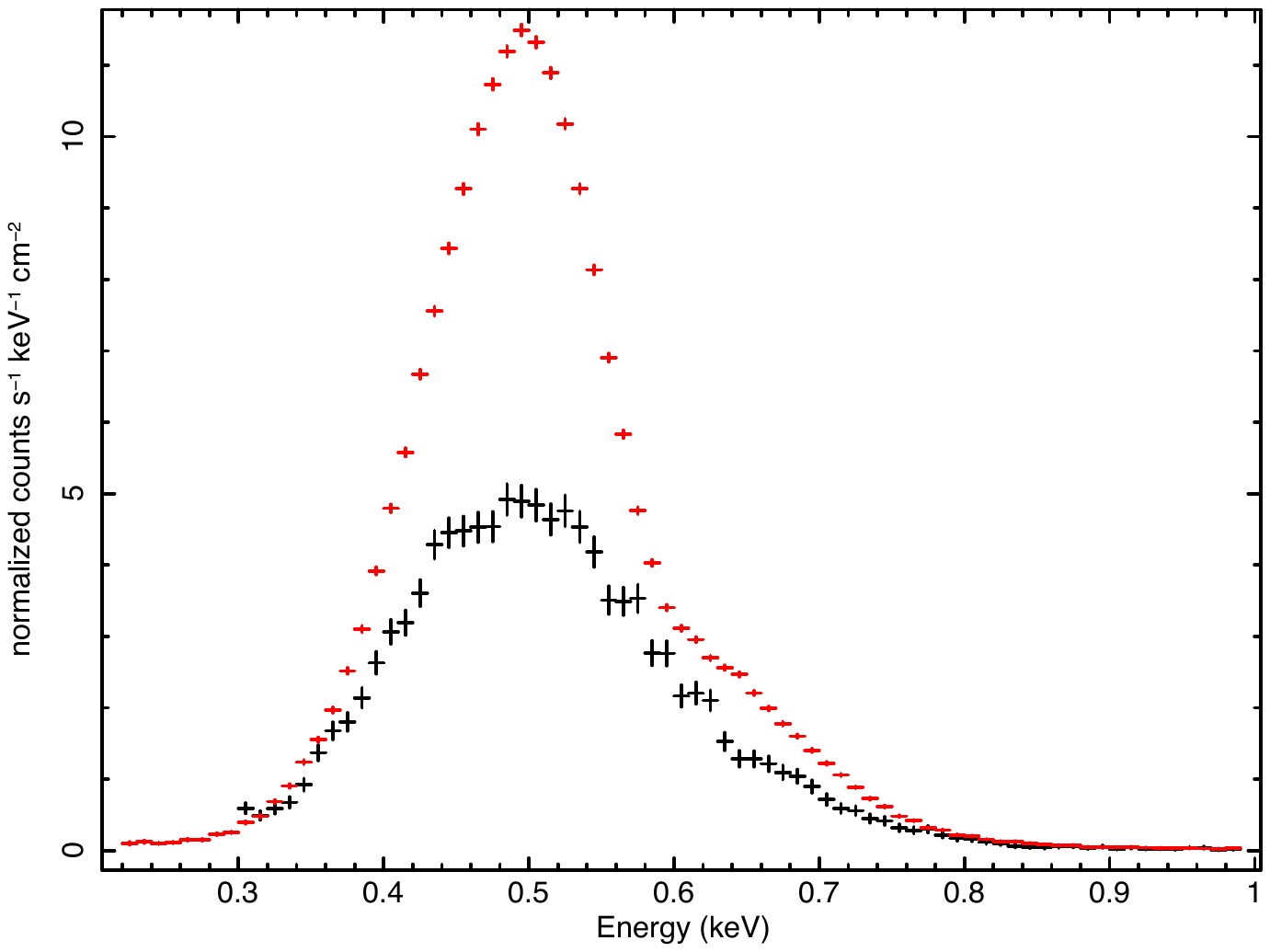}
\end{center}
\caption{Comparison between the RS Oph spectra below 1 keV obtained
 on day 57 (2021 October 5) with {\sl NICER} at 1:37:27 (red) and
 and the pile-up corrected Swift XRT spectrum of the same day during
 a 1 ks GTI starting at 4:37:22 (in black),
 extracted from an annulus around the center of the PSF with
 the \url{https://www.swift.ac.uk/user_objects/} webtool.}
\end{figure}
\begin{figure}
\begin{center}
\includegraphics[width=83mm]{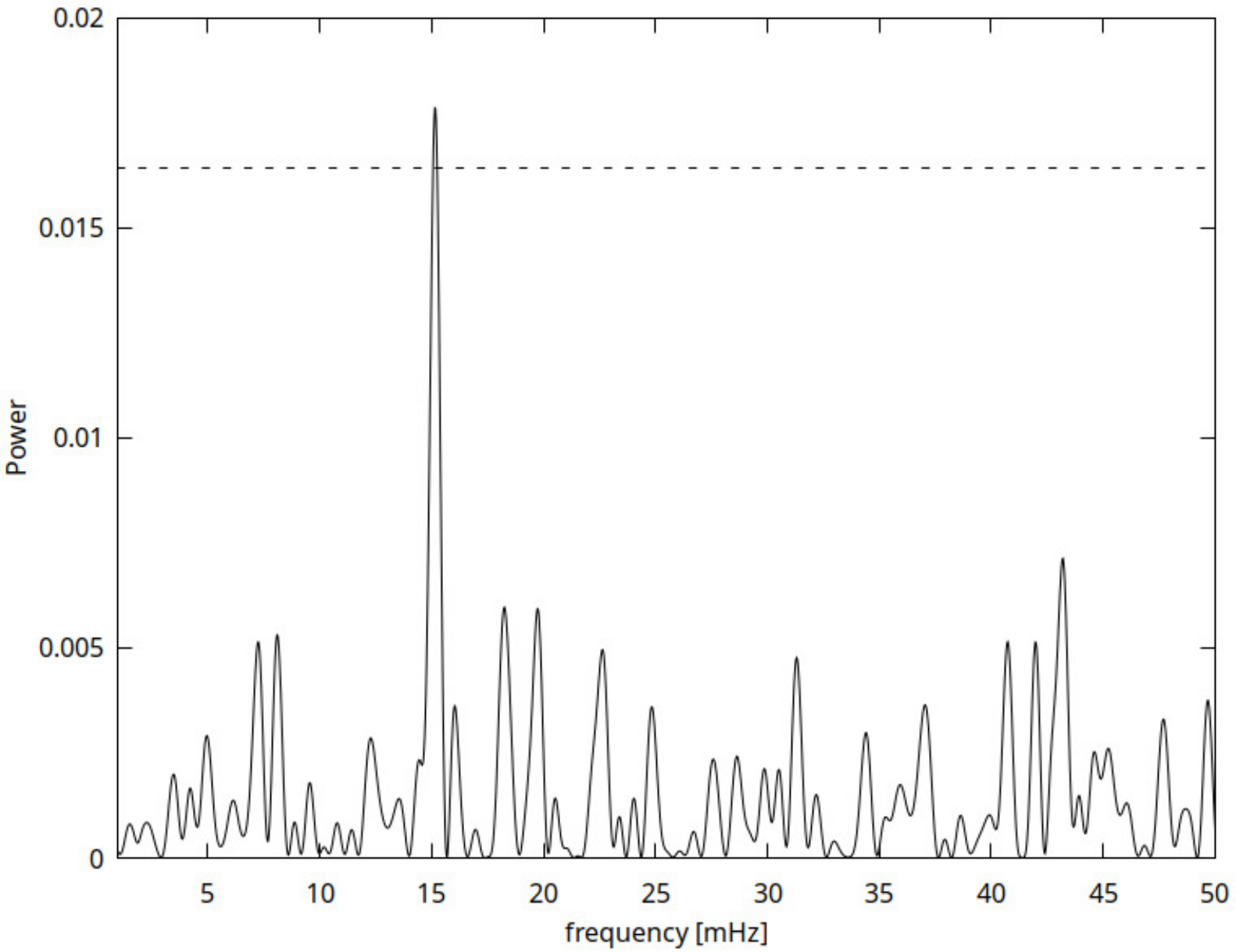}
\end{center}
\caption{Lomb-Scargle periodogram of the light curve of the 7th GTI of observation 4202300113, 2021 August 22 (day 13).
  The dashed line shows the 99.99\% confidence
 level.}
\label{15mHz_frequency}
\end{figure}
\begin{figure}
\begin{center}
\includegraphics[width=83mm]{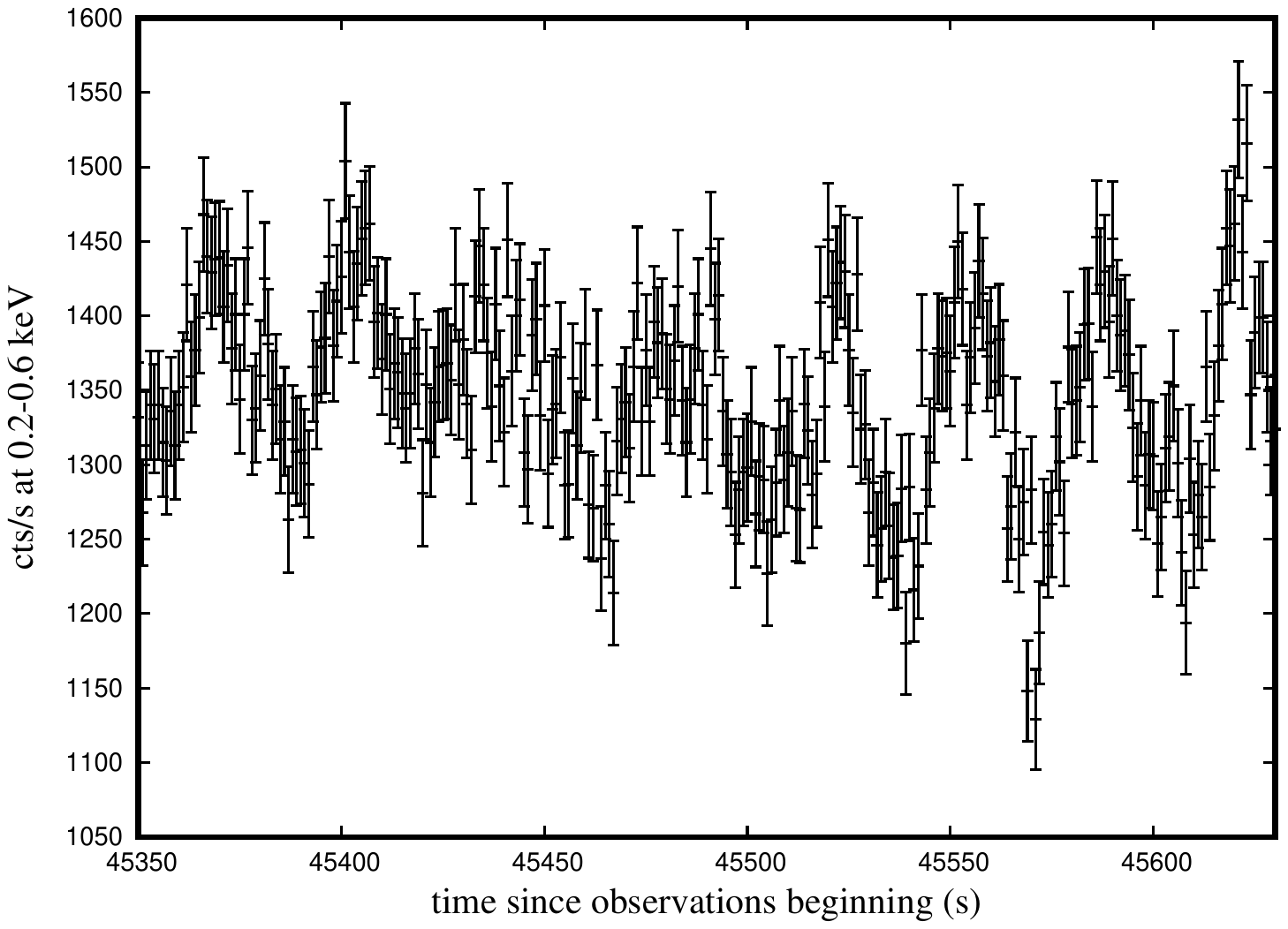}
\includegraphics[width=83mm]{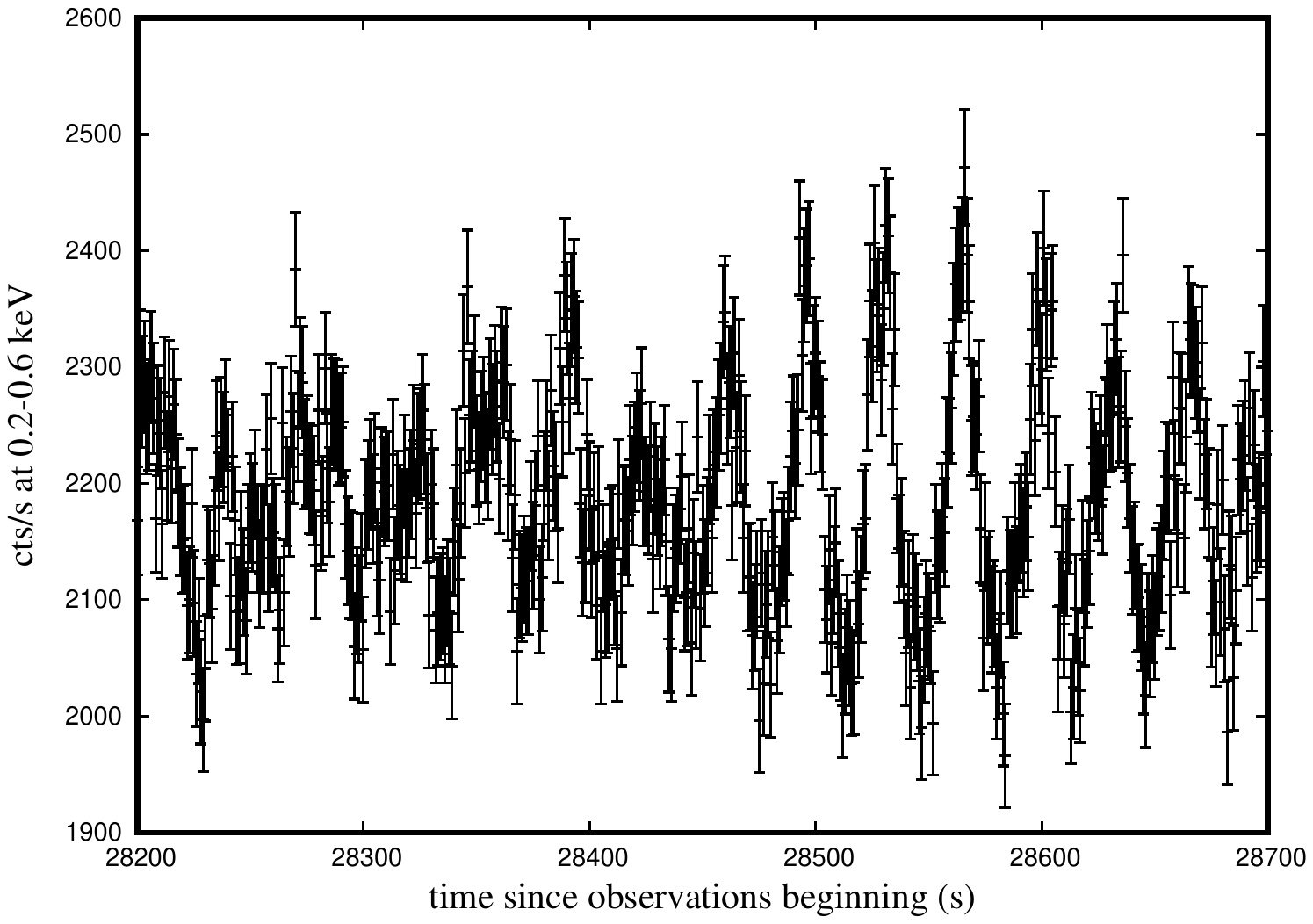}
\end{center}
\caption{Two examples illustrating how
 the modulation with the $\simeq$35 s period is evident even 
visually. On the left, a portion of the RS Oph NICER
 lightcurve on 2021/9/18 (day 40) and on the right, a
portion of the NICER lightcurve on 2021/9/19 (day 41), both
measured  in the 0.2-0.6 keV range with 1 s bins.
}
\end{figure}

We also investigated another possibility: that dust may have caused the variability
 and contributed to the observed
X-ray spectrum. RS Oph is not known to have ever formed a significant 
amount of dusti in previous oitbursts, but there was evidence of dust 
2 days after the 2021 optical maximum (Y. Nikolov, private communication).
 In any case, with such a large
supersoft luminosity, even a small amount of dust may affect the X-ray light curve and 
spectrum. The variability time scale is hours, and the flux amplitude variation is
 very large, up to 100\%.
We refer to work done a few years ago by \citet{Corrales2016, Smith2016, Heinz2016}.
 Dust scattering may cause variability, because the scattered X-rays traverse a
 longer path than the X-rays directly received from the source,
 creating an echo of the X-ray flux of the source. However, amplitude
variations of order 100\% are not caused by interstellar dust scattering
 (i.e., scattering from dust not intrinsic to
the source), since scattering cross sections are generally very small and the
 scattered component will emit little flux
compared to the flux from the source itself, given the column density towards the source.
Dust scattering on time scales of hours for reasonable scattering angles of order 
$\theta_{\rm sc} \lesssim 300\arcsec$ \citep[above which the
scattering cross section at soft energies drops precipitously, see][]{draine03}
 would imply that the dust is at distance of several
hundred parsecs or more, given the relation between distance $d$ 
from source to dust and time delay $\Delta t$:
\begin{equation}
d \approx \frac{c \Delta t \cos{\theta_{\rm sc}}}{1 - \cos{\theta_{\rm
sc}}}
\end{equation}
In addition, by virtue of being an echo, if the source
 is variable, dust scattering by intervening
 static clouds cannot produce variability on
time scales shorter than the intrinsic source variability.

The observed variability time scale can be created if the scatterer itself 
is moving rapidly enough, relative to the
line of sight, for the scattering angle to change by more than $\sim 300 \arcsec$.
 In this case, the variability does not arise
from a dust scattering echo but from the dynamically changing scattering geometry.
 In this scenario, dust clouds
in the outflow or in the circumstellar environment of RS Oph would need to 
move rapidly enough, be numerous enough, and
each scatter enough of the source flux to create not only the observed time scale,
 but also the repeated nature and
amplitude of the observed variability, respectively. To produce the observed amplitude
 of the variability (about 100\%),
a scattering cloud would need to have a scattering optical depth of order unity:
 if the scattering optical depth is too small, the scattering amplitude cannot be
 as large as 100\%, but if it is too large, multiple scattering and photo-electric
absorption would only attenuate the signal. Furthermore, the clouds would need 
to subtend a solid angle of order $(300")^2$ beyond which the scattering 
cross section decreases rapidly and below which too small a fraction of the source
flux would be scattered. To produce the frequent rapid changes would require
 many such clouds passing close to the line-of-sight within the observation.
 The required velocities transverse to the line of sight (and thus perpendicular to
the nova outflow) would need to exceed
\begin{equation}
v_{\perp} \gtrsim 60\,{\rm km\,s^{-1}} \frac{d}{1\,{\rm AU}}
\end{equation}
Dust scattering, in concert with photo-electric absorption (i.e., X-ray extinction) due to short-term occultations by
intervening clouds, could explain strong variability along with associated spectral changes, but there are also other
considerations. The time scale of hours implies a dust cloud at a distance of the order of a parsec, but on the other
hand, at such a distance the dynamical time scale of a dust cloud is too long to cause changes in the overall scattering
geometry within hours. If we assume that the absorber must pass in front of the source, it would be at a distance of
the order of 1 AU, matching the orbital separation, but at such a distance the variability time scale would be even longer, of the order of the orbital period.

In summary, the combination of the observed variability and amplitude time
scales is difficult to reconcile with
stochastic dust extinction by dynamically intervening clouds, and is inconsistent with the signatures of dust scattering
halos and echoes.

\section{The $\simeq$35 s pulsation (and a possible precursor)}
On day 13 (2021 August 22, observation id. 4202300113, 7th GTI)
 we measured a periodic modulation with a 15 mHz frequency.
 The duration of the exposure 
is 1831 seconds ($\approx $27 cycles of a 66.7 s period). The
power of the peak has high confidence, 
 with a false alarm probability (FAP)  $< 0.0001$. 
Given the confidence and exposure duration, the signal is real.
 The Lomb-Scargle (LS) periodogram \citep{Scargle1982} is shown in  
Fig. \ref{15mHz_frequency}. This was the first and only case
 of a clearly measured periodicity until day 26 (2021 September 4). 
The measured frequency is just a little more than half the
 frequency corresponding to the $\simeq$ 35 s period measured during 
 the SSS phase, discussed
 in the rest of this Section, so it is definitely intriguing. 

A quasi-periodic modulation with a $\simeq$35 s was measured in 2006
 for a large portion of the supersoft X-ray phase.  This modulation
 appeared again on 2021 September 4 \citep{Pei2021}, and
 a drift in the period was evident from the beginning. As announced in
 Astronomer's Telegram 14901 \citep{Pei2021}, the period was 36.7$\pm$0.1 s on
 September 4, but on day 27 (2021 September 5
 it was  instead measured  to be 34.88$\pm$0.02 s, with a larger amplitude
 (up to 10\%) 
  than on the previous day.  The errors in the periods were 
 derived from the statistical uncertainty on the frequency, estimated  
 in this case  and
 in the following text by fitting a Gaussian to the periodogram peak,
 and considering 1$\sigma$.
 It is remarkable that, like in 2006, the pulsation appeared for the
 first time during significant flaring activity (see Fig. 9). High spectral
 resolution obtained with the {\sl XMM-Newton} RGS gratings in 2006
 revealed that a soft flare on day 26 of the outburst was still due mainly
 to flux in emission lines, not in the stellar continuum \citep{Nelson2008}.

 Initially, this oscillation was not always measurable in all exposures, 
 but we suggest that this was due to varying amplitude.
 However, the modulation became very evident two days later, and
 after September 17,
 it was measured during all the supersoft X-ray phase until
 mid October 2021. Fig. 16 illustrates how definite the modulation was.
\begin{table}
\begin{center}
\caption{Continuous exposure intervals used for the power spectra:
 beginning, end (modified Julian date) and duration of exposure.}
\begin{tabular}{ccc}
\hline \\
 MJD start      & MJD end          &  exp. time (s) \\
 \hline \\
59475.89615288  &  59475.90808478  &  1030.917 \\
59476.02445090  &  59476.03723922  &  1104.911 \\
59476.28358148  &  59476.29551339  &  1030.916 \\
59476.86411679  &  59476.87704398  &  1116.909 \\
59476.92874031  &  59476.94163277  &  1113.909 \\
59476.99323651  & 59477.00617528   &  1117.909 \\
59477.05788317  & 59477.07055575   &  1094.911 \\
59477.31609947  & 59477.32937384   &  1146.906 \\
59477.96195267  & 59477.97454424   & 1087.911 \\
59478.09104925  & 59478.10425418   &  1140.906 \\
59478.15563804  & 59478.16843792   & 1105.909 \\
59478.28476934  & 59478.29761551   &  1109.909 \\
59479.38282513  & 59479.39565972   & 1108.908 \\
59480.22254893  & 59480.23435352   & 1019.916 \\
59481.46442242  & 59481.47946711   & 1299.861 \\
59481.59649327  & 59481.60871416  & 1055.886 \\
59482.16387018 & 59482.17855635  & 1268.885 \\
59482.22881776 & 59482.24243922 & 1176.894 \\
59482.69076498 & 59482.70643463 & 1353.858 \\
59482.82094946 & 59482.83561227 & 1266.866 \\
59483.59621204 & 59483.61063182 & 1245.869 \\
59483.65833582 & 59483.67520907 & 1457.849 \\
59483.72554014 &  59483.73979790 & 1231.871 \\
59483.77891468 &  59483.80438673 & 2200.785 \\
\hline \\
\end{tabular}
\end{center}
\end{table}
\subsection{Power spectra of the longest continuous exposure} 
We first performed a barycentric correction, and since in order to study the 
timing properties it is useful to have as
long an uninterrupted interval as possible, we performed
 the analysis on good time intervals (GTI) that were at least 1000 s
long (there were 24 such intervals). These GTIs are reported in Table 5.
 We considered events in the 0.3 - 0.9 keV range. We generated light curves sampled
 at 8 Hz.  We note that the mean count rate of each interval 
ranged from a low of 409 s$^{-1}$ to a high of 3526 s$^{-1}$.
 We padded the light curves to a length of 2400 s, using the mean
value, so that the power spectra are all defined on the same frequency grid.
 We then computed the average power
spectrum of all the intervals. The fundamental and 1st overtone frequencies of the quasi-periodic oscillation (QPO) are
clearly seen in the power spectrum in Fig. 17. We fitted a model including a power-law 
and two Lorentzian functions
to model the QPO (fundamental and 1st overtone) and found a good fit. Fig. 17
 also shows this best fitting model (red curve). The ratio of the overtone to to
 fundamental frequency is 1.9989$\pm$0.0087, consistent with a factor of 2.
The coherence (Q=$\nu_0/\delta \nu$)  (where $\nu_0$
 is the measured frequency and $\delta \nu$ the relative 
 statistical uncertainty, defined above) for the fundamental 
 is 10.53$\pm$0.22,
 and for of the overtone it is 9.53$\pm$1.1. The individual
power spectra show a good deal of variation: some have a single peak, 
while others show a multi-peaked structure,
sometimes with two or three clear peaks. Fig. 18 compares two power spectra
 (9th and 22nd GTIs in Table 5), one with
a single main peak and another with three. The frequency can wander from about 
0.025 to 0.033 Hz over periods of 1200 s.

\begin{figure}
\begin{center}
\includegraphics[width=87mm]{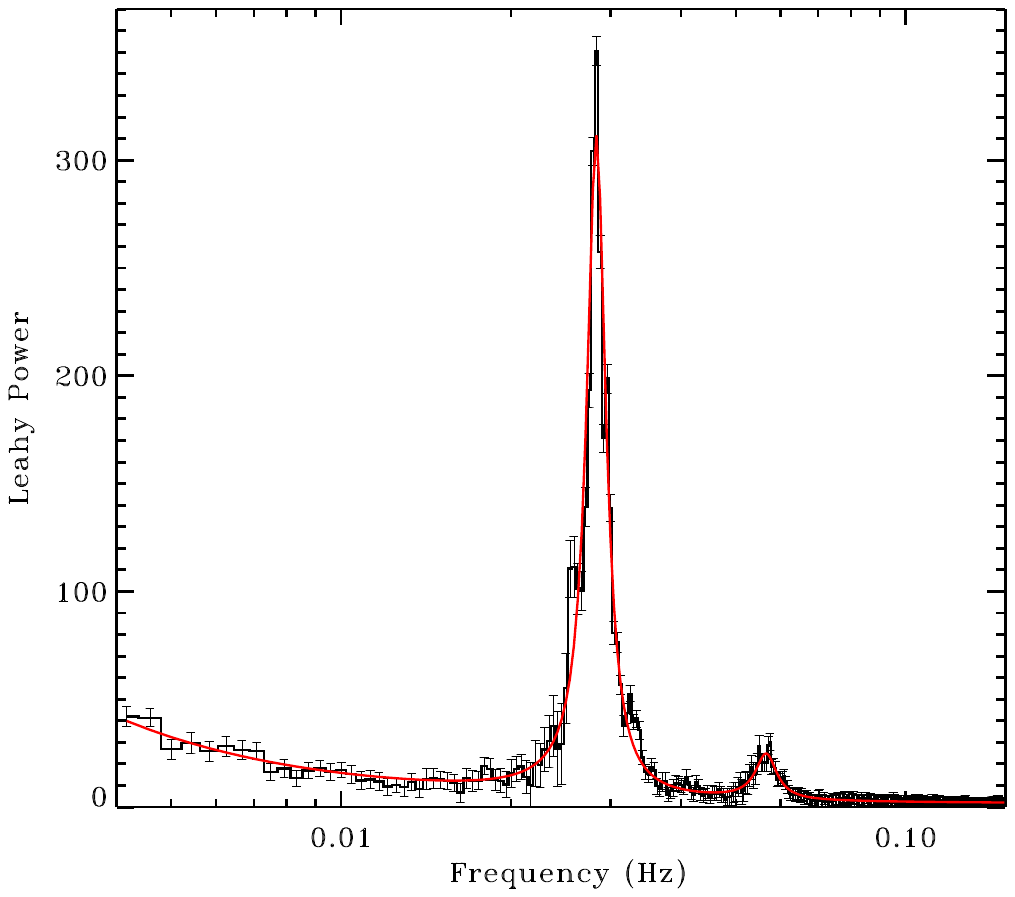}
\includegraphics[width=87mm]{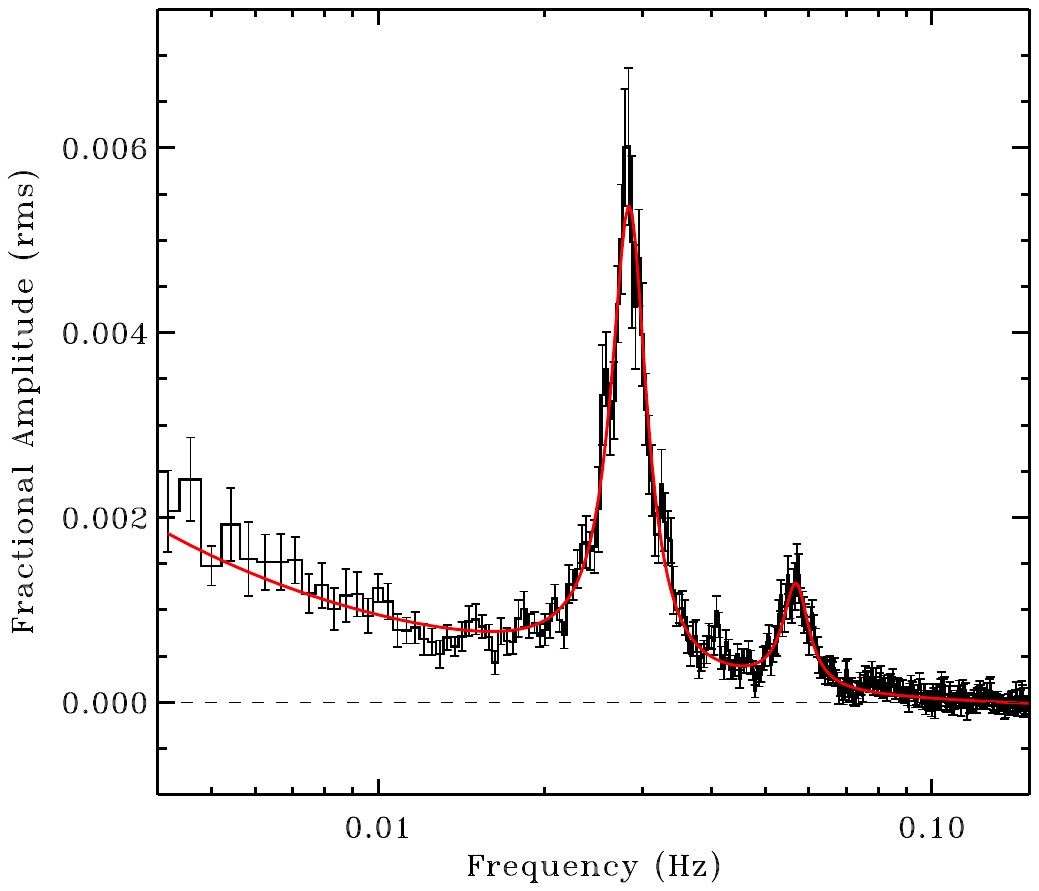}
\end{center}
\caption{
On the left, the power spectrum of the 24 continuous exposures longer than 1000 s (those in Table 5), and the best
fit with the model parameters in Table 6. On the right, the average rms spectrum, computed by dividing each power spectrum
by the mean count rate in each GTI and fitted with a two Lorentzian model.}
\end{figure}
\begin{figure}
\begin{center}
\includegraphics[width=140mm]{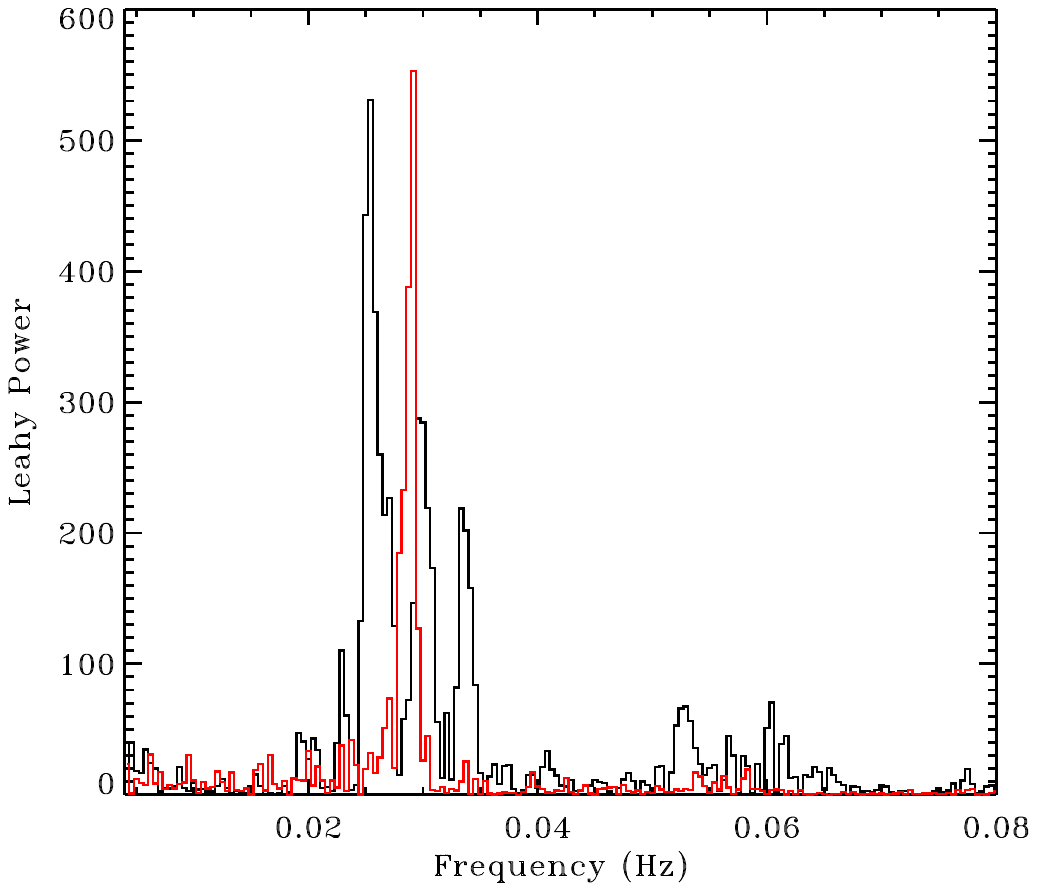}
\end{center}
\caption{Comparison of the power spectra of the 9th GTIs (black) and the
 22nd (red) of Table 5. Note that for purposes of comparison
the spectrum for GTI 22 has been reduced by a factor of 2.5 to fit within the plotted range.
}
\end{figure}

The light curves also suggest that the oscillation amplitude changes over periods
 of a few hundred seconds, ranging
from a fractional amplitude (the sine wave amplitude relative to the mean)
 from less than 2\% to a maximum of about 10\%. In Fig. 17 we also show in the 
right-hand panel the average rms spectrum, computed by dividing each
power spectrum by the mean count rate in each GTI and fitted with a two Lorentzian model. The parameters of
the Lorentzian and first overtone are reported in Table 6. The rms amplitudes obtained from integrating the two
Lorentzians are 9.41\% and 3.08\% (fractional rms) for the fundamental and 1st 
overtone, respectively. For several of
the intervals with relatively ample oscillations with a single main peak in the power spectrum, the light curve can
be folded at the best frequency, and an average QPO pulse profile is obtained. 
An example of this for the 17th GTI of Table 5 is shown
in Fig. 19. The phase folded light curve shape slightly deviates from a simple sine wave, as a component at the first
overtone frequency is required to adequately fit the data. In this case the relative amplitude of the fundamental and first overtone is 7.3.
\begin{table*}[t]
\caption{Model parameters for the fundamental and 1st overtone QPO peaks
obtained from a fit to the average power spectrum shown in Fig. 17 (left). }
\label{tab:QPO properties}
\begin{center}
\begin{tabular}{cccc}
\tableline\tableline
QPO Component & $P_{max}$ (Leahy Power) & $\nu_0$ (mHz) & $\delta\nu$ (mHz) \\
\tableline\tableline
Fundamental & $306.14 \pm 4.31$ & $28.316 \pm 0.021$ & $2.690 \pm 0.055$ \\
1st Overtone & $21.14 \pm 1.70$ & $56.600 \pm 0.242$ & $5.936 \pm 0.679$ \\
\tableline\tableline
\end{tabular}
\end{center}
\end{table*}
\begin{figure}
\begin{center}
\includegraphics[width=140mm]{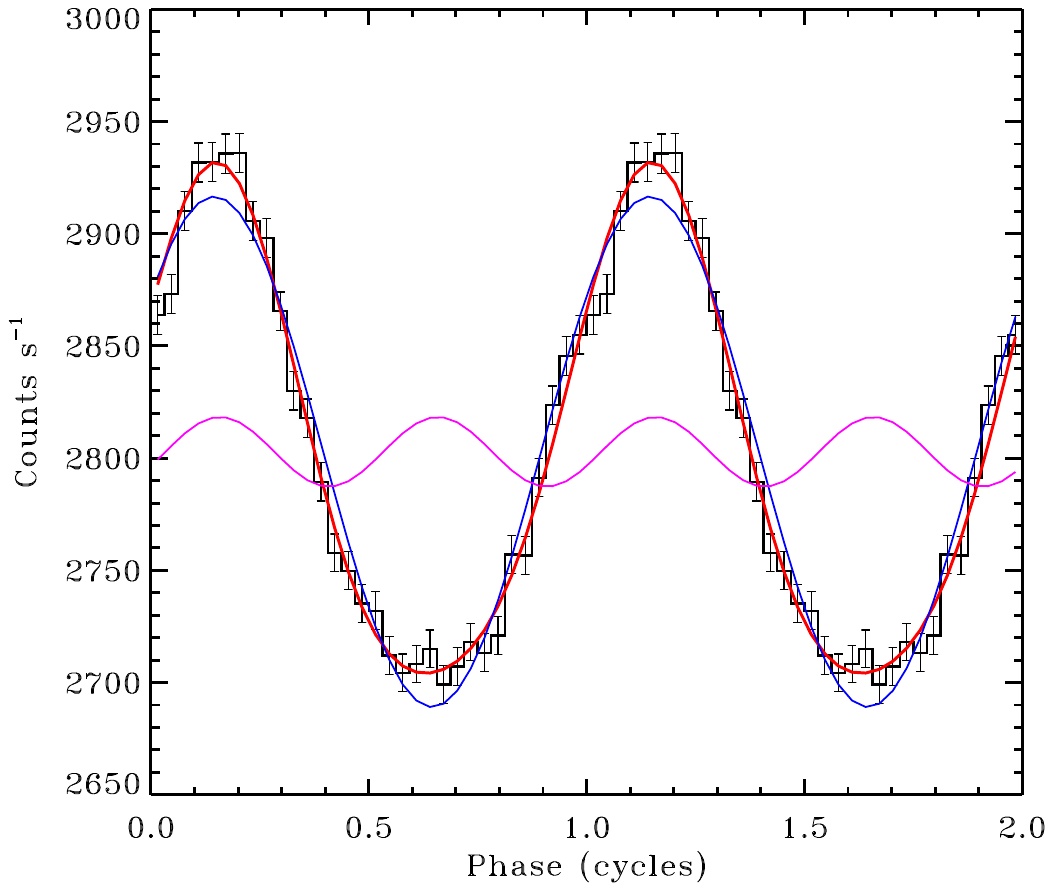}
\end{center}
\caption{
Folded QPO pulse profile for the 17th GTI  of Table 5 fitted with a model that includes a sine function at the
fundamental and the first overtone with relative amplitude of 7.3. The full model (red), fundamental (blue), and first overtone
components (magenta) are also shown.}
\end{figure}

\subsection{Statistical modeling of the short-term variability evolution}
We have described above how in some GTIs the frequency seems stable, 
but in the majority of them, we measure a QPO rather than a stable period, given the apparent wandering of the frequency. Is it really a ``QPO''?
Sometimes the drift of the measured frequency is not real, but it is an artifact of the changing amplitude as in the
case shown in \citet{Dobrotka2017}. In order to evaluate this possibility, we performed a statistical analysis, like
previously done in \citet{Orio2021a} for N LMC 2009 and in \citet{Orio2022b} 
for CAL 83.

For this analysis we considered  
the light curves in the 0.2-0.6 keV range, in which the amplitude of the oscillation is the largest, and
we used all continuous exposure times after the SSS emerged, longer than 700 s (that is, GTIs with at least 20 cycles).
 To detect and remove outlier points, we used the Hampel filter\footnote{Python Hampel library \url{https://github.com/MichaelisTrofficus/hampel_filter}}. 
For the light curve of each exposure, we computed an LS periodogram. 
 We simulated the light curves using a sinusoidal function:
\begin{equation}
\label{eq:flux_eq}
\psi = \phi + P_a\sin(2\pi t / P_p),
\end{equation}
where $\phi$ is the mean GTI count rate, $P_a$ and $P_p$ are polynomials 
representing the amplitude and the period, respectively.
Amplitude and period were obtained by generating a distribution of 
Gaussian points
 around their mean values in the selected GTIs.
The mean modulation amplitude for each GTI was estimated from the phase-folded light curve by simple sine fitting,
while the period was chosen randomly in the 20-40 mHz interval. We fitted the Gaussian random points with 25 degree
polynomials, our input functions $P_a$ and $P_p$. 
When the modulated amplitude function $P_a$ turned out to be negative, we
assumed P$_a$ = 0 (a negative amplitude is meaningless), 
indicating that the modulation was below detection threshold
for a while. 
\begin{figure}
\begin{center}
\includegraphics[width=150mm]{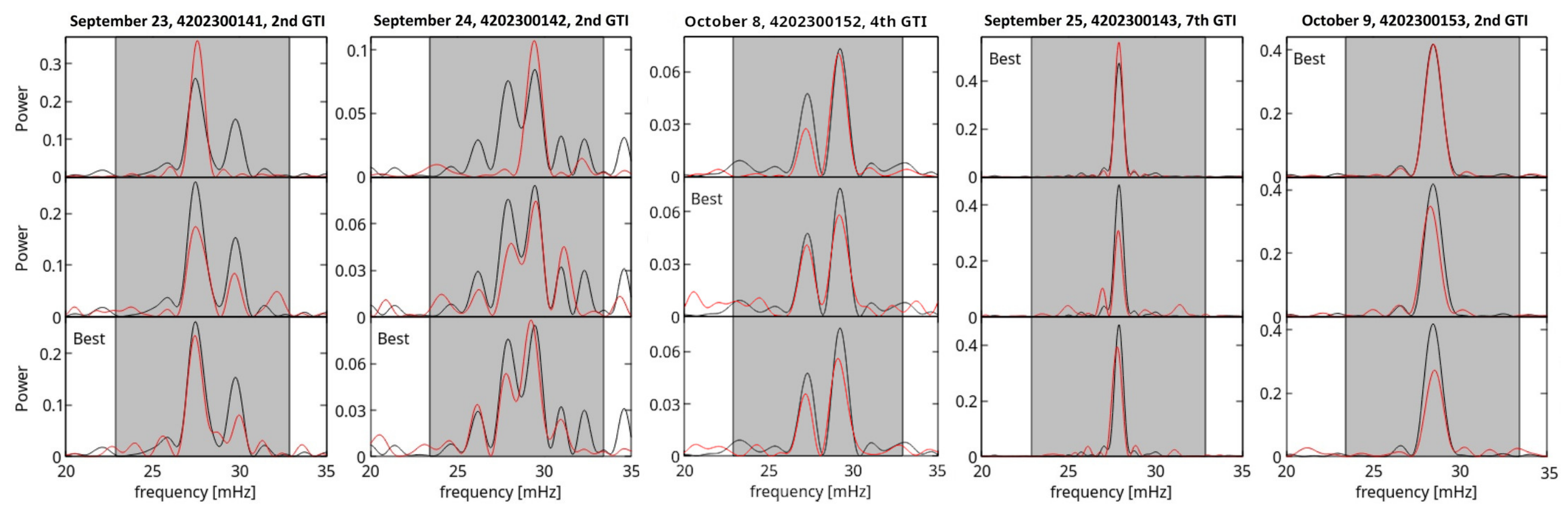}
\caption{Examples of each type of variability with the
 best simulations (red lines) for the corresponding class of models. 
 the top panels show stable periodicity with variable amplitude, the
 middle panels variable periodicity with stable amplitude, 
and in the bottom panels both parameters are assumed
 to be variable. For each observations, the best type
 of simulation is the one with the label "Best". 
We compare these simulations with the LS periodograms with $FAP > 0.01$,
 traced in black. The sum of residual squares was calculated over the 10 mHz
 shaded interval around the peak.}
\label{LSs_comparition}
\end{center}
\end{figure}
\begin{figure}
\begin{center}
\includegraphics[width=140mm]{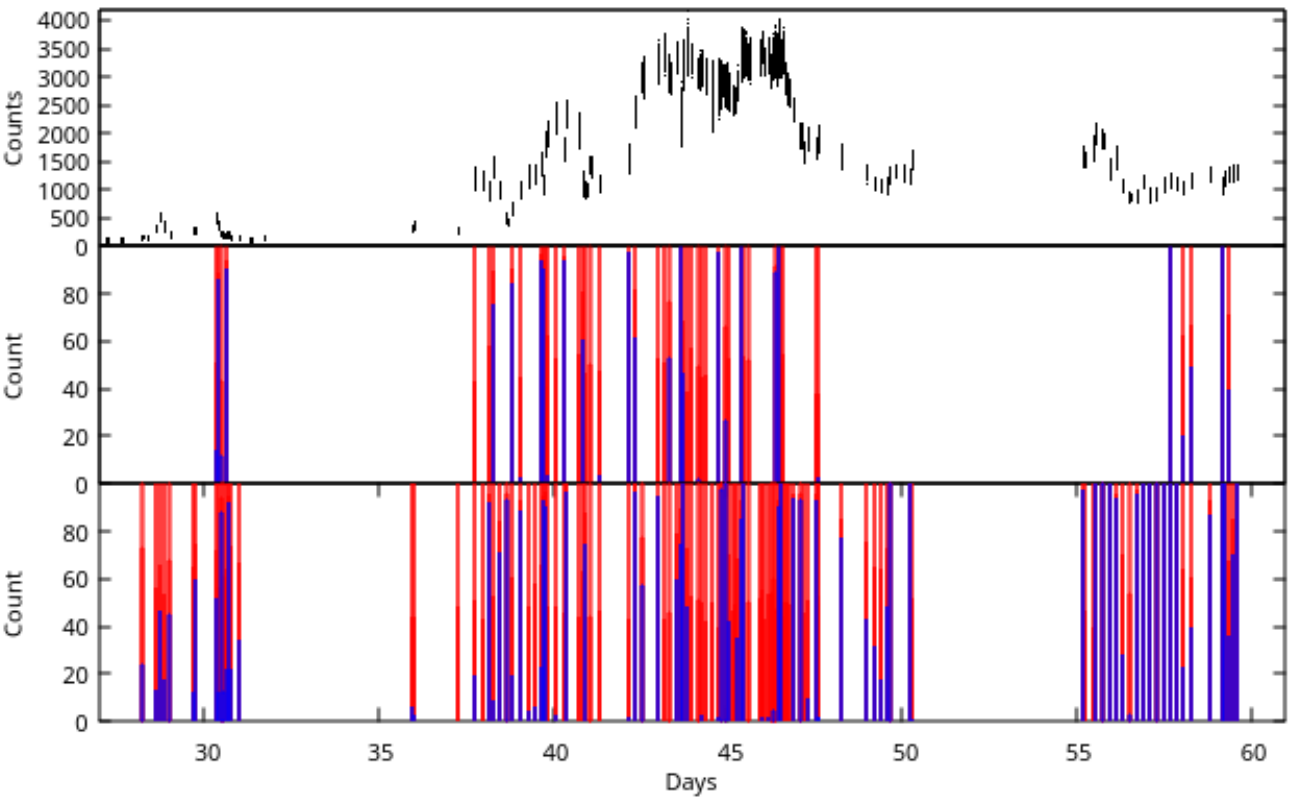}
\caption{To visualize how the variability patterns changed during the {\sl NICER} monitoring, the stacked bars show the number of best simulations 
(chosen from the best hundred out of 100,000) with each type of variability for each GTI: the number of best simulations with stable period and variable amplitude is plotted in blue,
 while the number of best simulations with variable period 
(or variable period and amplitude) is plotted in red. 
The upper panel represents the light curve for reference. 
 The bottom panel shows the results for GTIs
 in which the signal was detected with $FAP < 0.1$ and 
 the middle panel shows the result for GTI exposures with $FAP < 0.001$. }
\label{Stacked_bar}
\end{center}
\end{figure}

For three modes of variability (namely constant period with variable amplitude, 
variable period with
constant amplitude, and variable both amplitude and period), and for each GTI,
 we run 100,000 simulations with
relative LS periodograms to compare with the actual LS periodogram of each GTI exposure.
 We selected the best
simulations for each GTI, by calculating the sum of square residuals in a 10 mHz interval around the highest peak
(see shaded area in Fig. 20, that shows 
relevant examples). 
We chose GTI exposures done between days 30 and 60, in which the variability was
well detected, and we selected those for which the periodogram has false alarm probability
 FAP$<$0.001. We chose
the 100 best simulations (with lower residual squares) for every GTI, and plotted a stacked bar chart to visualize the
fraction of simulations with and without varying frequency.
 This is shown in the middle panel of Fig. 21.

We considered that focusing only on the high confidence detections can lead to artificial selection of cases with
non-varying amplitude, and we wanted to avoid a possible selection effect due to the fact that, if the amplitude varies,
the peak power and its confidence may also decrease. Thus, we repeated the statistical analysis for GTI exposures
done between day 26 and 60 with a minimal duration of 350 s (at least 10 cycles) and a less confident detection of the
signal, FAP$<$0.1, increasing the number of examined GTIs. Because the GTIs with low confidence detection can
be misleading due to random ``noisy'' peaks, we excluded those with periodograms that did not show the dominant
feature around the 35 s periodicity. At the beginning, around day 30, there are solutions with both stable and variable
period. Including the lower confidence data, the variability of the period is more obvious in the period of highest SSS
flux, with practically no stable period solutions around days 41 and 44. Between days 55 and 60 the period seems to
have stabilized, while the amplitude of the oscillation may have remained variable. We concluded that the 35 s period
was becoming stable over the short time scale of minutes of
 the single GTIs in the late outburst phase, but it was indeed drifting
over such scales early as the
 SSS emerged, and at peak SSS flux. Fig. 20 shows examples of the simulations for different dates. Periodograms
with stable periodicity have a typical single peak\footnote{Even if these simulations were made with variable amplitude they comprise also solutions with relatively stable amplitude (due to the
randomness of the process)} (see September 25th and October 9th). These simulations generate
blue bars in Fig. 21.
 We see in the top panel for October 8th that the simulations with stable period and variable
amplitude split the signal into two peaks, but the solution with variable frequency (middle panel) is a better match.
Both simulated solutions are very similar, and the difference is in the amplitude of the peaks. Both models, with
variable or stable frequency, may describe the observed periodograms relatively well and can be counted among the
selected best 100 cases. For this reason there are partially blue and partially red bars for most dates in Fig. 21, but a GTI like the one whose power spectrum is plotted in black in Fig. 18  
generates almost only blue bars in Fig. 21.

\subsection{Long-term variability of the 35 s signal}
In addition to testing the drift of the period on time scales of minutes of the individual GTIs, we also wanted to
investigate whether the mean period appears to be stable on long term time scales and/or whether there is any trend
towards longer or shorter periods. We selected the most stable GTIs, namely those with at least 90 (out of 100)
stable-period-simulations fitting the measured periodogram, defined by almost entirely blue bars in the middle panel
of Fig. 21.
 This left us with only 8 GTIs (out of the 24 in Table 5) for which we can assume that there was no significant frequency drift. We evaluated the statistical error of the frequency by fitting Gaussians to the periodogram
peaks and considering 1$\sigma$\footnote{This may overestimate the statistical error, but in this case the important issue is not to underestimate it}. The result is plotted in the top panel of Fig. 22. There is no clear trend: we notice that there is still variability
on this time scale of days, although there is only a significantly different (larger) frequency at the beginning, on day
30. We also calculated a periodogram (to compare with the average power spectrum of the 24 GTIs of Table 5, shown
above in Fig. 17) using only these "most stable, bluest" GTIs, and the result is shown in the middle panel. The
frequency measured using all the 8 stable-period GTIs is 27.8862$\pm$0.0012 mHz,
 corresponding to an average period
of 35.86 s. Using instead all the GTIs of Table 5 (all long, uninterrupted ones), as reported above we obtained a 
fundamental frequency of 23.316$\pm$0.021 mHz, corresponding to a 35.316 s period.
 The frequency of the combined
``mostly red/drifting'' GTIs is instead 28.5062$\pm$0.0009 mHz, corresponding to an average period of 35.08 s. When
we compared the phased light curves of the stable GTIs with the ``red''
 ones with large period drifts on time scales
of minutes, in the bottom panel for Fig. 21, we found a much smaller amplitude (and a less definite shape) of the
modulation for the ``red'' GTIs, the ones with a significant period drift, supporting the conclusion that the drift in
period in those GTIs is real and we can really define this modulation a QPO. To summarize, we suggest that the
period drift on short time scales of minutes is real, and that it was also not stable over time scales of days. The only
long term trend we could assess as the days passed is that fewer GTIs showed the
period drift. Although comparing the frequency obtained from the power spectrum of the single GTIs does not yield
conclusive results, this decreasing drift may mean that the QPO was becoming a fixed period at the later epochs.
\begin{figure}
\begin{center}
\includegraphics[width=90mm]{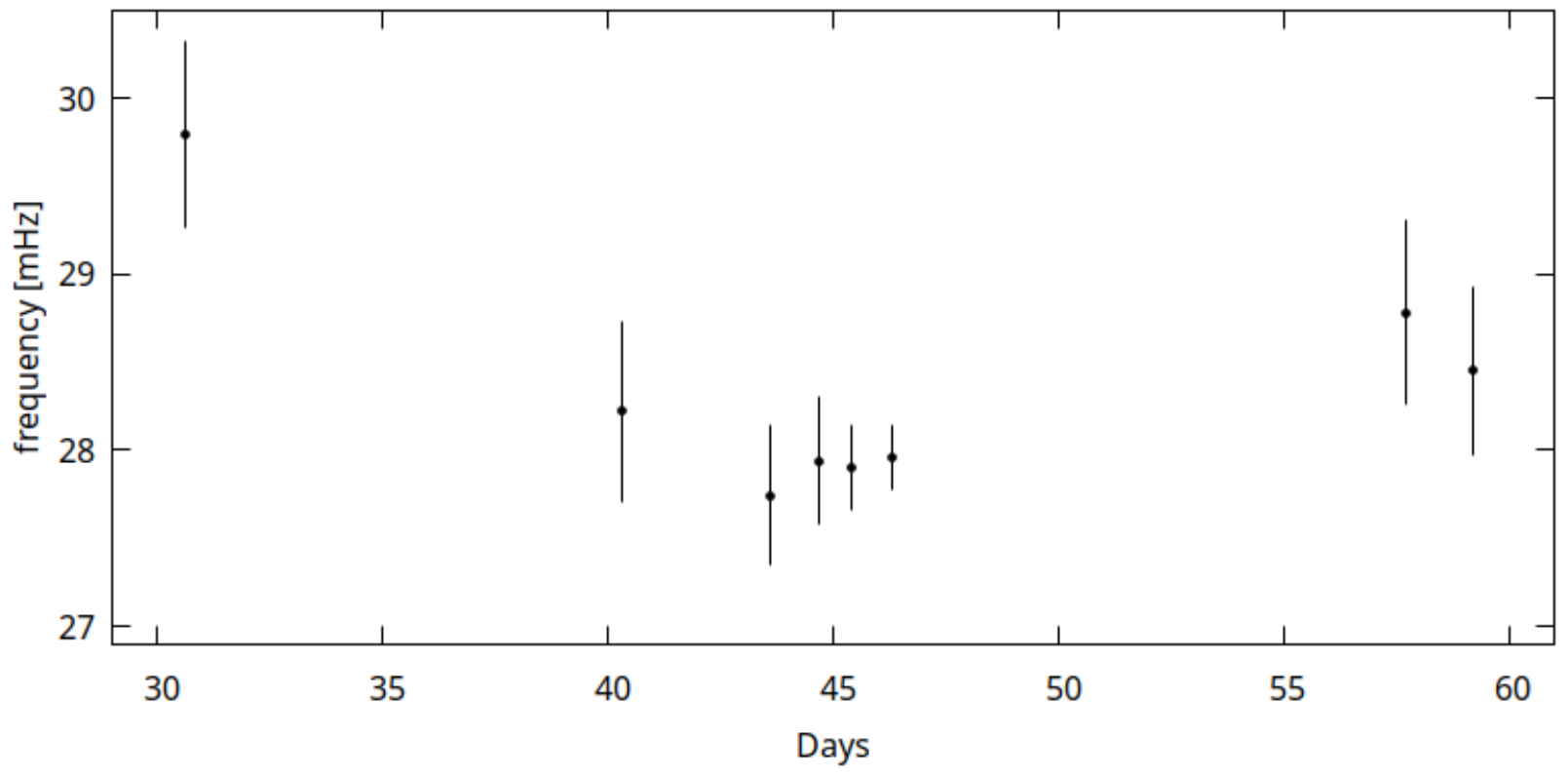}
\includegraphics[width=90mm]{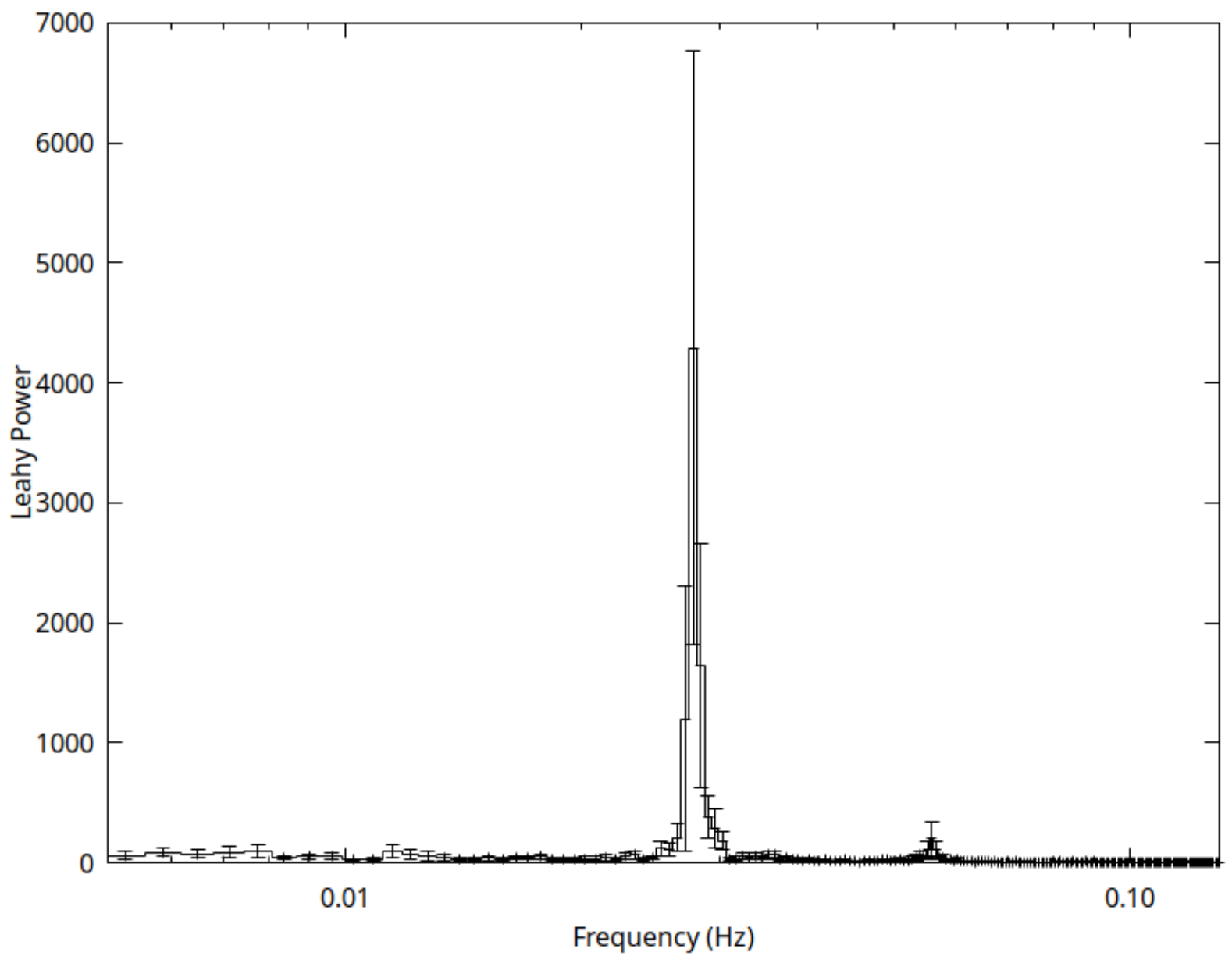}
\includegraphics[width=90mm]{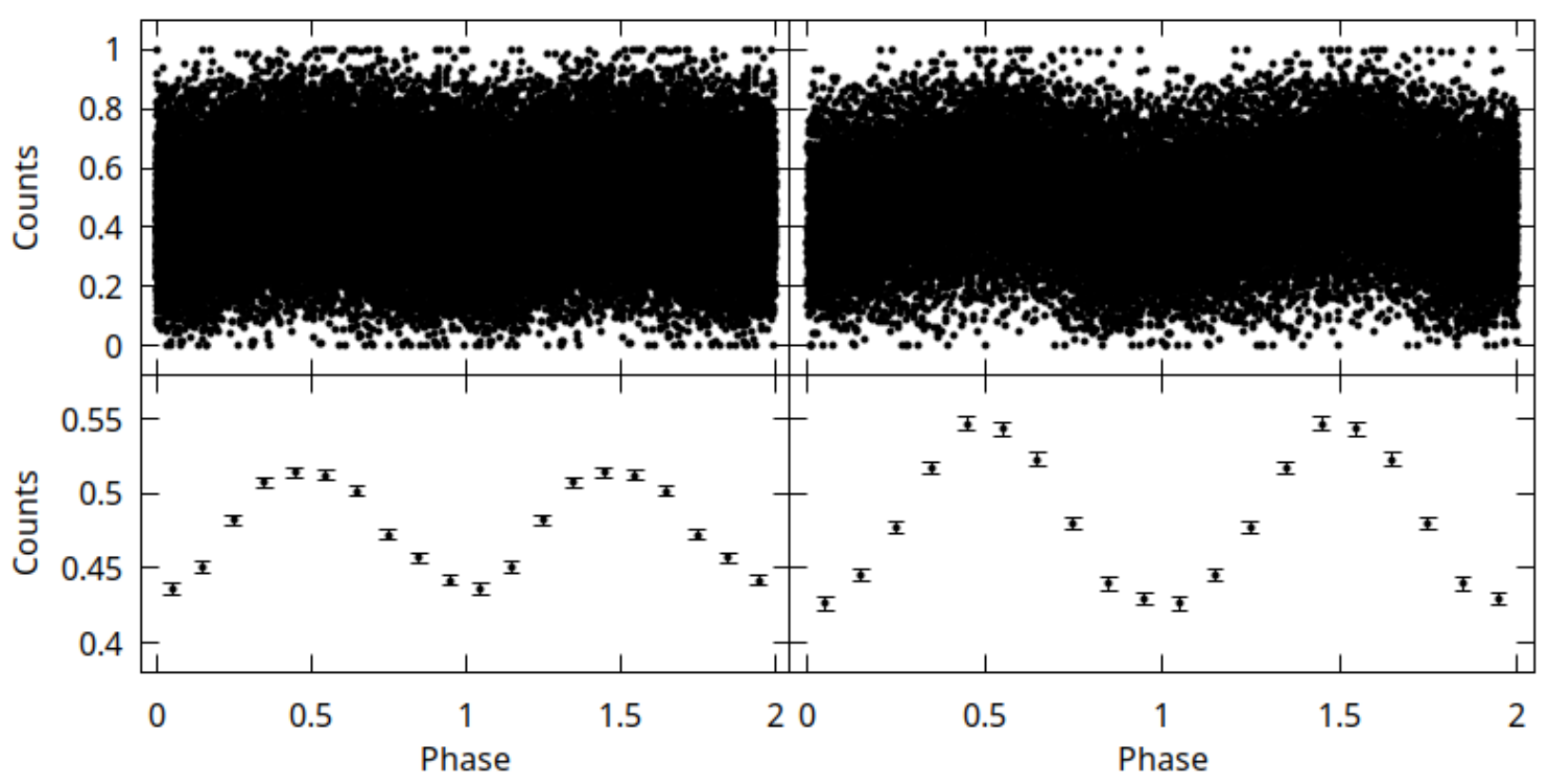}
\end{center}
\caption{
The top panel compares the frequency measured for the 8  GTIs without significant
 period drift. The middle
panel shows the periodogram measured for these 8 GTIs, yielding a period of 35.87 s (see text). The bottom panel shows the
phased light curves, with a 1 s bin, of the drifting-period ``red GTIs on the left, and that of the 8 stable-period ``blue'' GTIs on
the right. The data points were folded in the bottom panel with the 35.08 s period and with the 35.87 s periods, respectively. The count rate was
normalized by the highest measured, and below we binned the data in bins of 0.1 times the phase
(10 per cycle) and plotted the average count rate in each bin.
}
\end{figure}
\section{The final decline}
 The around-peak count rate was measured between day 45 and 59, still
 with significant variability, and after day 55 it decreased. 
 Daily observations had to be interrupted twice for technical reasons,
 once between days 51 and 57 (September 29 and October 5) and again
 between days  64 and 67 (October 12 and October 15). Although on day
 51 a new moderate rise was measured, after day 67 there was a
 constant decrease with lower irregular variability amplitude. 
 
 As the angular distance to the Sun decreased ($\leq 60^{\rm o}$),
 there was
 significant optical loading in the softer energies. After 
 day 67 we had to disregard the 0.2-0.3 range, and
 towards the end of October the optical loading became more and more
 severe with spurious flux leaking up to 0.4-0.5 keV. The exposure
 ceased early in November, with the last one on November 6 (day 89). 
 Thus, like the optical monitoring, X-ray monitoring
 had to end sooner in 2006 than in 2021 for the Sun constraint.
 
 As the luminous supersoft X-ray source was fading, 
emission lines were still clearly detected with
{\sl NICER}, with a spectrum indicating decreasing plasma
 temperature.
 As the SSS was becoming less and less luminous, like
 in the initial phases of the rise, discriminating between 
 the contribution of
 an atmosphere and a luminous component of very soft plasma was not possible,
 also because we could not resolve and measure
  emission lines in the softest range above the still strong
 continuum.
 While on day 67 in 2021 the {\sl NICER} estimated flux was
 only 2.15 $\times 10^{-10}$ erg cm$^{-2}$ s$^{-1}$, on day 67
 of the 2006 outburst (2006 April 20) the {\it measured} X-ray flux
 with the {\sl Chandra} LETG grating was still 
 3.8 $\times 10^{-9}$ erg cm$^{-2}$ s$^{-1}$, a factor of 47 higher
\citep{Nelson2008}.
 Our model fits to the {\sl NICER} data (see Table 2) 
 show that this very large difference is not explained with larger
 column density in 2021. Indeed, the absolute flux must have been much lower
 at this stage in 2021 than in 2006.
 
 Two models for October 28 (day 80) are shown in Fig. 23
We constrained the column density value to be at or above
 2 $\times 10^{21}$ cm$^{-2}$, and thus the fit to the softest
 portion (0.3-0.5 keV) is not perfect, but we attribute the discrepancy
 mainly to the optical loading contamination.  
 Another imperfection of the fit may be due to strong  He-like
 lines of Na X (the Na abundance
 is not included vary in the XSPEC BVAPEC model).
 However, both fits are statistically reasonably good, with $\chi$/d.o.f=1.5.
 At a distance of 2.4 kpc distance.
 the thermal plasma would have a very large absolute luminosity of 
6.1 $\times 10^{35}$ erg s$^{-1}$, close to the value obtained for
 the early days of
 the outburst, when the emission was much ``harder''. This seems
 hardly consistent with the previously observed flux decrease
 of the cooling shocked plasma, and it would imply considerable renewed
  mass ejection should have occurred
 long after after the initial outburst, which seems unlikely,
 also because there are no signatures of such an event in the optical
 spectra,

 On the other hand, the atmospheric component would
 emit  2.3  $\times 10^{35}$ erg s$^{-1}$, but the 
  effective temperature resulting from the fit is 564,000 K.
 With this temperature,
 most of the WD bolometric luminosity is emitted  the X-rays, but 
 the radius of  the emitting region with
 this luminosity would be only 5.7 $\times 10^7$ cm, only
 a small fraction of a WD radius.  Recent observational estimates of
 WD radii are found in \citet{Bedard2017}: for  instance, the  
 radius of a WD of 1.13 M$_\odot$
 is about 5.3 $\times 10^8$ cm$^{-2}$}. The WD of RS Oph
 may be even more massive, but it is very unlikely that 
 its radius is so small.  This may imply that we are not seeing
 the whole area of the WD emitting as an SSS, either because
 of new, large clumps that are opaque to soft X-rays, which
 seems unlikely at this stage, or because only part of the surface 
 is emitting as an SSS. 
 
 Above $\simeq$ 1 keV, where the SSS continuum is not dominant and
 the spectral resolution of {\sl NICER} is quite good,
 the spectrum still shows prominent emission lines. 
 There is evidence that in 2006, the emission due to shocks
 faded much later than the SSS.  In fact, 
 A late-epoch  high resolution X-ray spectrum  measured
 with the {\sl Chandra} LETG and HRC-S camera on 2006 June 4  (day 104 of the
 outburst) \citep{Nelson2008} is shown in Fig. 24. The figure also shows a fit,
 that constrains the column density to be  N(H)$\leq 2 \times 10^{21}$
 cm$^{-2}$. Two components of thermal
 plasma in equilibrium are assumed, at $\simeq$240 and  $\simeq$630 eV 
 (the fit with two regions explains most, but not
 all emission lines, however we did not want
 to introduce many free parameters). It is clear that  
 there was still a considerable region of 
 shocked material even at such post-outburst epochs in 2006,
 so the emission of the shocks faded much later than the SSS.
\section{Discussion}
 We summarize and discuss in this section some important points
 resulting from our data analysis.

 $\bullet$ The rise of the X-ray light curve in the 0.2-10 keV range
due to the optically thin thermal emission of shock-heated plasma lasted 
 until the 5th post-outburst day, 
 and a slow decline followed until the beginning of a ``mixed''
 phase in the fourth week, when supersoft flux started to emerge. 
 Our spectral fits imply that the peak of the plasma temperature 
 was close to 27 keV already at the end of the second day,
as the unabsorbed and absorbed flux were still rising. 

 $\bullet$
 In the first week, the spectrum can always be fitted with temperature
 higher than 5 keV. The total flux in He-like features
 is quite higher than that in the H-like lines,
 while  at such high temperature
 the emitting plasma should be almost completely ionized if it is
 in CIE. We experimented with several
 combination of models and found that we cannot fit the early
 X-ray spectra with two or more regions in CIE. We suggest that 
 the plasma was not in equilibrium
 until at least day 5. Between days 6 and 12, we could fit
 the spectra equally well with a combination of two thermal plasma regions
 at different temperature, either in CIE or NEI, but we
 find the BVAPEC model more likely to be realistic,
 because the NEI assumption requires a very high column density parameter,
so  the absolute flux turns out to be 
 super-Eddington. Moreover, we obtain inconsistent
 values for the electron density. We also found that until day 12, the best fits  are
 obtained assuming also a partially
 covering absorber with large column density, covering from 70\%
 to 40\% of the surface, slowly decreasing in time. 

 $\bullet$
 Towards the end of the third week, three components are necessary
 for spectral fits.
 Two of these
components are still modeled with the BVAPEC thermal plasma in CIE. 
In the following two weeks, the emerging SSS flux can be fitted equally well
 by adding a third component, which may be stellar
(blackbody or atmospheric model), or a thermal plasma at temperature kT$\leq$100 eV.
 In the initial SSS  rise, either we did 
 not observe emission from all the WD surface, or only 
 new layers of less hot, either shocked or photoionized ejected material 
 near the WD were becoming visible before the WD itself.
 Our data do not have the spectral resolution
 to distinguish between the two cases. 

  $\bullet$ The large SSS variability at peak
 luminosity, with soft flares lasting 
 for a few hours, is not explained with varying column density.
 The best fits to the initial in-flare and out-of-fare spectra seem to imply
 that the SSS flux suddenly increased for short periods because a larger
 emission region became visible. 
 After a large rise around day 37, the circumstellar material became
 permanently optically thin for soft X-rays  
  and the soft X-ray flux was consistent with an origin from all
 the WD surface. However, aperiodic
 variability persisted. We refer to \citet{Ness2022a},
 who  compared a high resolution
 X-ray SSS spectrum of 2021 with two taken in 2006,  
 concluding that the SSS variability  is due to 
 changes in the ionization stage of the absorbing medium near 
 the source. 
 Variations in an O I absorption edge explain most spectral
 differences; they may be due to a inhomogeneous outflow from
 the nova. The {\sl NICER} data confirm that we cannot attribute 
 the flux variations to other phenomena, specifically not to 
 variations in blackbody/atmospheric temperature, changes
 in column density, or a dust scattering halo. 

$\bullet$ \citet{Page2022} suggested that 
  the {\sl Swift} XRT spectra indicate 
 that after day 60 were {\it ``hotter, with smaller effective radii and lower
 wind absorption''}.
 The model by \citet{Page2022} does not yield a good
 fit to the more structured {\sl NICER} spectra, so we
 cannot support this conclusion.
 On the other hand, the example of a comparison shown in Fig. 12 
of a {\sl NICER} spectrum and a pile-up corrected {\sl Swift} XRT one for  
 day 57 of the outburst demonstrates that the spectrum extracted from a smaller
 region than the whole PSF, omitting the PSF center, results in a fit 
 (either with an atmospheric or a blackbody model)  
 converging at higher column density and temperature than
 the fit the {\sl NICER} spectrum of the same day.
 Thus, the parameters derived from the {\sl Swift} XRT spectra in this
phase may be an artifact of the ``partial'' extraction region method,
 that is necessary to avoid pile-up, but
does not provide an estimate of the actual flux for the normalization
 of the adopted model.

$\bullet$ 
 The {\sl NICER} spectra indicate a general trend of 
 SSS cooling, albeit a moderate one, on day 67 after an interruption
 in the monitoring.   The decrease in luminosity
 does not appear in the model fits to be only due to cooling, 
 and we inferred that the SSS emitting region was 
 shrinking. 
 There is also no indication that the decline in luminosity was due to a sudden
 increase in column density, a possibility discussed - but not
 found to be supported by the data - by \cite{Page2022}. 
    We note that several novae, particularly magnetic ones (intermediate polars) 
 which may already have hotter polar regions at
 the onset of the outburst, have been found to fade
 at almost constant temperature in the X-rays: V2491 Cyg \citep{Page2012},
 V4743 Sgr \citep{Zemko2016}, V407 Lup \citep{Aydi2018}, V1674 Her 
\citep{Orio2022b}. 

 \citet{Page2022, Ness2022a} discussed how  the [Fe X] 
 coronal emission line detected in optical mimicked the light curve
 and the decay of the SSS in 2006 (but there are only sparse
 data for the actual trend of this line in 2006) and how  this
 may indicate the same cooling time and
 duration of  the central photoionizing source in
 2006 and 2021, despite the faster decay of the SSS.
 The line is due to a $^2P_{1/2} => ^2P_{3/2}$ transition of Fe$^{9+}$,
 a ion with a ionization potential of 233.6 eV. 
 It indicates
 high temperature and it can be either formed by photoionization very
 close to the central source, or by shock ionization anywhere in
 the ejecta.  The evolution of the [Fe X] line flux was accurately monitored 
 in 2021 \citep[see][]{Page2022} and does  not
  follow the SSS light curve of 2021. 
 \cite{Ness2022a} thus suggested that the SSS decline of the in 2021 may 
 {\sl not} have been more rapid than in 2006, and that part of the SSS emitting
 region may have been blocked from the line of sight at late epochs.
 The reasoning is that the [Fe X]
 line is due to photoionization by the central source in the ejecta, while 
 occulting blobs of matter obscured only the supersoft X-rays. The complex
 geometry that would be required for the SSS flux to be photoionizing
 (visible) ejecta that should be very close to it, without being 
 observable any more along the line of sight, is not discussed in the paper.
   We suggest instead a simpler interpretation: 
 still  shocked ejecta present
 during the SSS turn-off in 2021 (as well as even later in 2006)
 indicate shock ionization, rather than photoionization  also for the 
 [Fe X] line at this stage in the outburst. 
  We do not rule out a large contribution of photoionization in the
 formation of this line in earlier phases, 
 but we suggest that the [Fe X] line is not due photoionization by a still
 hot central source (due to hidden, ongoing nuclear burning) during the late decline.

 We concluded that the SSS did actually turn off after day 60.
 In Fig. 24, we supported our conclusion by presenting the {\sl Chandra} HRC-S+LETG 
 high resolution spectrum of RS Oph on day 114 of the outburst, 2006 June 6
 \citep[see also][]{Nelson2008}. At the same epoch in the outburst, RS Oph
 was close to the Sun in 2021 and could not be observed.
 Several emission lines in the X-ray range were still consistent with
 being due to shock ionization and we suggest that the [Fe X] line most likely
 originated in the same medium.
 Furthermore, still as late as on day 201 in the 1985 
 outburst, when the [Fe X] line had already disappeared, many
 other lines in the optical range were still due to shock ionization
 \citep{Contini1995}.

On day 67 of the 2021 outburst, October 15, the absorbed flux
  obtained in our fits is about 2.5 $\times 10^{-10}$ erg cm$^{-2}$ s$^{-1}$,
  while on day 67 of 2006 (April 20 2006) the {\it measured}
 flux with the LETG {\sl Chandra} X-ray grating was 
 still 3.8 $\times 10^{-9}$ erg cm$^{-2}$ s$^{-1}$ \citep{Nelson2008}, 
 a factor of 15 higher than at
 earlier epoch in 2006, too large a difference to be explained 
 with partial obscuration by an opaque cloud or clump. 
  Furthermore, even if we did not achieve rigorous spectral fits
 for the SSS in the most luminous period, 
 the ``qualitatively best'' fits are consistent 
 with cooling  after day 67. 
 We could not prolong the observations 
 during complete turn off of the SSS, as it was possible back in 2006,
 but the constant decrease rate in SSS flux and the large difference with respect
 to the 2006 flux at the same post-outburst stage 
 indicate that,
 when our {\sl NICER} observations ended at day 89, the SSS was already
 switching off and  
 the SSS phase was quite shorter in 2021 than in 2006.

$\bullet$ We would like to compare RS Oph with the "sister system" V3890 Sgr.
 V3890 Sgr is a symbiotic at a distance of 4.2 kpc, with a later spectral type
 M giant than RS Oph, M5 III \citep{Anupama1999}. Its
recurrent nova outbursts were observed in 1962, 1990 and 2019.
 The main known differences from RS Oph are the
longer recurrence time,  about 29 years, and the more rapid development and
 turn-off in X-rays, despite an optical
light curve that was very similar to RS Oph. Given the shorter recurrence time
 of RS Oph, we may expect it had a 
larger average mass accretion rate. The X-ray spectrum of the shocked
 plasma of V3890 Sgr on day 7 was well fitted
with two thermal components in collisional ionization equilibrium at temperatures of
 1 keV and 4 keV \citep{Orio2020}, close to values obtained for RS Oph around day 10-11.
 The SSS in V3890 Sgr rose to maximum already on
the 9th day and declined within 26 days (Singh et al. 2021; Ness et al. 2022b);
 the effective temperature was between
820,000 K and 950,000 K (Singh et al. 2021). Thus the SSS was shorter lived and 
probably hotter than in RS Oph, and the models predict that these parameters
 indicate an even higher mass WD than the (already high) mass of the
RS Oph WD. Both should be quite close to the Chandrasekhar mass (Yaron et al. 2005)
 \begin{figure}
\begin{center}
\includegraphics[width=88mm]{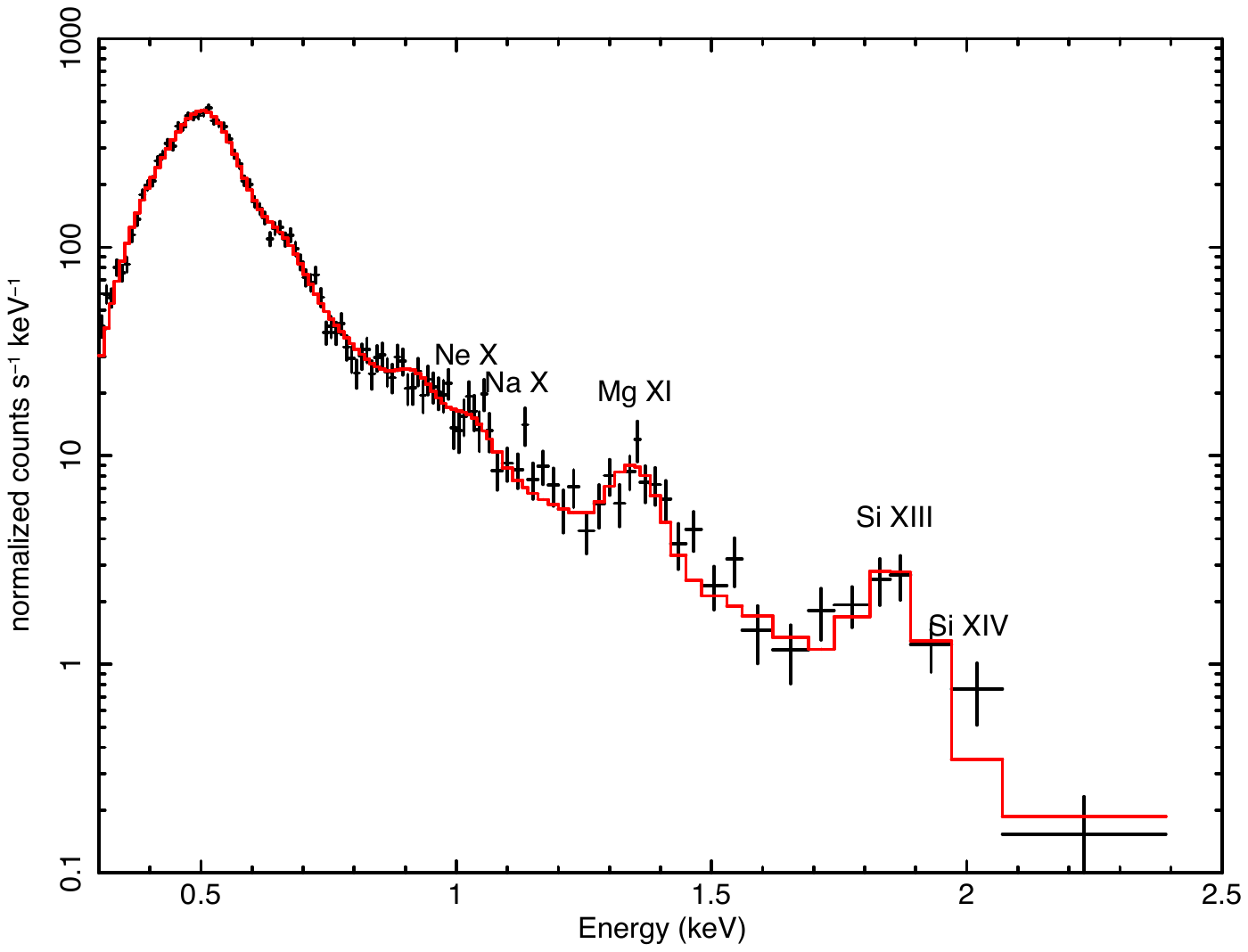}
\includegraphics[width=88mm]{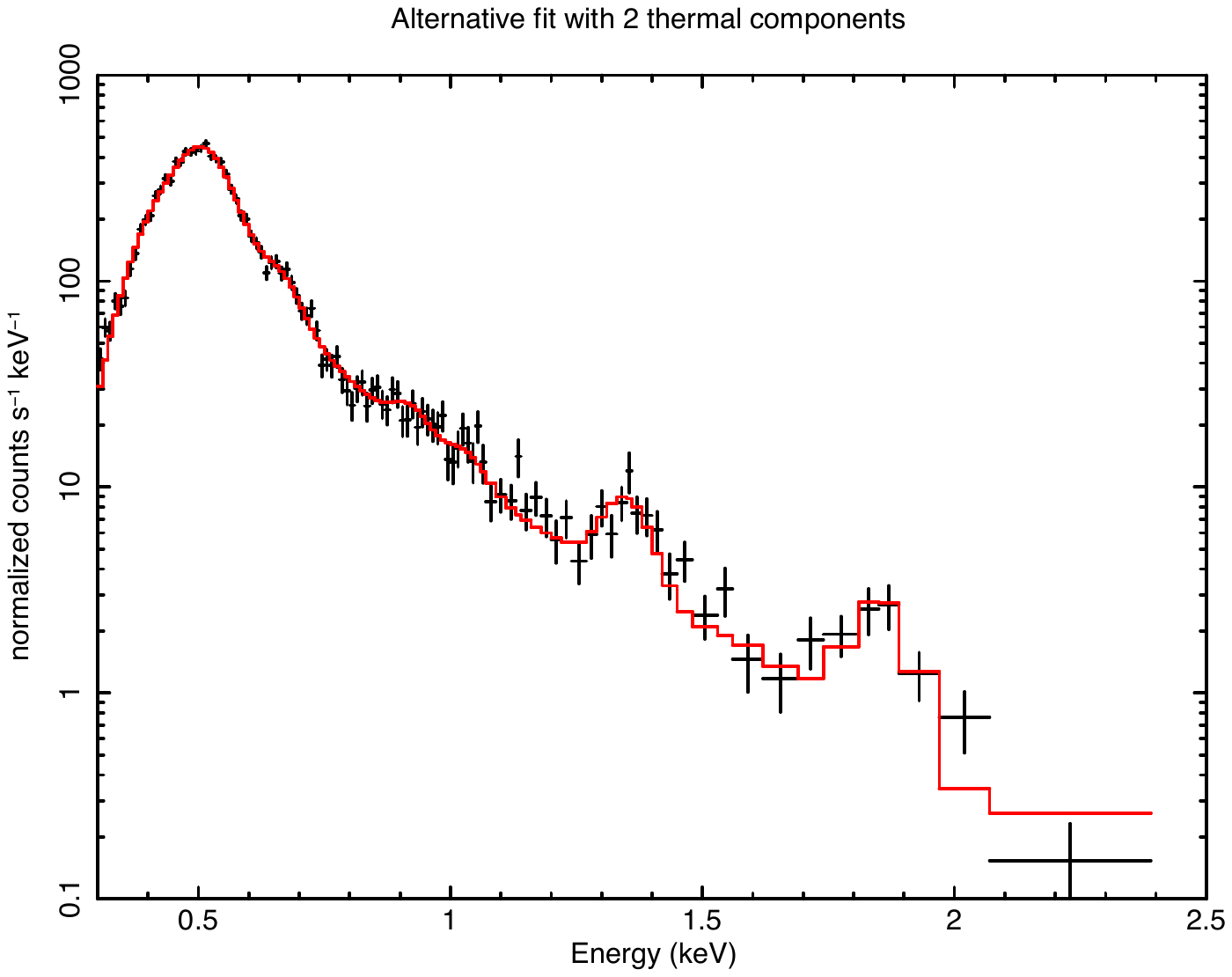}
 \end{center}
\caption{The average spectrum observed on 2021 October 28, and the fit with two
composite model: a stellar atmosphere and a BVAPEC thermal component (left),
 and two BVAPEC thermal components (left). Both fits yield $\chi^2$/d.o.f.=1.4,
 and are virtually indistinguishable.}
\end{figure}
\begin{figure}
\begin{center}
\includegraphics[width=149mm]{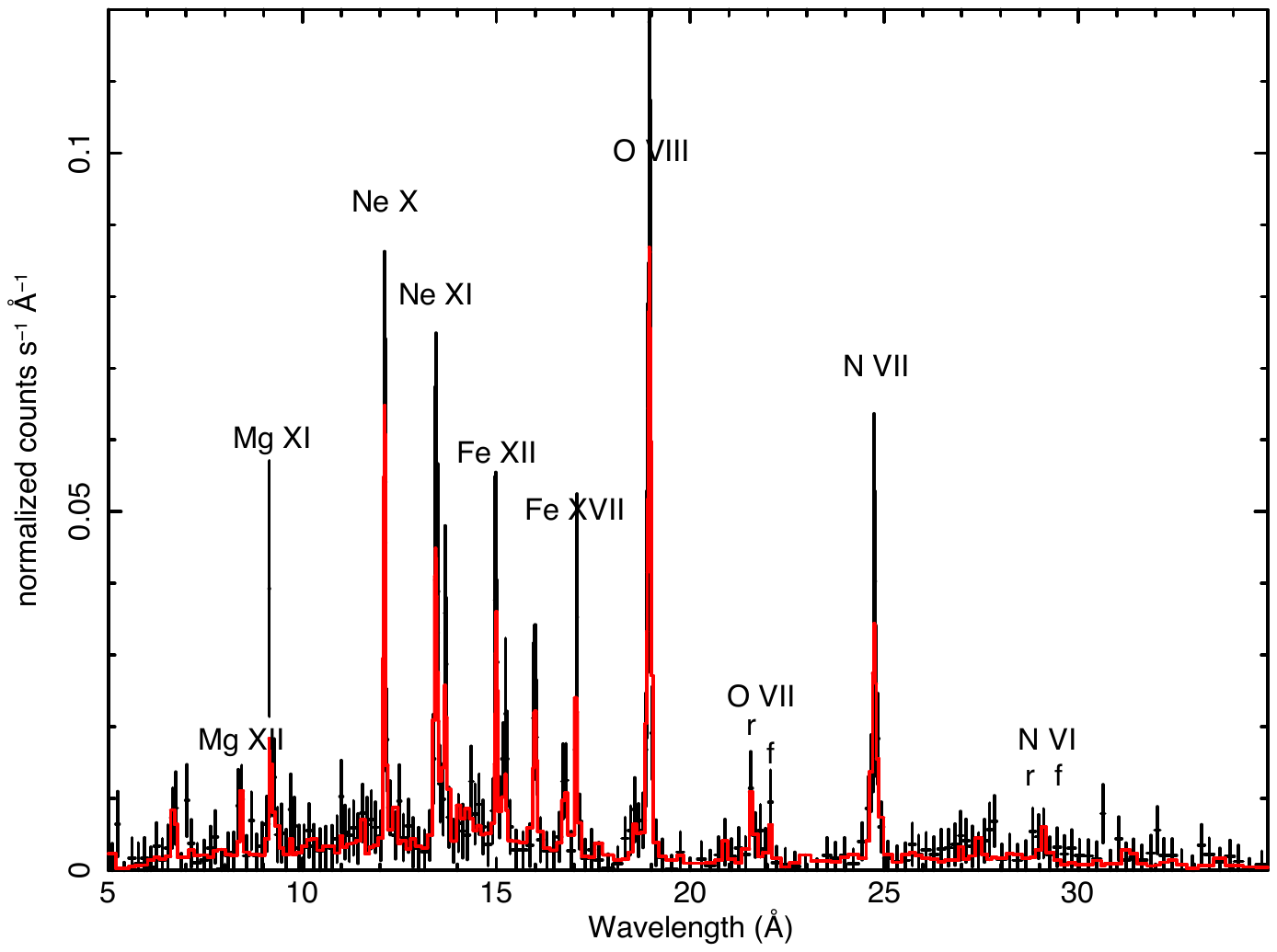}
 \end{center}
\caption{The spectrum observed with the Chandra LETG grating and the HRC-S camera
 on 2006 June 6, and a fit with two components at thermal plasma at 240 eV
 and 630 eV, with N(H)=2.2 $\times 10^{21}$ cm$^{-2}$ and moderately
 enhanced 
 abundances except for iron (which is only 15\% the solar abundance).}
\end{figure}
\section{Conclusions}
RS Oph was monitored with {\sl NICER} in its
 2021 nova outburst from the second day of the outburst until day 89, when it was too close to the Sun.
Monitoring occurred in $\simeq$1 ks exposures almost every
 day, and often several times a day, 
producing an exquisite dataset of high quality.
Because this was the second outburst of this recurrent nova monitored
 in X-rays, we can also learn much from the comparison between outbursts.
 Moreover, other observations were done at all wavelengths, from
 high energy gamma rays to radio.
 
{\bf 1. With RS Oph, we obtained for the first time simultaneous
 data with high S/N in both X-rays and gamma rays.} 
In the first three weeks, 
 the X-ray rise to maximum took about four days, while in optical and in 
 gamma rays in the Fermi range \citep[peaking around 1 GeV, see][]{Cheung2022}
the peak was observed at the end of the second day.  
 The AAVSO light curve shows an extremely rapid rise from 
optical magnitude V$\simeq$9 to  almost V$\simeq$5 
 magnitude within few hours, and a subsequent rise to V$\simeq$4.2 within
 the next  day.
 The flux observed with the H.E.S.S. Cherenkov telescope, on the other hand,
 reached maximum on the fourth day, with peak energy
about 100 GeV \citep{HESS2022}.   
 If we consider the shock temperature instead
 of the X-ray flux, the best model fits indicate 
that the plasma temperature was already peaking at $\simeq$27 keV  
 at the end of day 2.  

The gamma rays in novae have been attributed to
either inverse Compton effect or proton acceleration following the
 powerful shocks  between
 different ejecta, or between the ejecta and the circumstellar environment. 
 Typical nova outflow velocities are such that the regions
 of the shocks emit X-rays. 
 It is reasonable to assume radiative shocks \citep[see][]{Diesing2022},
 in which the velocity of the shock speed is proportional to
 the temperature as 
 kT $\simeq$ 1.2 (v/1000 km s$^{-1})^2$ \citep{Mabey2020}.
 The maximum temperature of 27 keV that we obtained,
 corresponds to a shock velocity around 4750 km s$^{-1}$. 
 \citet[][]{Diesing2022},
 among others, suggested that  in symbiotic novae like RS Oph with dense
 pre-existing red giant wind are likely
  to be due to proton accelerations. This
 is the so-called ``hadronic'' mechanism in which protons collide
 with neutrons or ions, and pions are generated. The neutral
 pions decay, producing gamma rays. 
 Secondary X-ray emission, namely the X-ray tails of the phenomena emitting gamma rays,
 is predicted with a non-thermal component, but 
 \citet{Vurm2018} find that only a small fraction,
 of order 10$^{-4}-10^{-3}$ 
  the gamma ray luminosity is radiated in X-rays above 10 keV
 (and even much less is radiated below 10 keV). 
\citet{Nelson2019}, among others, pointed out that all X-rays emission
 (primary and secondary) must
 have been in a very absorbing medium because very little
 or no X-ray flux was detected simultaneously  with the gamma rays
 before this recent  outburst of RS Oph.

\citet{Diesing2022} found that the Fermi and H.E.S.S. light curves
 are not compatible with an origin in the same shock, and that two
 or more shocks must have occurred. Given the time scales, the X-rays
 we measured with {\sl NICER} below 12 keV are more
 compatible with H.E.S.S.  We notice that the flux ratio 
of the maximum {\sl NICER} flux  to the peak
 H.E.S.S. flux (integrated from about 250 GeV and 2.5 TeV)
 is about 100,  and the ratio of soft X-rays to 
 the Fermi flux 
 (integrated between 60 MeV and 500 GeV)  is almost an order
 of magnitude higher than the flux measured with {\sl NICER}.
However, there is no signature for RS Oph of the predicted, accompanying
 non-thermal emission in
the {\sl NICER} and {\sl NuSTAR} ranges \citep[see][for the latter,]{Luna2021} namely the
 ``tail'' of the gamma-ray flux at
lower energy. This
implies that, despite the high thermal X-ray flux compared to the gamma-ray
flux, the lower limit ratio of non-thermal X-rays to gamma rays flux (measured 
in the Fermi energy range) is orders of magnitude lower than indicated by upper limits
 and detections measured for many classical novae
 \citep{Metzger2015, Nelson2018a, Nelson2018b,
 Nelson2019, Sokolovsky2023} and even less than the factor of 10$^{-4}-10^{-3}$ 
 expected for the secondary  emission above 3 keV.  
 It is thus possible  that the X-rays we observed were
 altogether from a completely different region than the
 original shocks that caused particle acceleration.

 {\bf 2. Modeling the shocks observed in X--rays, their 
 the nature and sites 
 is critically dependent on accurate estimates of the plasma temperature.} 
  For most GTIs in the first 5.5 days, we find a good fit only by modeling
 the spectrum as a thermal plasma that has not yet reached
 CIE. We obtain fits that are compatible with the ones to the {\it SWIFT} 
XRT spectra, requiring two thermal components, only from day 6
 and can completely rule out the NEI models only from day 12. 
  However, we caution that the NEI models yield values
of the electron density obtained from the ionization time scale, that are not
 consistent with the lower
limits given by the emission measures, even assuming the
 extremely high velocity inferred in the radio
in the first days (the optical spectra indicate a lower peak velocity, 
around 2800 km s$^{-1}$).
 With {\sl Swift} the emission lines were either not resolved, or resolved
 with a small S/N ratio,  so CIE
 models yielded a good fit
 from the beginning,  but {\sl NICER} presents a more complex picture. 
 
 {\bf 3. It is not surprising that  the spectra of the shocked gas cannot be
 fitted with only one region at uniform temperature}
already from day 2, because two or more plasma temperatures have often been found
 necessary to fit the early X-ray spectra of both recurrent and classical novae
\citep{Drake2016, Peretz2016, Orio2020}. For classical novae, the shocks are thought to be due to colliding winds from the nova at different velocity; perhaps at
the intersection of a fast polar outflow with a slower expanding torus of material around the WD \citep[e.g.][]{Chomiuk2021}, while for symbiotic novae the shock has been mainly attributed to the impact of the fast nova wind with the
less dense and much slower pre-existing red giant wind, especially in the equatorial plane 
\citep[see][]{Orlando2009}.
However, in RS Oph the scenario may be even much more complex, with multiple shock sites. In symbiotics, in fact,
shocks may also occur near the red giant, which subtends a large angle, 
and/or in the impact with the accretion disk (when present). 
\citet{Diesing2022} invoke two different shocked region to explain the gamma rays of RS Oph in
different energy ranges, that of Fermi and that of the Cherenkov telescopes.
 At the same time, as outlined above, the
absence of even a low-flux non-thermal component in the 3-79 keV range of {\sl NuSTAR},
 namely the ``low energy tail''
of the flux measured in the gamma rays exposures (done while high gamma ray flux was still detected Luna et al.
2021), suggests that the shocks we observed with {\sl NICER} may even not be related to those that caused the gamma ray
emission. It is also very interesting to note that, by examining the coronal lines of [Fe X], [Fe XIV] and [Ca XV] in
published optical spectra of RS Oph in the outbursts of 1958 and 1969, 
Cassinelli (private communication) found that
the ratios of the fluxes in these forbidden lines are not consistent with each other, indicating out at an origin in at
least two different region of shocks, with material at different velocity
 and temperature.

{\bf 4.  Towards the end of the third week a
 third plasma component is necessary} for the fit, and the spectrum became
 difficult to interpret as  the supersoft flux increased. For a period of
 at least 10 days, fits with a blackbody (or a WD atmosphere) as third
 component are equally good as the fits with a third thermal component in 
 the softest range, around 80-90 eV. This outlines the possibility
 of confusion and related uncertainty in determining an exact SSS turn-on
 time with broad-band X-ray spectra.   
 X-ray grating spectra of RS Oph published by \citet{Orio2022a}
 show that supersoft flux on day 21 was due to an emission line
 spectrum. On day 26(September 4), the 35 s pulsation
 appeared, probably indicating that at this stage the stellar surface 
 became visible, at
 least partially. We thus conclude that the ejecta became optically
 thin to the WD emission between these two post-outburst dates,
day 21 and 26 (August 30 and September 4).  
 The general assumption is that the source of the supersoft
 flux is the WD atmosphere,
 or layers being detached from it, but still very close to
 the WD \citep[e.g.][]{Rauch2010}. The material is extremely hot because nuclear
 burning is still ongoing at the bottom of the accreted envelope,
 with only a thin layer above it.  The time it takes
 for the SSS to become detectable may be interesting quantity because,
 at least in an ``ideal world'' in which the ejecta
 are spherically symmetric and are outflowing with constant velocity, 
 is  proportional to the mass of the ejecta, which cause 
 intrinsic absorption before they expand enough to  be transparent to the SSS.

 {\bf 5. The turn-off time is probably an even more important quantity,}
 because the models predict that it 
 should be directly proportional to the leftover hydrogen mass, which
 is roughly directly proportional to the initially accreted envelope. 
 Moreover, the models also predict that WD mass is inversely
 proportional to the effective temperature, because a thinner layer is
 left on a more massive WDs \citep{Yaron2005, Starrfield2012, Wolf2013}. 
  We find that the blackbody temperature for the SSS fits with variable abundance
 in the interstellar medium is in the 35-40 eV 
 ($\approx$ 405,000-470,000 K) range \citep[consistently also
 with {\sl Swift}, for which of ad-hoc absorption edges were superimposed
 on the blackbody][]{Page2022}.  According to \citet{Wolf2013}
 this is  the peak temperature of a 1 M$_\odot$ WD after the outburst.  
 However, the effective temperature of the WD atmosphere
  is quite higher than a blackbody of
 with the same X-ray luminosity \citep[see][]{Heise1994, Starrfield2012,
 Wolf2013}; although we could not
 obtain a rigorous fit with the (limited) public model grid,   
  the fits tend to converge at about $\simeq$750,000 K
 ($\simeq$65 eV), predicted
 to be the peak temperature of a 1.2  M$\odot$ WD \citep{Wolf2013}.

 We also established that between days 75 and 89 the SSS was
 already cooling off
 and its flux was rapidly decreasing. We presented evidence that this
 has to be interpreted as a real switching-off of nuclear burning,
 rather than a temporary obscuration by increased, large column density
 \citep[a possibility hypothesized by][]{Page2022, Ness2022a}.  
 A turn-off time around 85 days in the models is associated with 
 a WD mass close to 1.3 M$_\odot$ \citep{Wolf2013}. 
After day day 67, we also found evidence that  
 the cooling was accompanied by a shrinking of
 the emitting region. At this stage, we did not observe irregular
 variability any more, and there was only a constant decrease in
 X-ray luminosity. The fact that shocked plasma was still detected until
 the last day (day 89) and was present even later in outburst phase in 2006,
 points out at the [Fe X] coronal emission line in the optical region
 as having its origin - at least in the late outburst phase - in shocked
  ejecta, instead 
 of being an indicator of a central photoionizing source. 

{\bf 6. Another example of very similar optical light curves, yet with quite
 a different development in X-rays, is the one of V3890 Sgr}, described in Section 10.
 Therefore, it is not entirely surprising that the X-ray
lightcurves of RS Oph in 2006 and in 2021 were different.
 The previous RS Oph outburst occurred after 21 years, the recent one after 15:
 probably, the bottom shell of the WD envelope was sufficiently hot and 
electron-degenerate to allow the thermonuclear runaway with a smaller envelope mass.
 It is possible that recurrent novae explode with
lower and lower accreted mass as they evolve in time. In fact, we note that
 also the recurrent nova U Sco in its 2022
outburst had a more rapid SSS phase than in 2010 (Zhang et al. 2023,
 article in preparation).

The fact that the optical light curves of RS Oph in 2006 and 2021 were virtually ``clones'' of each other, while the
X-ray light curve shows a much shorter SSS phase, is extremely interesting because the SSS is the only indication
we have of thermonuclear burning. The mass ejected following the thermonuclear runaway in RS Oph may have been
quite smaller than in 2006, but the almost identical optical peak luminosity and decay rate indicate that there must
be another important mechanism powering the optical luminosity. This mechanism is likely to be the reprocessing of
the X-ray flux into optical light, inferred for nova V906 Car by 
\citet{Aydi2020} and previously also suggested for nova
 V1324 Sco \citet{Metzger2014}. Although detailed calculations are not
 available yet, \citet{Metzger2015} presented a phenomenological
 estimate, confirming this scenario. We suggest that also in RS Oph  
 powerful shocks, like those
we probed in X-rays but not necessarily in the same sites, are fundamental in shaping also the optical
characteristics and parameters. This implies that the optical decay rate (specifically, the time for optical decay by 2 and 3 magnitudes, t$_2$ and t$_3$) may not be
 indicative of the nova absolute luminosity (hence, its distance) and of the
WD mass as previously thought. Furthermore, mass loss via Roche Lobe overflow in the common envelope phase may
have contributed to the optical phenomenology, even if the accumulated envelope was smaller (possibly, because the
temperature on the WD surface increased after the previous outbursts), while the initial radiation driven wind carried away
a lower amount of mass in 2021 \citep[see nova mass loss models by][]{Shen2022}.

 {\bf 7. The $\simeq$35 s  quasi-periodic pulsation, identical
 in the 2006 and 2021 outbursts, must be an intrinsic property 
 and is most likely  related to the WD} and/or its shell thermonuclear burning.
 The coincidence of a QPO of exactly the same duration of $\simeq$35 s
 in another nova in the SSS phase \citep[KT Eri, see][]{Ness2015}
 is suggestive of a stellar pulsation.
 While in other X-ray data \citep[of RS Oph in 2006 and other novae,
 see][]{Ness2015}  the period of the SSS flux oscillations
does not appear to be clearly 
  detectable in the whole SSS phase, {\sl NICER} has demonstrated
 that although the amplitude varies (yielding
 a non-detection with less sensitive instruments), the semi-periodicity
 is always present as long as the SSS is luminous.
With a statistical analysis, we also found evidence that the
 period drift decreased toward the end
 of the SSS phase and the period tended to stability. 
 
   Finally, we would like to conclude pointing out
 the characteristics of {\sl NICER} that
 allowed gathering this beautiful
 and very useful data set:  high S/N without pile-up problems that make it very suitable to study the SSS, the possibility
 to study some emission line  ratios, and the timing capabilities.
 In the future, {\sl NICER} will continue to prove an invaluable
 instrument to examine the nova mysteries connected to fundamental
 physical processes.
%
\facilities{NICER, Swift, Chandra, ADS, HEASARC} 
\software{HEASOFT 6.31.1 \url{https://ascl.net/1408.004},
 CIAO v4.14.0 \citep{Fruscione2006}}
%



\begin{acknowledgements}
{\sl NICER} is a 0.2-12 keV X-ray telescope operating on the International Space Station. The {\sl NICER} mission and portions
of the {\sl NICER} science team activities are funded by NASA. This work made also use of data supplied by the UK {\sl Swift}
Science Data Centre at the University of Leicester and {\sl Chandra}
 public data in the HEASARC archive.
M. Orio thanks Jay Gallagher and Ehud Behar for many useful and
 interesting conversations. 
Dr. Ferrara is supported by NASA under award number 80GSFC17M0002. GJML is member of the CIC-CONICET (Argentina) and acknowledges support from grant ANPCYT-PICT 0901/2017.
AD and JM were supported by the Slovak grant VEGA 1/0408/20, and by the European Regional Development Fund, project No. ITMS2014+: 313011W085.
\end{acknowledgements}
\bibliography{rsoph2021ap.bib}{}
\bibliographystyle{aasjournal}
\end{document}